\input epsf
%\magnification=\magstep1
\documentstyle{amsppt}
%\pagewidth{5.4truein}\hcorrection{0.55in}
%\pageheight{7.5truein}\vcorrection{0.75in}
\pagewidth{6.4truein}\hcorrection{0in}
%\pageheight{8.1truein}\vcorrection{0.35in}
\pageheight{9truein}\vcorrection{0.00in}

%\pagewidth{5.8truein}\hcorrection{0.0in}
%\pageheight{8.8truein}\vcorrection{0.0in}
\TagsOnRight
\NoRunningHeads
%\NoBlackBoxes
\catcode`\@=11
\def\logo@{}
\footline={\ifnum\pageno>1 \hfil\folio\hfil\else\hfil\fi}
\topmatter
\title %Part B \\ \\  The correlation of collinear gaps on the square lattice under various circumstances
%The interaction of collinear gaps of arbitrary charge in a two dimensional dimer system, Part II
%Macroscopic gaps in dimer coverings of Aztec rectangles
Macroscopically separated gaps in dimer coverings of Aztec rectangles
%Long long-range correlation of gaps in dimer coverings of Aztec rectangles
%A finer analysis of the interaction of diagonal clusters in a two dimensional dimer system
%The interaction of collinear gaps of arbitrary charge in a two dimensional dimer system
\endtitle
\author Mihai Ciucu\endauthor
\thanks Research supported in part by NSF grants DMS-0801625 and DMS-1101670.
\endthanks
\affil
  Department of Mathematics, Indiana University\\
  Bloomington, Indiana 47405-5701
\endaffil
%\date   \enddate
\abstract 
In this paper we determine the interaction of diagonal defect clusters in regions of an Aztec rectangle that scale to arbitrary points on its symmetry axis (in earlier work we treated the case when this point was the center of the scaled Aztec rectangle). We use the resulting formulas to determine the asymptotics of the correlation of defects that are macroscopically separated from one another and feel the influence of the boundary. In several of the treated situations this seems not to be accomplishable by previous methods.
%These seem to be the first instances of this kind treated in the literature. 
Our applications include the case of two long neutral strings, which turn out to interact by an analog of the Casimir force, two families of neutral doublets that turn out to interact completely independently of one another, a neutral doublet and a very long neutral string, a general collection of macroscopically separated monomer and separation defects, and the case of long strings consisting of consecutive monomers. 
%and the asymptotics of probabilities of some special dimer configurations along the symmetry axis.

%in several situations where this seems not to be accomplishable by previous methods.

\endabstract 
\endtopmatter
\document

\def\mysec#1{\bigskip\centerline{\bf #1}\message{ * }\nopagebreak\par\bigskip}

\def\myref#1{\item"{[{\bf #1}]}"} 
 
\def\pf{{\it Proof.\ }} 

\def\epf{\hfill{$\square$}\smallpagebreak}

\def\cite#1{\relaxnext@
  \def\nextiii@##1,##2\end@{[{\bf##1},\,##2]}%
  \in@,{#1}\ifin@\def\next{\nextiii@#1\end@}\else
  \def\next{[{\bf#1}]}\fi\next}
\def\proclaimheadfont@{\smc}

\def\pf{{\it Proof.\ }}

\define\Z{{\Bbb Z}}

\define\R{{\Bbb R}}

\define\M{\operatorname{M}}

\define\q{\operatorname{q}}

\define\twoline#1#2{\line{\hfill{\smc #1}\hfill{\smc #2}\hfill}}
\define\threeline#1#2#3{\line{\hfill{\smc #1}\hfill{\smc #2}\hfill{\smc #3}\hfill}}
\define\ltwoline#1#2{\line{{\smc #1}{\smc #2}}}

\def\mypic#1{\epsffile{figs/#1}}

%\def\epsfsize#1#2{0.80#1}
%\topinsert
%\centerline{\mypic{1-2.eps}}
%\centerline{{\smc Figure~1.3.} {\rm $H_{(2,4),(3,5)}(7,8,3)$.}}
%\endinsert

%\topinsert
%\twoline{\mypic{1-0.eps}}{\mypic{1-0b.eps}}
%\twoline{Figure 1.1(b){\rm (a). The hexagon $H$ for $a=5$, $b=5$ and $k=3$.}}
%{Figure~1.1{\rm (b). A tiling of $H$.}}
%\endinsert

% ref nos
%  \define\Baikone{1}
  \define\Baiktwo{1}
%  \define\FT{3}
%  \define\ri{2} 
  \define\sc{2}
%  \define\ppone{6}
  \define\ec{3}
  \define\ef{4}
  \define\ov{5}
  \define\gd{6}
  \define\CEP{7}
%  \define\CKP{8}
  \define\FS{8}
  \define\MF{9}
%  \define\FisherIsing{12}
  \define\Glaish{10}
%  \define\Hart{14}
  \define\KOS{11}
  \define\Olver{12}
%  \define\SM{15}
%  \define\MRR{16}
%  \define\ZI{17}

% eq nos

\define\Eba{2.1}
\define\Ebb{2.2}
\define\Ebc{2.3}
\define\Ebd{2.4}
\define\Ebdd{2.5}
\define\Ebddd{2.6}
\define\Ebddda{2.7}
\define\Ebdddb{2.8}
\define\Ebdddc{2.9}
%\define\Ebdddd{2.10}
%\define\Ebddde{2.11}
\define\Ebe{2.10}
\define\Ebf{2.11}
\define\Ebg{2.12}
\define\Ebh{2.13}
\define\Ebi{2.14}
\define\Ebj{2.15}
\define\Ebk{2.16}
\define\Ebl{2.17}
\define\Ebm{2.18}
\define\Ebn{2.19}
\define\Ebo{2.20}
\define\Ebp{2.21}
\define\Ebpp{2.22}
\define\Ebq{2.23}
\define\Ebr{2.24}
\define\Ebs{2.25}
\define\Ebt{2.26}
\define\Ebu{2.27}
\define\Ebuu{2.28}
\define\Ebv{2.29}
\define\Ebw{2.30}
\define\Ebx{2.31}
\define\Eby{2.32}

\define\Eca{3.1}
\define\Ecb{3.2}
\define\Ecc{3.3}
%\define\Ecd{3.4}
\define\Ece{3.4}
\define\Ecee{3.5}
\define\Ecf{3.6}
\define\Ecg{3.7}
\define\Ech{3.8}
\define\Eci{3.9}
\define\Ecj{3.10}
\define\Eck{3.11}
\define\Ecl{3.12}
\define\Ecm{3.13}
\define\Ecn{3.14}
\define\Eco{3.15}
\define\Ecp{3.16}
%\define\Ecpp{3.17}
\define\Ecq{3.17}
\define\Ecr{3.18}
\define\Ecs{3.19}
\define\Ect{3.20}
\define\Ecu{3.21}
\define\Ecv{3.22}
\define\Ecw{3.23}
\define\Ecww{3.24}
\define\Ecx{3.25}
\define\Ecy{3.26}
\define\Ecz{3.27}
\define\Ecza{3.28}
\define\Eczb{3.29}
\define\Eczc{3.30}

\define\Eda{4.1}
\define\Edaa{4.2}
\define\Edaaa{4.3}
\define\Edb{4.4}
\define\Edc{4.5}
\define\Edd{4.6}
\define\Ede{4.7}
\define\Edf{4.8}
\define\Edg{4.9}

\define\Eea{5.1}
\define\Eeb{5.2}
\define\Eec{5.3}
\define\Eed{5.4}
\define\Eee{5.5}
\define\Eef{5.6}
\define\Eeg{5.7}
\define\Eeh{5.8}
\define\Eei{5.9}
\define\Eej{5.10}

\define\Efa{6.1}
\define\Efb{6.2}
\define\Efc{6.3}
\define\Efd{6.4}

\define\Ega{7.1}
\define\Egb{7.2}
\define\Egc{7.3}
\define\Egd{7.4}
\define\Ege{7.5}
\define\Egf{7.6}
\define\Egg{7.7}
\define\Egh{7.8}
\define\Egi{7.9}
\define\Egj{7.10}
\define\Egk{7.11}
\define\Egl{7.12}
\define\Egm{7.13}
\define\Egn{7.14}
\define\Ego{7.15}

\define\Eha{8.1}
\define\Ehb{8.2}
\define\Ehc{8.3}
\define\Ehd{8.4}
\define\Ehe{8.5}
\define\Ehf{8.6}
\define\Ehg{8.7}
\define\Ehh{8.8}
\define\Ehi{8.9}
\define\Ehj{8.10}
\define\Ehk{8.11}
\define\Ehl{8.12}
\define\Ehm{8.13}
\define\Ehn{8.14}
\define\Eho{8.15}
\define\Ehp{8.16}
\define\Ehq{8.17}
\define\Ehr{8.18}
\define\Ehs{8.19}
\define\Ehrr{8.20}
\define\Ehss{8.21}
\define\Eht{8.22}
\define\Ehu{8.23}
\define\Ehv{8.24}
\define\Ehw{8.25}
\define\Ehx{8.26}
\define\Ehy{8.27}
\define\Ehz{8.28}
\define\Ehza{8.29}
\define\Ehzb{8.30}
\define\Ehzc{8.31}
\define\Ehzd{8.32}
\define\Ehze{8.33}
\define\Ehzf{8.34}
%\define\Ehzg{8.33}
%\define\Ehzh{8.34}
\define\Ehzi{8.35}
\define\Ehzj{8.36}
\define\Ehzk{8.37}
\define\Ehzl{8.38}
\define\Ehzm{8.39}
\define\Ehzn{8.40}
\define\Ehzo{8.41}
\define\Ehzp{8.42}
\define\Ehzq{8.43}
\define\Ehzr{8.44}
\define\Ehzs{8.45}
\define\Ehzt{8.46}
\define\Ehzu{8.47}
\define\Ehzv{8.48}
\define\Ehzw{8.49}
\define\Ehzx{8.50}
\define\Ehzy{8.51}
\define\Ehzz{8.52}
\define\Ehzza{8.53}

\define\Eia{9.1}
\define\Eib{9.2}
\define\Eic{9.3}
\define\Eid{9.4}
\define\Eie{9.5}
\define\Eif{9.6}
\define\Eig{9.7}
\define\Eih{9.8}
\define\Eii{9.9}
\define\Eij{9.10}
\define\Eik{9.11}
\define\Eil{9.12}
\define\Eim{9.13}
\define\Ein{9.14}
\define\Eio{9.15}
\define\Eip{9.16}
\define\Eiq{9.17}
\define\Eir{9.18}
\define\Eis{9.19}
\define\Eit{9.20}
\define\Eiu{9.21}
\define\Eiv{9.22}
\define\Eiw{9.23}
\define\Eix{9.24}
\define\Eiy{9.25}
\define\Eiz{9.26}
\define\Eiza{9.27}
\define\Eizb{9.28}
\define\Eizc{9.29}
\define\Eizd{9.30}
\define\Eize{9.31}

% th nos

\define\Tba{2.1}
\define\Tbb{2.2}
\define\Tbc{2.3}
\define\Tbd{2.4}
\define\Tbe{2.5}
\define\Tbf{2.6}

\define\Tca{3.1}
\define\Tcb{3.2}
\define\Tcc{3.3}
\define\Tcd{3.4}
\define\Tce{3.5}

\define\Tda{4.1}

\define\Tea{5.1}
\define\Teb{5.2}

\define\Tfa{6.1}

\define\Tga{7.1}
\define\Tgb{7.2}
\define\Tgc{7.3}
\define\Tgd{7.4}

\define\Tha{8.1}
\define\Thb{8.2}
\define\Thc{8.3}
\define\Thd{8.4}
\define\The{8.5}
\define\Thf{8.6}
\define\Thg{8.7}
\define\Thh{8.8}
\define\Thi{8.9}
\define\Thj{8.10}
\define\Thk{8.11}
\define\Thl{8.12}
\define\Thm{8.13}
\define\Thn{8.14}

\define\Tia{9.1}
\define\Tib{9.2}
\define\Tic{9.3}
\define\Tid{9.4}
\define\Tie{9.5}
\define\Tif{9.6}

% fig nos

\define\fab{1.1}

\define\Fba{2.1}

\define\Fca{3.1}

\define\Fea{5.1}

\define\Fha{8.1}
\define\Fhb{8.2}
\define\Fhc{8.3}

\define\Fia{9.1}

% remark nos

\define\ra{1}
\define\rb{2}
%\define\rc{3}
\define\rd{3}
\define\re{4}
\define\rf{5}
\define\rg{6}
\define\rh{7}
\define\ri{8}
\define\rj{9}
\define\rk{10}
\define\rl{11}
\define\rmm{12}

\vskip-0.05in
\mysec{1. Introduction}

%The correlation of gaps in dimer systems was introduced in 1963 by Fisher and Stephenson \cite{\FS}, who looked at the interaction of two monomers generated by the rigid exclusion of dimers completely covering the rest of the square lattice. In previous work (see \cite{\ri}, \cite{\sc}, \cite{\ec}, \cite{\ef} and \cite{\ov}) we considered the analogous problem on the hexagonal lattice, and we extended the set-up to include the correlation of any finite number of monomer clusters. For fairly general classes of monomer clusters we proved that the asymptotics of their correlation is given, for large separations between the clusters, by a multiplicative version of Coulomb's law for 2D electrostatics. However, our previous results required that the monomer clusters consist (with possibly one exception, see \cite{\sc}) of an even number of monomers. 

Consider a $(2m+1)\times(2n+1)$ rectangular chessboard and suppose the corners are black. 
The {\it Aztec rectangle} $AR_{m,n}$ is the graph whose vertices are the white squares and whose edges connect precisely those pairs of white squares that are diagonally adjacent. 

Let $k,l \geq 0$ be integers, and consider the Aztec rectangle $AR_{2n,2n+k-l}$. Let $\ell$ be its horizontal symmetry axis, and label the vertices on $\ell$ from left to right by $1,2,\dotsc,2n+k-l$. Set $[m]:=\{1,2,\dotsc,m\}$. For any disjoint subsets $H,S\subseteq [2n+k-l]$ so that $|H|=k$ and $|S|=l$, define $AR_{2n,2n+k-l}(H,S)$ to be the graph obtained from $AR_{2n,2n+k-l}$ by deleting the vertices on $\ell$ whose labels belong to $H$, and creating separations\footnote{ A separation is a place where a vertex $v$ had ``split'' in two: one vertex above the symmetry axis $\ell$, incident to the neighbors of $v$ above $\ell$, and the other below the symmetry axis, incident to the neighbors of $v$ below $\ell$ (three such instances are shown in Figure {\fab}).} at those vertices on $\ell$ whose labels belong to $S$ (see Figure {\fab} for an example). Clearly, $AR_{2n,2n+k-l}(H,S)$ is bipartite (i.e., its vertices can be colored black and white so that each edge has oppositely colored endpoints), and one readily checks that it is also balanced (i.e., the two color classes have the same number vertices), an obvious necessary condition for it to have perfect matchings.

In the closely related paper \cite{\gd} we have seen that the number of perfect matchings of $AR_{2n,2n+k-l}(H,S)$ is given by a simple product formula (see \cite{\gd, Theorem 1.1}). Using this formula as our starting point, we showed that the correlation in the bulk of an arbitrary collection of defect clusters\footnote{ A defect cluster is an arbitrary finite union of monomers (unit holes, or missing vertices) and separations on $\ell$.}  along $\ell$ has asymptotics given by Coulomb's law for two dimensional electrostatics\footnote{In this analogy, monomers are regarded as having charge $+1$, and separations as having charge $-1$. The charge of a defect cluster is the sum of the charges of its constituent monomers and separations.}, thus confirming in this instance our general ``electrostatic hypothesis'' presented in \cite{\ov, Conjecture 1}.

In this paper we use the product formula provided by \cite{\gd, Theorem 1.1} to determine the asymptotics of the correlation of defect clusters in several situations away from the bulk and involving macroscopic defect clusters, results that seem not to be accomplishable by previous methods. 

In Section 2 we consider the
interaction of two families of neutral doublet defects, in the limit as their mutual distances and their distances from the boundary approach fixed ratios. The two families turn out to behave completely independently of each other, each doublet interacting only with members of its own family. Sections 3--5 consider the case of long neutral defect clusters, and present the asymptotics of their correlation under various circumstances. 
In Section 6 we consider the situation when we have two defect clusters, one being much longer than the other. Section 7 treats the case when the scaling limit of the defect clusters is taken so that they all shrink to a common point different than the origin.
In Section 8 we work out the asymptotics of an arbitrary family of monomers and separations, in the limit when their mutual distances and their distances from the boundary approach fixed ratios (this could be called the ``long long-range correlation,'' in a suggestive term that Michael Fisher attributes to Elliott Lieb \cite{\MF}). The interesting feature of this set-up is that in the scaling limit the portion of the Aztec rectangle around each defect is governed by its own Gibbs measure (cf. \cite{\KOS}), a case not treated previously in the literature. The effect of the boundary is also reflected in the formula we obtain. 
%Section D shows how the results on doublets can be used to compute the asymptotics of the probabilities of some dimer configurations along the symmetry axis of the Aztec diamond. Our calculations confirm the values predicted by a conjecture of Cohn, Kenyon and Propp (see \cite{\CKP, Conjecture 13.5}) for the probabilities of these dimer configurations.
Section 9 treats the case of a finite number of monomer clusters of macroscopic length, each consisting of a string of consecutive monomers, and shows that in the scaling limit there is a unique equilibrium position that the clusters tend to occupy, in accordance with physical intuition based on the parallel to electrostatics developed in our earlier work (see \cite{\sc}\cite{\ec}\cite{\ef}\cite{\ov}).

\topinsert
\centerline{\mypic{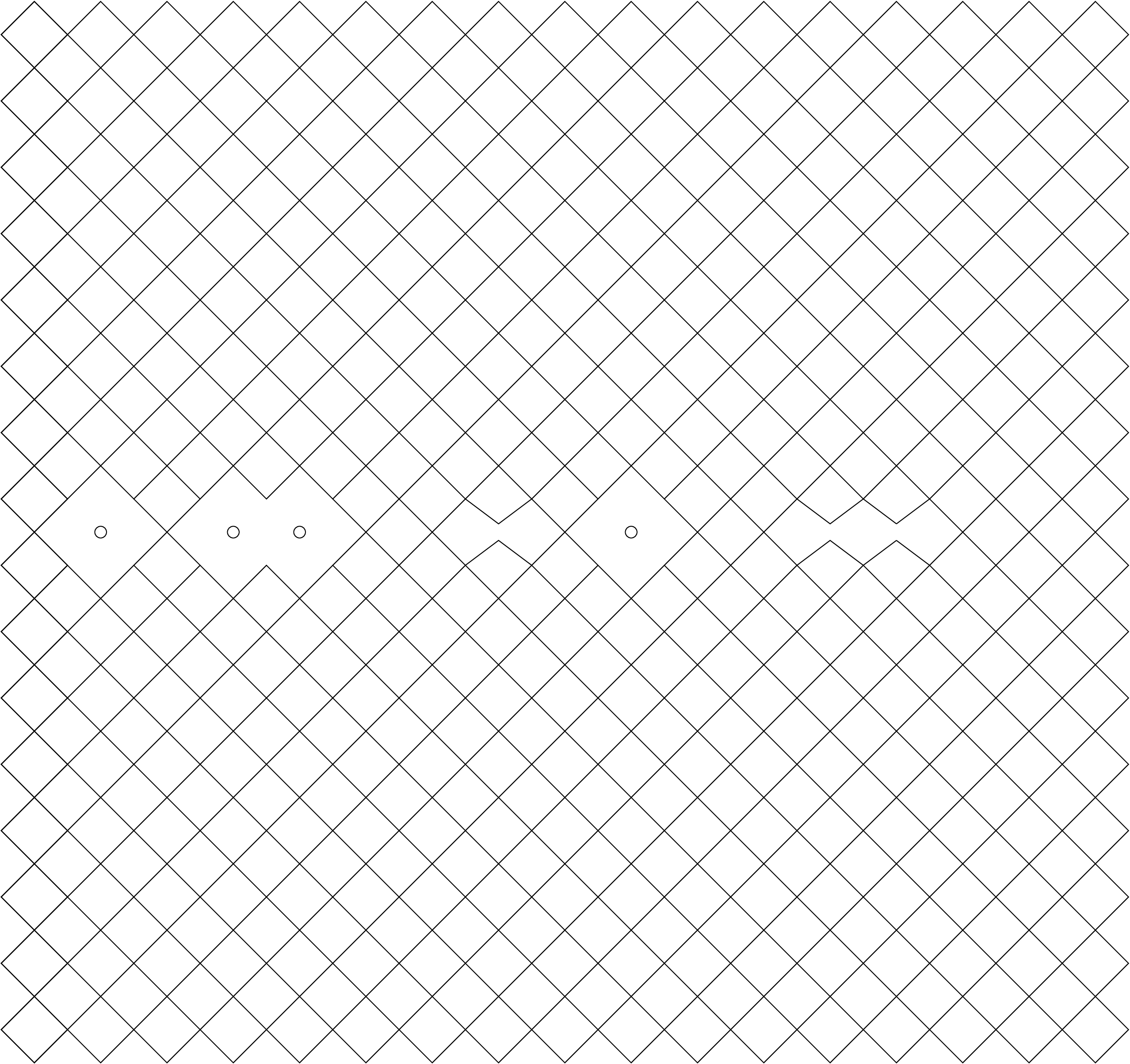}}
\centerline{{\smc Figure~{\fab}.} $AR_{16,17}(\{2,4,5,10\},\{8,13,14\})$.}
\endinsert

Note that for $m=n$, the Aztec rectangle $AR_{m,n}$ becomes the Aztec diamond graph $AD_n$. In view of this, if $|H|=|S|$ (which will be the case throughout Sections 2--6), we denote the graph $AR_{2n,2n+k-l}(H,S)$ by $AD_{2n}(H,S)$.

\mysec{2. Dipole correlations}

We define a {\it dipole} to be the union of two consecutive defects of opposite kind on $\ell\cap AD_{2n}$: either a monomer followed by a separation, or a separation followed by a monomer. If the hole in a dipole $D$ is the $i$th vertex from the left on $\ell\cap AD_{2n}$, and $i$ is odd, we say that $D$ is an {\it odd dipole}; if $i$ is even, we call it an {\it even dipole}.

%We call the dipole {\it even} or {\it odd} according as the number of vertices on $\ell\cap AD_{2n}$ to to the left of its hole is even or odd.

Let us define the {\it finite size correlation} of a collection $\Cal D$ of dipoles on $\ell\cap AD_{2n}$ as follows. Let $H$ and~$S$ be the subsets of $[2n]$ recording the positions of the holes and separations in $\Cal D$, respectively. Then we define the finite size correlation of the collection of dipoles $\Cal D$ by
$$
\omega_{2n}(\Cal D):=\frac{\M(AD_{2n}(H,S))}{\M(AD_{2n})},\tag\Eba
$$
where $\M(G)$ denotes the number of perfect matchings of the graph $G$.

Interestingly, it turns out that the joint interaction of even and odd dipoles, already at the finite size level, is dictated solely by the interaction of even dipoles among themselves and the boundary, and that of odd dipoles among themselves and the boundary: There is absolutely no interaction between an even and and odd dipole. More precisely, the following result holds.

As in \cite{\gd}, the function $E$ is defined by
$$
E(a_1,a_2,\dotsc,a_k;b_1,b_2\dotsc,b_l):=\frac
{\prod_{1\leq i<j\leq k}|a_i-a_j|^\frac12\prod_{1\leq i<j\leq l}|b_i-b_j|^\frac12}
{\prod_{i=1}^k\prod_{j=1}^l|a_i-b_j|^\frac12}.\tag\Ebb
$$
If $\Cal C$ is a collection of defect clusters, $E(\Cal C)$ is obtained from the above formula by letting $H$ and $S$ be the sets of integers that record the positions of the holes and separations involved in $\Cal C$. 

Denote by $\circ_{j}$ a unit hole at location $j$, and by $\times_{j}$ a separation\footnote{ We note that in the interest of better readability and typesetting, we use in this paper  $\circ_{j}$ and $\times_j$ for what was denoted in \cite{\gd} by $\boxdot_j$ and ${\vee\atop\wedge}_j$, respectively.} at location $j$ on $\ell\cap AD_{2n}$.

\proclaim{Theorem \Tba} $(${\rm a}$)$. Let $\Cal D$ be a collection consisting of odd dipoles $D_1,\dotsc,D_k$ and even dipoles $D'_1,\dotsc,D'_l$. Then the finite size correlation of $\Cal D$ is given by
$$
\spreadlines{2\jot}
\align
\omega_{2n}(\Cal D)&=\omega_{2n}(D_1,\dotsc,D_k)\,\omega_{2n}(D'_1,\dotsc,D'_l)
\\
&=
\left(\prod_{D\in\Cal D}\omega_{2n}(D)\right)
E^2(D_1,\dotsc,D_k)\,E^2(D'_1,\dotsc,D'_l).\tag\Ebc
\endalign
$$

$(${\rm b}$)$. The finite size correlation of single dipoles is given by\footnote{ Recall that $(a)_k$ is the Pochhammer symbol, defined to be equal to $a(a+1)(a+2)\cdots(a+k-1)$.}
$$
\spreadlines{4\jot}
\spreadmatrixlines{4\jot}
\align
\omega_{2n}(D)=\left\{\matrix \dfrac{\left(\frac12\right)_{s+1}\left(\frac12\right)_{n-s-1}}
{(1)_s(1)_{n-s-1}}, & \ \ \,D=\{\circ_{2s+1},\times_{2s+2}\}\\
\dfrac{\left(\frac12\right)_{s}\left(\frac12\right)_{n-s}}
{(1)_{s-1}(1)_{n-s}}, & D=\{\circ_{2s},\times_{2s+1}\}\\
\dfrac{\left(\frac12\right)_{s}\left(\frac12\right)_{n-s}}
{(1)_s(1)_{n-s-1}}, & \ \ \,D=\{\times_{2s+1},\circ_{2s+2}\}\\
\dfrac{\left(\frac12\right)_{s}\left(\frac12\right)_{n-s}}
{(1)_s(1)_{n-s-1}}, & D=\{\times_{2s},\circ_{2s+1}\}\ \\
\endmatrix\right.\tag\Ebd
\endalign
$$

\endproclaim

\medskip
\flushpar
{\smc{Remark \ra.}} If one views dipoles on $\ell\cap AD_{2n}$ as neutral objects consisting of two unit charges of opposite sign (see \cite{\gd, Theorem\,2.1}), the above theorem shows that, from the point of view of their joint dimer-mediated interaction, they naturally come in two flavors --- odd and even --- so that we have (exact!) Coulomb-squared interaction among the objects of the same flavor, and absolutely no interaction between objects of opposite flavor! It is remarkable that this equality is exact already at the finite size level. The multiplicative constant can also be conceptually understood as the product of the individual dipole correlations.

\medskip
\flushpar
{\smc{Remark \rb.}} Theorem 2.1 of \cite{gd} gives the asymptotics of the correlation of general defect clusters. However, the Coulomb product on the right hand side of \cite{\gd,(2.10)} becomes simply 1 when all defect clusters have charge zero. Thus, if we regard each dipole as a defect cluster, the statement of Theorem 2.1 of \cite{gd} becomes simply that, in the limit of large distances between the dipoles, their joint correlation is equal to the product of their individual correlations. 

Theorem {\Tba} refines this by providing the exact value of the correlation of dipoles. Furthermore, it provides this exact value already for finite size correlations. Therefore the formula it provides controls a variety of set-ups, for instance we can deduce from it the exact value of the correlation of dipoles around any point on the symmetry axis of the scaling limit of the Aztec diamond, even at finite distance between the dipoles.

\medskip
In our proof we employ the following preliminary result, which is a generalization of Lemma 6.2 in \cite{\gd}.

\proclaim{Lemma \Tbb} $(${\rm a}$)$. For any integers $0\leq s_1<\cdots<s_k\leq n-1$ and 
$0\leq t_1<\cdots<t_l\leq n-1$, we have
$$
\spreadlines{4\jot}
\align
&
\!\!\!\!\!\!\!\!\!\!
\frac{\M(AD_{2n}(\{2s_1+1,\dotsc,2s_k+1,2t_1+1,\dotsc,2t_l+1\},\{2s_1+2,\dotsc,2s_k+2,2t_1,\dotsc,2t_l\})}
{\M(AD_{2n})}
=
\\
&\ \ \ \ \ \ \  
\prod_{i=1}^k\frac{\left(\tfrac12\right)_{s_i+1}\left(\tfrac12\right)_{n-s_i-1}}{(1)_{s_i}(1)_{n-s_i-1}}
\prod_{j=1}^l\frac{\left(\tfrac12\right)_{t_j}\left(\tfrac12\right)_{n-t_j}}{(1)_{t_j}(1)_{n-t_j-1}}
%\prod_{1\leq i<j\leq k} \frac{(2s_j-2s_i)^2}{(2s_j-2s_i-1)(2s_j-2s_i+1)}
\\
&
\ \ \   
\times
E^2(\{2s_1+1,\dotsc,2s_k+1,2t_1+1,\dotsc,2t_l+1\},\{2s_1+2,\dotsc,2s_k+2,2t_1,\dotsc,2t_l\}).
%\prod_{1\leq i<j\leq l} \frac{(2t_j-2t_i)^2}{(2t_j-2t_i-1)(2t_j-2t_i+1)}.
\tag\Ebdd
\endalign
$$

$(${\rm b}$)$. For any integers $1\leq s_1<\cdots<s_k\leq n$ and 
$0\leq t_1<\cdots<t_l\leq n-1$, we have
$$
\spreadlines{4\jot}
\align
&
\!\!\!\!\!\!\!\!\!\!
\frac{\M(AD_{2n}(\{2s_1,\dotsc,2s_k,2t_1+2,\dotsc,2t_l+2\},\{2s_1+1,\dotsc,2s_k+1,2t_1+1,\dotsc,2t_l+1\})}
{\M(AD_{2n})}
=
\\
&\ \ \ \ \ \ \  
\prod_{i=1}^k\frac{\left(\tfrac12\right)_{s_i}\left(\tfrac12\right)_{n-s_i}}{(1)_{s_i-1}(1)_{n-s_i}}
\prod_{j=1}^l\frac{\left(\tfrac12\right)_{t_j}\left(\tfrac12\right)_{n-t_j}}{(1)_{t_j}(1)_{n-t_j-1}}
%\prod_{1\leq i<j\leq k} \frac{(2s_j-2s_i)^2}{(2s_j-2s_i-1)(2s_j-2s_i+1)}
\\
&
\ \ \   
\times
E^2(\{2s_1,\dotsc,2s_k,2t_1+2,\dotsc,2t_l+2\},\{2s_1+1,\dotsc,2s_k+1,2t_1+1,\dotsc,2t_l+1\}).
%\prod_{1\leq i<j\leq l} \frac{(2t_j-2t_i)^2}{(2t_j-2t_i-1)(2t_j-2t_i+1)}.
\tag\Ebddd
\endalign
$$

\endproclaim

\pf (a). Consider the dipoles $D_i=\{\circ_{2s_i+1},\times_{2s_i+2}\}$, $i=1,\dotsc,k$, and $D'_j=\{\times_{2t_j},\circ_{2t_j+1}\}$, $j=1,\dotsc,l$. Then the left hand side of (\Ebdd) is precisely the finite size correlation of the collection of dipoles $D_1,\dotsc,D_k,D'_1,\dotsc,D'_l$.

Let $i\in\{1,\dotsc,k\}$. In the fraction
$$
\frac{\omega_{2n}(D_1,\dotsc,D_i)}{\omega_{2n}(D_1,\dotsc,D_{i-1})},\tag\Ebddda
$$
apply Theorem 1.1 of \cite{\gd} both at the numerator and denominator, but with each element $i$ of ${\Cal O}$ and ${\Cal E}$ replaced by an indeterminate $a_i$ (this modification helps keep track of the simplifications that occur between the factors at the numerator and denominator). One readily sees that after simplifications this yields 
$$
\frac{\omega_{2n}(D_1,\dotsc,D_i)}{\omega_{2n}(D_1,\dotsc,D_{i-1})}
=
\frac{1}{2}
\prod_{j=0 \atop j\neq s_1,\dotsc,s_{i-1}}^{n-1}\frac{a_{2s_i+2}-a_{2j+1}}{a_{2s_i+1}-a_{2j+1}}
\prod_{1\leq j<i}\frac{a_{2s_i+2}-a_{2s_j+2}}{a_{2s_i+1}-a_{2s_j+2}}.
\tag\Ebdddb
$$
For any $i\in\{1,\dotsc,l\}$, the same considerations lead to 
$$
\spreadlines{4\jot}
\align
\frac{\omega_{2n}(D_1,\dotsc,D_k,D'_1,\dotsc,D'_i)}{\omega_{2n}(D_1,\dotsc,D_k,D'_1,\dotsc,D'_{i-1})}
=
\frac{1}{2}
&
\prod_{j=0 \atop j\neq s_1,\dotsc,s_k,t_1,\dotsc,t_{i-1}}^{n-1}\frac{a_{2t_i}-a_{2j+1}}{a_{2t_i+1}-a_{2j+1}}
\\
& 
\prod_{1\leq j\leq k}\frac{a_{2t_i}-a_{2s_j+2}}{a_{2t_i+1}-a_{2s_j+2}}
\prod_{1\leq j<i}\frac{a_{2t_i}-a_{2t_j}}{a_{2t_i+1}-a_{2t_j}}.
\tag\Ebdddc
\endalign
$$
It is routine to check that by specializing back $a_i=i$ for all indices above, and multiplying together equations (\Ebdddb) for $i=1,\dotsc,k$ and (\Ebdddc) for $i=1,\dotsc,l$, one obtains (\Ebdd).

(b). The left hand side is the finite size correlation of the collection consisting of dipoles $\{\circ_{2s_i},\times_{2s_i+1}\}$, $i=1,\dotsc,k$, and $\{\times_{2t_j+1},\circ_{2t_j+2}\}$, $j=1,\dotsc,l$. By performing a $180^\circ$ rotation, this is seen to be the same as the finite size correlation of the collection consisting of dipoles $\{\circ_{2n-2t_j-1},\times_{2n-2t_j}\}$, $j=1,\dotsc,l$, and $\times_{2n-2s_i},\circ_{2n-2s_i+1}\}$, $i=1,\dotsc,k$, which is of the type addressed in part (a). Applying formula (\Ebdd) to it one arrives at (\Ebddd).~\epf

{\it Proof of Theorem \Tba.} The first formula in (\Ebd) follows from Lemma {\Tbb}(a) by choosing $k=1$, $l=0$. Similarly, the fourth formula in (\Ebd) follows from Lemma {\Tbb}(a) by choosing $k=0$, $l=1$. The middle two formulas in (\Ebd) follow analogously from Lemma {\Tbb}(b). This proves part (b) of the Lemma.

For part (a), notice first that, by Theorem 1.1 of \cite{\gd}, the denominator of the fraction (\Eba) defining $\omega_{2n}(\Cal D)$  is equal to 
$$
\frac{2^{n^2+2n}}{(0!\,1!\cdots(n-1)!)^2}\Delta(\Cal O)\Delta(\Cal E),\tag\Ebe
$$ 
where $\Cal O$ and $\Cal E$ are the odd and even integers in $[2n]$, respectively, and $\Delta(T):=\prod_{1\leq i<j\leq n}(t_j-t_i)$, for the set $T$ consisting of integers $t_1<\cdots<t_n$ (the elements of $\Cal O$ and $\Cal E$ are indicated by $O$'s and $E$'s, respectively, in the bottom half of Figure \Fba).

\topinsert
\centerline{\mypic{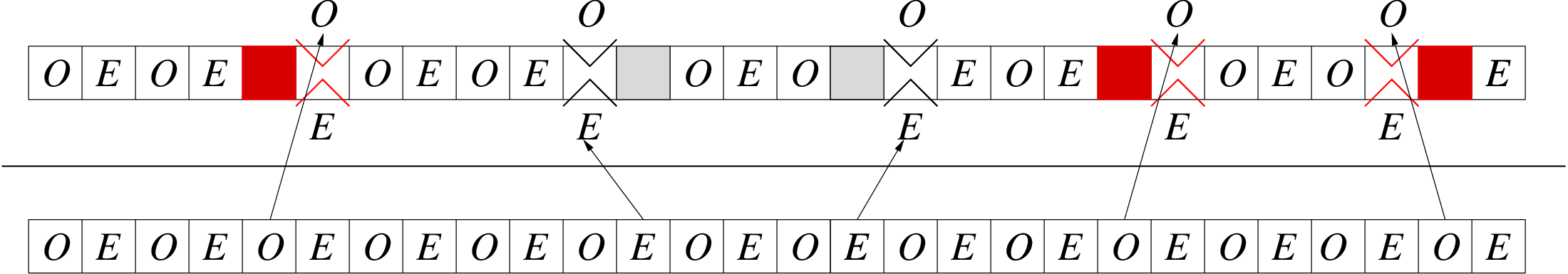}}
\medskip
\centerline{{\smc Figure~{\Fba}.}  The disturbance in the $O$-$E$ pattern created by even and odd dipoles (odd} 
\centerline{\ \ dipoles have their monomer marked by a dark square, even dipoles by a light square).}
\endinsert

By the same theorem, the numerator of the fraction (\Eba) defining $\omega_{2n}(\Cal D)$ is equal to 
$$
\frac{2^{n^2+2n-k-l}}{(0!\,1!\cdots(n-1)!)^2}\Delta(\Cal O')\Delta(\Cal E'),\tag\Ebf
$$ 
where $\Cal O'$ and $\Cal E'$ are the sets obtained from $\Cal O$ and $\Cal E$ by disturbing the positions of the $O$'s and $E$'s as follows: The $O$'s corresponding to holes of odd dipoles are shifted one unit to the right for dipoles of type $\circ\times$ and one unit to the left for dipoles of type $\times\circ$, while the $E$'s corresponding to holes of even dipoles are shifted one unit to the right for dipoles of type $\circ\times$ and one unit to the left for dipoles of type $\times\circ$. The top half of Figure {\Fba} illustrates this for a collection $\Cal D$ consisting of three odd and two even dipoles; $\Cal O'$ and $\Cal E'$ consist of the positions of the $O$'s and $E$'s in the top row.

Taking the ratio of (\Ebe) and (\Ebf), it follows that
$$
\omega_{2n}(\Cal D)=\frac{1}{2^{k+l}}\frac{\Delta(\Cal O')\Delta(\Cal E')}{\Delta(\Cal O)\Delta(\Cal E)}.\tag\Ebg
$$

Next, note that the sole effect on the $O$-$E$ pattern of including an odd dipole is to cause the $O$ that used to be in it to be shifted one unit to the right or left. Inclusion of an even dipole has the analogous effect on the $E$ that used to be in it (this is indicated by arrows in Figure \Fba). Therefore, if $\Cal D_o$ is the subcollection of $\Cal D$ consisting of the odd dipoles, the arguments that led to (\Ebg) give
$$
\omega_{2n}(\Cal D_o)=\frac{1}{2^{k}}\frac{\Delta(\Cal O')}{\Delta(\Cal O)}.\tag\Ebh
$$
Indeed, odd dipoles do not disturb the pattern of $E$'s, so the $\Delta(\Cal E)$ contributions from \cite{\gd, Theorem 1.1} cancel out. Furthermore, the pattern of $O$'s resulting by consideration of just the odd dipoles is the same as the pattern of $O$'s generated by the entire collection $\Cal D$, as even dipoles have no effect on it. 

The same reasoning shows that if $\Cal D_e$ is the subcollection of $\Cal D$ consisting of the even dipoles, then we have
$$
\omega_{2n}(\Cal D_e)=\frac{1}{2^{l}}\frac{\Delta(\Cal E')}{\Delta(\Cal E)}.\tag\Ebi
$$
Multiplying (\Ebh) and (\Ebi) and comparing the result to (\Ebg) we obtain
$$
\omega_{2n}(\Cal D)=\omega_{2n}(\Cal D_o)\,\omega_{2n}(\Cal D_e),\tag\Ebj
$$
which proves the first equality in (\Ebc).

To prove the second equality in (\Ebc), note that the first and fourth formulas in (\Ebd), together with Lemma {\Tbb}(a), imply that
$$
\omega_{2n}(\Cal D_o)=\left(\prod_{D\in\Cal D_o}\omega_{2n}(D)\right)
E^2(D_1,\dotsc,D_k).
$$
Similarly, the middle two equalities formulas in (\Ebd), together with Lemma {\Tbb}(b), imply that
$$
\omega_{2n}(\Cal D_e)=\left(\prod_{D\in\Cal D_e}\omega_{2n}(D)\right)
E^2(D'_1,\dotsc,D'_k).
$$
Multiplying together the above two equations one obtains the second equality in (\Ebc). \epf

\medskip
In agreement with Fisher and Stephenson's point of view when they studied the interaction of two dimers in \cite{\FS}, given two dipoles $D_1$ and $D_2$ let us consider the quantity
$$
\Omega_{2n}(D_1,D_2):=\omega_{2n}(D_1,D_2)-\omega_{2n}(D_1)\omega_{2n}(D_2),
\tag\Ebk
$$
which gives the difference between the probability that the dipoles are observed jointly and the product of the probabilities that they occur separately. The exact values of this quantity are given by the following result.

\medskip
\proclaim{Corollary \Tbc} For odd dipoles $D_1$ and $D_2$ we have
$$
\spreadlines{4\jot}
\spreadmatrixlines{4\jot}
\align
\Omega_{2n}(D_1,D_2)=\left\{\matrix 
\dfrac{\omega_{2n}(D_1)\omega_{2n}(D_2)}{(2t-2s-1)(2t-2s+1)}, & \ \ \,D_1=\{\circ_{2s+1},\times_{2s+2}\},\ D_2=\{\circ_{2t+1},\times_{2t+2}\}\\
-\dfrac{\omega_{2n}(D_1)\omega_{2n}(D_2)}{(2t-2s-1)^2}, & \, D_1=\{\circ_{2s+1},\times_{2s+2}\},\ D_2=\{\times_{2t},\circ_{2t+1}\}\\
\endmatrix\right.\tag\Ebl
\endalign
$$
The analogous formulas for even dipoles are given by incrementing both $s$ and $t$ in $(\Ebl)$ by one unit. If~$D_1$ and $D_2$ are dipoles of opposite parity, $\Omega_{2n}(D_1,D_2)=0$.

\endproclaim

\pf Apply Theorem {\Tba}(a) in the special case when $\Cal D$ consists of the dipoles $D_1$ and $D_2$, and use the definition (\Ebb) of $E$.  \epf

%\medskip
%\flushpar
%{\smc Remark 3.} The above formulas are the counterparts for dipoles of the formulas given by Fisher and Stephenson in \cite{\FS} for the correlations of dimers. This shows that the electrostatic analogy continues to be meaningful even a step beyond the context described in \cite{\ov}, which only considered defects of non-zero charge. Namely, if we regard the neutral dipoles as being made up of two unit and opposite charges, then the dipoles tend to attract or repel as dictated by the 2D Coulomb interaction of the resulting system of charges (except, interestingly, that we need to take the {\it square} of this Coulomb interaction). Furthermore, the formulas we get are exact already at the level of the finite size correlation!

\medskip
From the above results it is easy to deduce the value of the correlation $\omega$ of \cite{\gd} for an arbitrary collection of dipoles. We in fact generalize this to the situation when the scaling limit is taken so that the dipoles shrink to some arbitrary fixed points on the symmetry axis of the scaled Aztec diamond (and not necessarily to its center).

We will now take limits as the size of the Aztec diamond goes to infinity, and therefore any vertex of the infinite square grid on $\ell$ can be the location of a defect. Label these vertices from left to right by the elements of~$\Z$. 

Let $\Cal D_i$ be a finite collection of dipoles, for $i=1,\dotsc,k$. Denote by $AD_{2n}(\Cal D)$ the graph obtained from $AD_{2n}$ by creating defects at its vertices on $\ell$ as specified by the collection of dipoles $\Cal D$. If $D=\{\circ_{s},\times_{t}\}$ is a dipole and $m\in\Z$, denote by $D(m)$ the dipole $\{\circ_{m+s},\times_{m+t}\}$. Let $\Cal D(m)=\{D(m):D\in\Cal D\}$.

Given  real numbers $0<\alpha_1<\cdots<\alpha_k<2$, we define the correlation $\omega_{\alpha_1,\dotsc,\alpha_k}$ of the dipole collections $\Cal D_1,\dotsc,\Cal D_k$ as $\Cal D_i$ shrinks in the scaling limit to the point on $\ell$ corresponding to $\alpha_i$, $i=1,\dotsc,k$, by
$$
\omega_{\alpha_1,\dotsc,\alpha_k}(\Cal D_1,\dotsc,\Cal D_k):=\lim_{n\to\infty}\frac{\M(AD_{2n}(\Cal D_1(a_n^{(1)})\cup\cdots\cup\Cal D_k(a_n^{(k)}))}{\M(AD_{2n})},\tag\Ebm
$$
where $(a_n^{(i)})_n$ is a sequence of even integers\footnote{ The integers $a_n^{(i)}$ are required to be even so that $D(a_n^{(i)})$ has the same parity as $D$.} so that $\lim_{n\to\infty} a_n^{(i)}/n=\alpha_i$, for $i=1,\dotsc,k$, and the Aztec diamond graphs are placed so that their vertices on $\ell$ have labels $1,2,\dotsc,2n$.

\medskip
\flushpar
{\smc Remark \rd.} Note that this is a generalization of the correlation $\omega$ of \cite{gd}: $\omega_{\alpha_1,\dotsc,\alpha_k}(\Cal D_1,\dotsc,\Cal D_k)$ measures the correlation of the dipole collections $\Cal D_i$ as they interact while each being in a patch whose dimer statistics is governed by a different translationally invariant ergodic Gibbs measure (cf. \cite{\KOS}); $\omega$ corresponds to the case $\alpha_1=\cdots=\alpha_k=1$.\ %In general, if some of $\alpha_i$'s are equal, specific subsets of the collection of dipoles can be in the same Gibbs regime.

\proclaim{Corollary \Tbd} Let $\Cal D^{(i)}$ be finite collections of dipoles, for $i=1,\dotsc,k$. Let $\Cal D_o^{(i)}$ and $\Cal D_e^{(i)}$ be the subsets of
$\Cal D^{(i)}$ consisting of odd and even dipoles, respectively.
Then for any real numbers $0<\alpha_1<\cdots<\alpha_k<2$ we have
$$
\spreadlines{3\jot}
\align
\omega_{\alpha_1,\dotsc,\alpha_k}(\Cal D_1,\dotsc,\Cal D_k)
&=
\omega_{\alpha_1,\dotsc,\alpha_k}(\Cal D_o^{(1)},\dotsc,\Cal D_o^{(k)})\,\,
\omega_{\alpha_1,\dotsc,\alpha_k}(\Cal D_e^{(1)},\dotsc,\Cal D_e^{(k)})
\\
&=
\prod_{i=1}^k\prod_{D\in\Cal D^{(i)}}\frac{1}{\pi}\left(\dfrac{\alpha_i}{2-\alpha_i}\right)^\frac{\epsilon(D)}{2}\ \prod_{i=1}^k E^2(\Cal D_o^{(i)})\, E^2(\Cal D_e^{(i)}),\tag\Ebn
\endalign
$$
where the sign $\epsilon(D)$ of the dipole $D$ is $1$ if it has the hole on its left, and $-1$ if it has the hole on its right.

\endproclaim

\pf By Theorem {\Tba}(a), the assertion will follow provided we prove that
$$
\lim_{n\to\infty}\omega_{2n}(D(a_n))=\frac{1}{\pi}\left(\dfrac{\alpha}{2-\alpha}\right)^\frac{\epsilon(D)}{2}\tag\Ebo
$$
for any dipole $D$ and sequence $(a_n)_n$ of even integers with $a_n/n\to\alpha$. Indeed, the contribution to the $E^2$ factors in ({\Ebc}) coming from the interactions of dipoles from different collections become $1$ in the limit as $n\to\infty$, as this causes them to become infinitely separated. 

To prove (\Ebo), recall that, for any fixed $a,b\in\R$, the asymptotics of the ratios of Gamma functions is given by (see e.g. \cite{\Olver, Ch.\,4,\,(5.02)})
$$
\frac{\Gamma(x+a)}{\Gamma(x+b)}= x^{a-b}\left\{1+\frac12(a-b)(a+b-1)\frac{1}{x}+O\left(\frac{1}{x^2}\right)\right\}, \ \ \ x\to\infty.
\tag\Ebp
$$
Let $D=\{\circ_{2s+1},\times_{2s+2}\}$. Using the formula 
$$
(a)_k=\frac{\Gamma(a+k)}{\Gamma(k)}\tag\Ebpp
$$ 
to express the Pochhammer symbols in (\Ebd) in terms of Gamma functions and the fact that $\Gamma(1/2)=\sqrt{\pi}$, we obtain from the first equation in (\Ebd) that
$$
\omega_{2n}(D)=\frac{1}{\pi}\frac
{\Gamma\left(s+\frac32\right)\Gamma\left(n-s-\frac12\right)}
{\Gamma(s+1)\Gamma(n-s-1)}.\tag\Ebq
$$
Since (\Ebo) involves the translation $D(a_n)$ of $D$, in which the hole is at location $a_n+2s+1=2(a_n/2+s)+1$, the correlation $\omega_{2n}(D(a_n))$ is obtained by replacing $s$ by $a_n/2+s$ on the right hand side of (\Ebq). Using~(\Ebp) one obtains
$$
\omega_{2n}(D(a_n))=\frac{1}{\pi}\left(\frac{a_n}{2}\right)^{\frac{1}{2}}\left(n-\frac{a_n}{2}\right)^{-\frac12}+O(n^{-1/2}),\ \ \ n\to\infty.\tag\Ebr
$$
Since by assumption $a_n/n\to\alpha$, the above implies (\Ebo) for dipoles of type $\{\circ_{2s+1},\times_{2s+2}\}$. 

For odd dipoles $D$ of type $\times\circ$, start with the fourth equation in (\Ebd) and use the same argument. The fact that this time the first Pochhammer symbols at the numerator and denominator have the same index, and the second ones differ by one, results in the main term in the asymptotics being given by the reciprocal of the main term in (\Ebr), in accordance with the fact that now the sign $\epsilon(D)$ is $-1$. The case of even dipoles follows in the same fashion. \epf

%\medskip
%\flushpar
%{\smc Remark 5.} Specializing $k=1$, $\alpha_1=1$ in Corollary {\Tbd} yields
%$$
%\omega  

Setting 
$$
\Omega_{\alpha_1,\dotsc,\alpha_k}(\Cal D_1,\dotsc,\Cal D_k)=
\omega_{\alpha_1,\dotsc,\alpha_k}(\Cal D_1,\dotsc,\Cal D_k)
-\prod_{i=1}^k\prod_{D\in\Cal D_i}\omega_{\alpha_i}(D),\tag\Ebs
$$
this implies the following.% extension of Corollary {\Tbc}.

\proclaim{Corollary {\Tbe}} With the same notation as in Corollary ${\Tbd}$, we have
$$
\Omega_{\alpha_1,\dotsc,\alpha_k}(\Cal D_1,\dotsc,\Cal D_k)=
\prod_{i=1}^k\omega_{\alpha_i}(\Cal D_i)\left\{\left(\prod_{i=1}^k E^2(\Cal D_o^{(i)})\, E^2(\Cal D_e^{(i)})\right) -1\right\}.\tag\Ebt
$$

\endproclaim

\pf By (\Ebo) it follows that 
$$
\omega_\alpha(D)=\frac{1}{\pi}\left(\frac{\alpha}{2-\alpha}\right)^{\frac{\epsilon(D)}{2}}.\tag\Ebu
$$
Using this, the statement follows directly from Corollary {\Tbd}. \epf

The following special case of Corollary {\Tbd} will be used several times later in this paper.

\proclaim{Corollary \Tbf} We have
$$
\spreadlines{3\jot}
\align
&
\omega(\circ_{2s_1+1}\times_{2s_1+2},\dotsc,\circ_{2s_k+1}\times_{2s_k+2},
\times_{2t_1}\circ_{2t_1+1},\dotsc,\times_{2t_l}\circ_{2t_1+l})
\\
&\ \ \ \ \ \ \ \
=
\omega(\circ_{2s_1+1}\times_{2s_1+2},\dotsc,\circ_{2s_k+1}\times_{2s_k+2})\,\,
\omega(\times_{2t_1}\circ_{2t_1+1},\dotsc,\times_{2t_l}\circ_{2t_1+l})
\\
&\ \ \ \ \ \ \ \ 
=
\frac{1}{\pi^{k+l}}
\prod_{1\leq i<j\leq k}\frac{(2s_j-2s_i)}{(2s_j-2s_i-1)(2s_j-2s_i+1)}
\prod_{1\leq i<j\leq l}\frac{(2t_j-2t_i)}{(2t_j-2t_i-1)(2t_j-2t_i+1)},\tag\Ebuu
\endalign
$$
where $\omega$ is the correlation at the center of the scaled Aztec diamond.

\endproclaim 

\pf As the listed dipoles are all of the same parity, the statement follows from Corollary {\Tbd} by specializing $k=1$, $\alpha_1=1$ and using (\Ebu). \epf

\medskip
\flushpar
{\smc Remark \re.} Specializing $k=1$, $\alpha_1=1$ in (\Ebt) and using (\Ebu) one obtains
$$
\align
\omega(\circ_{2s+1}\times_{2s+2},\circ_{2t+1}\times_{2t+2})- \omega(\circ_{2s+1}\times_{2s+2}) \omega(\circ_{2t+1}\times_{2t+2})&=
\frac{1}{\pi^2(2t-2s-1)(2t-2s+1)},\tag\Ebv
\\
\omega(\circ_{2s+1}\times_{2s+2},\circ_{2t}\times_{2t+1})- \omega(\circ_{2s+1}\times_{2s+2}) \omega(\circ_{2t}\times_{2t+1})&=
0,\tag\Ebw
\\
\omega(\circ_{2s+1}\times_{2s+2},\times_{2t}\circ_{2t+1})- \omega(\circ_{2s+1}\times_{2s+2}) \omega(\times_{2t}\circ_{2t+1})&=
-\frac{1}{\pi^2(2t-2s-1)^2},\tag\Ebx
\\
\omega(\circ_{2s+1}\times_{2s+2},\times_{2t+1}\circ_{2t+2})- \omega(\circ_{2s+1}\times_{2s+2}) \omega(\times_{2t+1}\circ_{2t+2})&=
0.\tag\Eby
\endalign
$$
These are in some sense analogous to the formulas of Fisher and Stephenson given in \cite{\FS} for the correlations of vertical dimers facing one another. The curious fact that the latter decreased as $d^{-2}$ and $-d^{-4}$ for even and odd separations $d$ between them, respectively, is paralleled by the fact that, as shown by the above formulas, the correlation of two dipoles of type $\circ\times$ either decreases as one over the square of the distance between them, or is equal to zero (exactly)! 

A notable difference --- and an even {\it more} curious fact --- is that in the analysis of Fisher and Stephenson changing the parity of the distance between the dimers causes the moving dimer to flip its ``polarization'', i.e. the dimer switches from having say the upper vertex white to having it black. That is not the case in our example with dipoles! So while the former phenomenon could be understood at least qualitatively (attraction vs. repellance) using the electrostatic picture pieced together by \cite{\sc}\cite{\ec}\cite{\ef}\cite{\ov}, the latter doesn't seem to be explainable the same way --- it is hard to see how in such an argument the parity of the distance between the dipoles could play such a radical role. This unusual behavior seems to be an effect of the dipoles interacting with the underlying lattice itself.

%\medskip
%We also note that there is a connection between (\Ebv) and (\Eby), namely the latter follows from the former via the exactness phenomenon of \cite{\gd, Section \,4} (a similar connection exists between (\Ebw) and (\Ebx)). Indeed, by \cite{\gd, Lemma 4.1} we have
%$$
%\spreadlines{4\jot}
%\align
%&
%\frac
%{\omega(\circ_{2s+1}\times_{2s+2},\times_{2t+1}\circ_{2t+2})}
%{\omega(\circ_{2s+1}\times_{2s+2},\circ_{2t+1}\times_{2t+2})}
%=
%\frac
%{\displaystyle \left[\frac{(2t-2s-1)(2t-2s+1)}{(2t-2s)^2}\right]^\frac12}
%{\displaystyle \left[\frac{(2t-2s)^2}{(2t-2s-1)(2t-2s+1)}\right]^\frac12}
%=
%\\
%&\ \ \ \ \ \ \ \ \ \ \ \ \ \ \ \ \ \ \ \ \ \ \ \ \ 
%\frac{(2t-2s-1)(2t-2s+1)}{(2t-2s)^2}
%=
%\frac{1}{E^2(\circ_{2s+1}\times_{2s+2},\circ_{2t+1}\times_{2t+2})},
%\endalign
%$$
%and therefore
%$$
%{\omega(\circ_{2s+1}\times_{2s+2},\times_{2t+1}\circ_{2t+2})}
%=
%\frac{\omega(\circ_{2s+1}\times_{2s+2},\circ_{2t+1}\times_{2t+2})}
%{E^2(\circ_{2s+1}\times_{2s+2},\circ_{2t+1}\times_{2t+2})}.
%$$
%Using the fact that
%$$
%\omega(\circ_{2s+1}\times_{2s+2},\circ_{2t+1}\times_{2t+2})
%=
%\frac{1}{\pi^2}\frac{(2t-2s)^2}{(2t-2s-1)(2t-2s+1)}
%=
%E^2(\circ_{2s+1}\times_{2s+2},\circ_{2t+1}\times_{2t+2})
%$$
%(which is equivalent to (\Ebv), and a special case of Corollary {\Tbf}),
%one obtains (\Eby)).

\mysec{3. The interaction of infinitely long neutral slits}

Denote by $[\circ\times]_n$ the cluster of defects consisting of $n$ monomers and $n$ separations occurring at $2n$ consecutive sites on $\ell$, so that monomers and separations alternate, and the leftmost defect is a monomer. Let $[\times\circ]_n$ be defined similarly, with the one difference that the leftmost defect is a separation. As a separation defect is equivalent to the superposition of four trimers (see \cite{\gd,Section\,1}), $[\circ\times]_n$ can be viewed as the superposition of $2^n$ slits, each consisting of $2n$ consecutive vertices on $\ell$, $n$ vertices on the diagonal immediately above $\ell$, and $n$ vertices on the diagonal immediately below $\ell$. With this interpretation in mind, we call $[\circ\times]_n$ and $[\times\circ]_n$ {\it fluctuating slits} (see Figure {\Fca} for a schematical illustration of two such slits; sample removed trimers are indicated; all four possible trimers are considered inside each big circle). 

In this section we study the asymptotics of the correlation of fluctuating slits by letting first the slits grow infinitely long, and then increasing the separation between them (see Theorem \Tcd). Sections~4 and~5 present two more points of view: letting first the separation go to infinity and then increasing the length of the slits (in Section 4), and letting the slits and the separation between them go to infinity at the same time (in Section 5).

Denote by
$$
{\omega([\circ\times]_a,[\circ\times]_b;d)}\tag\Eca
%{\omega([\circ\times]_a\underbrace{\phantom{aaaaa}}_{d}\ [\circ\times]_b)}\tag\Eca
$$
the correlation at the center\footnote{ In \cite{\gd} we denoted this correlation by $\bar\omega$; in the interest of notational simplicity, we will drop the bar from now on and simply write $\omega$ for this correlation.} (as considered in \cite{\gd}) of the monomer cluster consisting of the fluctuating slit $[\circ\times]_a$ followed, after $d$ unaffected sites on $\ell$, by the  fluctuating slit $[\circ\times]_b$; define $ {\omega([\circ\times]_a,[\times\circ]_b;d)}$ to have the obvious analogous meaning.

We start by giving an expression for the correlation (\Eca) for even $d$ as a product of ratios of Gamma functions.

\proclaim{Lemma \Tca} We have
$$
\spreadlines{4\jot}
\align
&
\frac
{\omega([\circ\times]_a,[\circ\times]_b;2d)}
{\omega([\circ\times]_{a+b})}
=
\prod_{i=1}^a
\left\{
\left[\frac{\Gamma(a+b+d-i+1)\Gamma(a-i+1)}
{\Gamma(a+b-i+1)\Gamma(a+d-i+1)}\right]^2
\right.
\\
&\ \ \ \ \ \ \ \ \ \ \ \ \ \ \ \ \ \ \
\left.
\times
\frac
{\Gamma\left(a+d-i+\frac12\right)\Gamma\left(a+d-i+\frac32\right)
\Gamma\left(a+b-i+\frac12\right)\Gamma\left(a+b-i+\frac32\right)}
{\Gamma\left(a+b+d-i+\frac12\right)\Gamma\left(a+b+d-i+\frac32\right)
\Gamma\left(a-i+\frac12\right)\Gamma\left(a-i+\frac32\right)}
\right\}.\tag\Ecb
\endalign
$$

\endproclaim

\pf Note that both defect clusters whose correlation is taken on the left hand side of (\Ecb) can be regarded as unions of odd dipoles. Apply Corollary {\Tbf} to both of them. Clearly, the resulting factors at the numerator and denominator that involve $\pi$ cancel out, and so do also those factors involved in the $E^2$ parts that come from interactions within $[\circ\times]_a$, and from interactions within $[\circ\times]_b$. One readily sees that the remaining factors can be organized so as to obtain the formula
$$
\frac
{\omega([\circ\times]_a, [\circ\times]_b;2d)}
{\omega([\circ\times]_{a+b})}
=
%\frac{1}{\pi^{a+b}}
\frac
{\dfrac
{\left[(a+d)_b(a+d-1)_b\cdots(d+1)_b\right]^2}
{\left(a+d+\frac12\right)_b\left(a+d-\frac12\right)_b\cdots\left(d+\frac32\right)_b
\left(a+d-\frac12\right)_b\left(a+d-\frac32\right)_b\cdots\left(d+\frac12\right)_b}}
{\dfrac
{\left[(a)_b(a-1)_b\cdots(1)_b\right]^2}
{\left(a+\frac12\right)_b\left(a-\frac12\right)_b\cdots\left(\frac32\right)_b
\left(a-\frac12\right)_b\left(a-\frac32\right)_b\cdots\left(\frac12\right)_b}}.
\tag\Ecc
$$
Using formula (\Ebpp) to express the Pochhammer symbols above in terms of Gamma functions one obtains~(\Ecb). \epf
%$$
%\frac
%{\omega([\circ\times]_a\underbrace{\phantom{aaaaa}}_{2d}\ [\circ\times]_b)}
%{\omega([\circ\times]_{a+b}}
%=
%%\frac{1}{\pi^{a+b}}
%\prod_{i=1}^a
%\left\{
%\left[\frac{\Gamma(a+b+d-i+1)}
%{\Gamma(a+d-i+1)}\right]^2
%\frac
%{\Gamma\left(a+d-i+\frac12\right)\Gamma\left(a+d-i+\frac32\right)}
%{\Gamma\left(a+b+d-i+\frac12\right)\Gamma\left(a+b+d-i+\frac32\right)}
%\right\}.\tag\Ecd
%$$
%The statement follows by taking the ratio of the right hand side of (\Ecd) with its 
%$d=0$ specialization. \epf

\medskip
This implies the following surprising exact equality.

\proclaim{Corollary \Tcb} For any non-negative integers $a$, $b$ and $d$ one has
$$
\omega([\circ\times]_a, [\circ\times]_b;2d)
=
\omega([\circ\times]_a, [\circ\times]_d;2b).\tag\Ece
$$

\endproclaim

\pf Note that the expression on the right hand side of (\Ecb) is invariant under interchanging the variables $b$ and $d$. \epf

We record for later use the following simple consequence of Corollary {\Tbf}. 

\proclaim{Lemma \Tcc} For any positive integer $a$ we have
$$
\omega([\circ\times]_a)=\omega([\times\circ]_a)=\frac{1}{2^a}\prod_{i=1}^a
\frac{\Gamma^2(i)}
{\Gamma\left(i-\frac12\right)\Gamma\left(i+\frac12\right)}.\tag\Ecee
$$

\endproclaim

\pf Apply Corollary {\Tbf}, and convert the resulting Pochhammer symbols to Gamma functions using formula (\Ebpp). \epf

The next result concerns the first point of view mentioned at the beginning of this section (namely, letting the fluctuating slits grow infinitely long first, and then studying the asymptotics of their correlation as the separation between them increases).

\proclaim{Theorem \Tcd} $(${\rm a}$)$. Consider two like-oriented fluctuating slits $[\circ\times]_a$ and $[\circ\times]_b$ separated by an even number $2d$ of sites. Then we have
$$
\spreadlines{1\jot}
\align
\lim_{a,b\to\infty}
\frac
{\omega([\circ\times]_a, [\circ\times]_b;2d)}
{\omega([\circ\times]_{a+b})}
&=2^d\omega([\circ\times]_d)
\\
&\sim
\frac{2^\frac13 e^\frac14}{A^3}(2d)^{-\frac14},\ \ \ d\to\infty,\tag\Ecf
\endalign
$$
where $A$ is the Glaisher-Kinkelin constant $($see \text{\rm \cite{\Glaish}}$)$, defined by the limit 
$$
\lim_{n\to\infty}
\dfrac
 {0!\,1!\,\cdots\,(n-1)!}
 {n^{\frac{n^2}{2}-\frac{1}{12}}\,(2\pi)^{\frac{n}{2}}\,e^{-\frac{3n^2}{4}}}
=
\dfrac
 {e^{\frac{1}{12}}}
 {A}.\tag\Ecg
$$

On the other hand, if they are separated by an odd number of sites, then
$$
{\omega([\circ\times]_a, [\circ\times]_b;2d+1)}
=
\omega([\circ\times]_a)\omega([\circ\times]_b)
=
\frac{1}{2^{a+b}}
{\displaystyle \prod_{i=1}^a\frac{\Gamma^2(i)}{\Gamma\left(i-\frac12\right)\Gamma\left(i+\frac12\right)}
\prod_{i=1}^b\frac{\Gamma^2(i)}{\Gamma\left(i-\frac12\right)\Gamma\left(i+\frac12\right)}},
\tag\Ech
$$
independently of $d$.

$(${\rm b}$)$. Consider two oppositely oriented fluctuating slits $[\circ\times]_a$ and $[\times\circ]_b$ separated by an odd number $2d+1$ of sites. Then we have
$$
\lim_{a,b\to\infty}
\frac
{\omega([\circ\times]_a, [\times\circ]_b;2d+1)}
{\omega([\circ\times]_{a}, [\times\circ]_{b};1)}
\sim
\frac{\pi^{\frac12} e^\frac14}{2^\frac16 A^3}(2d+1)^{\frac14},\ \ \ d\to\infty.\tag\Eci
$$

On the other hand, if they are separated by an even number of sites, then
$$
{\omega([\circ\times]_a, [\times\circ]_b;2d)}
=
\omega([\circ\times]_a)\omega([\times\circ]_b)
=
\frac{1}{2^{a+b}}
{\displaystyle \prod_{i=1}^a\frac{\Gamma^2(i)}{\Gamma\left(i-\frac12\right)\Gamma\left(i+\frac12\right)}
\prod_{i=1}^b\frac{\Gamma^2(i)}{\Gamma\left(i-\frac12\right)\Gamma\left(i+\frac12\right)}},
\tag\Ecj
$$
independently of $d$.

\endproclaim

\topinsert
%\vskip-0.2in
\centerline{\mypic{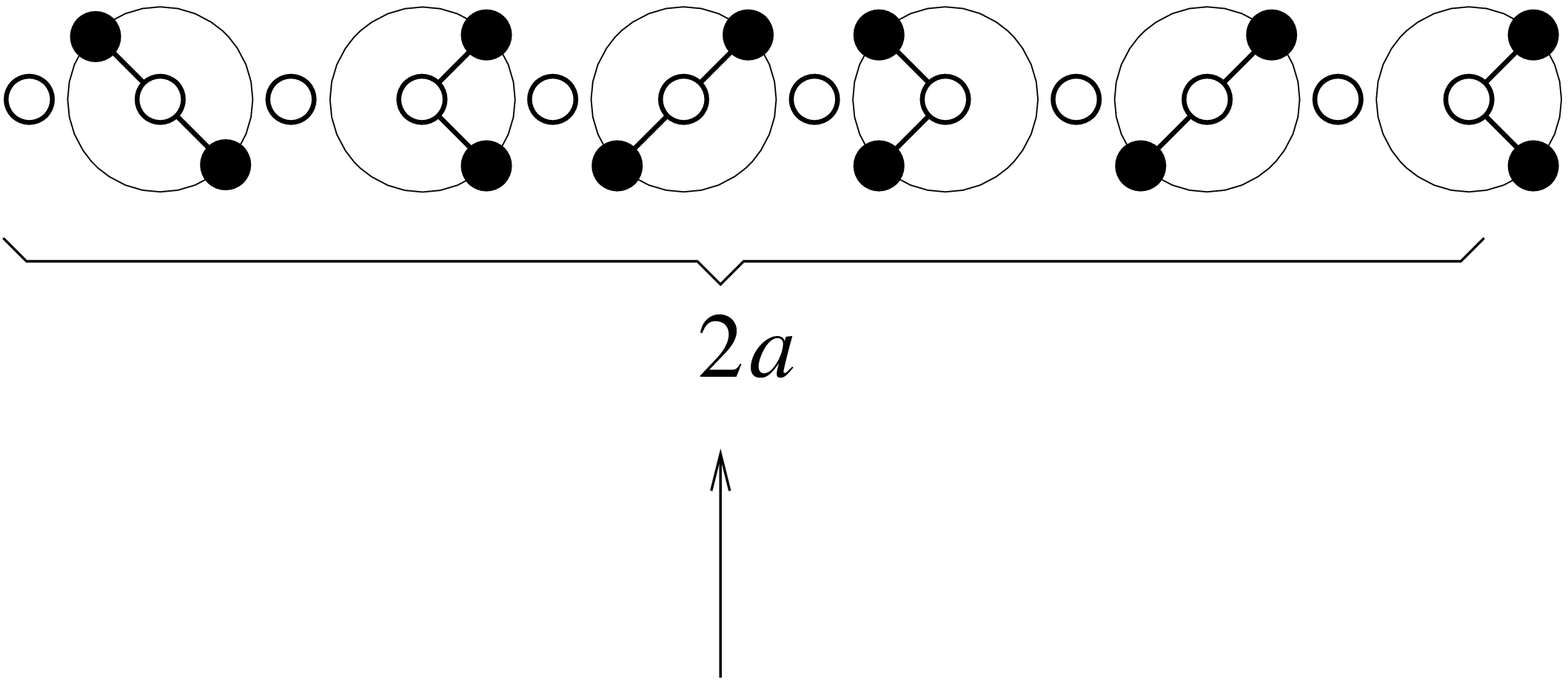}}
%\vskip-0.15in
%\endinsert

$
\ \ \ \ \ \ \ \ \ \ \ \ \ \ \ \ \ \ \ \ \ \ \ \ \ \ \ \ \ \ \ \ 
[\circ\times]_{a}
$
\medskip
\centerline{Figure \Fca. {\rm Illustration of the fluctuating slits involved in
$\omega([\circ\times]_a,[\circ\times]_b;2d)$.}}
\endinsert

\pf In addition to the equality of Corollary {\Tcb}, the correlation of fluctuating slits is also invariant under interchanging $a$ and $b$:
$$
\omega([\circ\times]_a, [\circ\times]_b;2d)
=
\omega([\circ\times]_b, [\circ\times]_a;2d).\tag\Eck
$$
Indeed, a rotation by $180^\circ$ shows that
$$
\omega([\circ\times]_a, [\circ\times]_b;2d)
=
\omega([\times\circ]_b, [\times\circ]_a;2d).\tag\Ecl
$$
However, regarding all defect clusters whose correlation is taken as unions of odd dipoles, one sees that Corollary {\Tbd} implies
$$
\omega([\times\circ]_b, [\times\circ]_a;2d)
=
\omega([\circ\times]_b, [\circ\times]_a;2d)\tag\Ecm
$$
(this is because upon changing all holes into separations and all separations into holes, each pair of likes remains a pair of likes, and each pair of unlikes remains a pair of unlikes). Using (\Ecl) and (\Ecm) one obtains (\Eck).

By (\Ece) and (\Eck) we obtain from Lemma {\Tca} that
$$
\spreadlines{4\jot}
\align
&
\frac
{\omega([\circ\times]_a, [\circ\times]_b;2d)}
{\omega([\circ\times]_{a+b})}
=
\prod_{i=1}^d
\left\{
\left[\frac{\Gamma(d+a+b-i+1)\Gamma(d-i+1)}
{\Gamma(d+b-i+1)\Gamma(d+a-i+1)}\right]^2
\right.
\\
&\ \ \ \ \ \ \ \ \ \ \ \ \ \ \ \ \ \ \
\left.
\times
\frac
{\Gamma\left(d+b-i+\frac12\right)\Gamma\left(d+b-i+\frac32\right)
\Gamma\left(d+a-i+\frac12\right)\Gamma\left(d+a-i+\frac32\right)}
{\Gamma\left(d+a+b-i+\frac12\right)\Gamma\left(d+a+b-i+\frac32\right)
\Gamma\left(d-i+\frac12\right)\Gamma\left(d-i+\frac32\right)}
\right\}.\tag\Ecn
\endalign
$$
We are interested in the limit of the above as $a,b\to\infty$, with $d$ kept constant. Taking out the factors involving only $d$, (\Ecn) becomes
$$
\spreadlines{4\jot}
\align
&
\frac
{\omega([\circ\times]_a, [\circ\times]_b;2d)}
{\omega([\circ\times]_{a+b})}
=
\prod_{i=1}^d
\frac{\Gamma^2(d-i+1)}
{\Gamma\left(d-i+\frac12\right)\Gamma\left(d-i+\frac32\right)}
\\
&\ \ \ \ \ \ \ \ \ 
\times
\prod_{i=1}^d
\frac
{\Gamma^2(a+b+d-i+1)
\Gamma\left(a+d-i+\frac12\right)\Gamma\left(a+d-i+\frac32\right)
\Gamma\left(b+d-i+\frac12\right)\Gamma\left(b+d-i+\frac32\right)}
{\Gamma\left(a+b+d-i+\frac12\right)\Gamma\left(a+b+d-i+\frac32\right)
\Gamma^2(a+d-i+1)\Gamma^2(b+d-i+1)}.
\\
\tag\Eco
\endalign
$$
Writing the factor in the second product above as
$$
\spreadlines{4\jot}
\align
\frac
{\Gamma^2(a+b+d-i+1)}
{\Gamma\left(a+b+d-i+\frac12\right)\Gamma\left(a+b+d-i+\frac32\right)}
&
\frac
{\Gamma\left(a+d-i+\frac12\right)\Gamma\left(a+d-i+\frac32\right)}
{\Gamma^2(a+d-i+1)}
\\
\times
&
\frac
{\Gamma\left(b+d-i+\frac12\right)\Gamma\left(b+d-i+\frac32\right)}
{\Gamma^2(b+d-i+1)}
\endalign
$$
and using (\Ebp), one sees that the factor in the second product in (\Eco) approaches
1 for any fixed $d$ as $a,b\to\infty$. It follows then from (\Eco) that
$$
\lim_{a,b\to\infty}\frac
{\omega([\circ\times]_a, [\circ\times]_b;2d)}
{\omega([\circ\times]_{a+b})}
=
\prod_{i=1}^d
\frac{\Gamma^2(d-i+1)}
{\Gamma\left(d-i+\frac12\right)\Gamma\left(d-i+\frac32\right)}
=
\prod_{i=1}^d
\frac{\Gamma^2(i)}
{\Gamma\left(i-\frac12\right)\Gamma\left(i+\frac32\right)}
.\tag\Ecp
$$
%Setting $k=1$, $\alpha_1=1$ in Corollary {\Tbc} one obtains, after converting the resulting Pochhammer symbols to Gamma functions, that
%$$
%\omega([\circ\times]_d)=\frac{1}{2^n}\prod_{i=1}^d
%\frac{\Gamma^2(i)}
%{\Gamma\left(i-\frac12\right)\Gamma\left(i+\frac12\right)}.\tag\Ecpp
%$$
Combining (\Ecp) and (\Ecee) one obtains the first equality in (\Ecf).

In order to work out the asymptotics, for non-negative integers $n$ set 
$$
H(n):=0!\,1!\,\cdots\,(n-1)!\tag\Ecq
$$
and 
$$
E(n):=[2!\,4!\,\cdots\,(2n)!]^2.\tag\Ecr
$$
Then (\Ecg) is equivalent to
$$
H(n)\sim\dfrac
 {e^{\frac{1}{12}}}
 {A}
n^{\frac{n^2}{2}-\frac{1}{12}}\,(2\pi)^{\frac{n}{2}}\,e^{-\frac{3n^2}{4}},
\ \ \ n\to\infty.\tag\Ecs
$$
One also sees, writing
$$
E^2(n)=2^n(0!\,1!\cdots(2n-1)!)n!\,(2n)!
$$
and using (\Ecs) and Stirling's approximation for the factorial, that
$$
E(n)\sim\dfrac
 {e^{\frac{1}{12}}}
 {A}
n^{2n^2-\frac{1}{12}+3n+1}\,\pi^{n+1}\,2^{2n^2-\frac{1}{12}+4n+\frac32}\,e^{-3n^2-3n},
\ \ \ n\to\infty.\tag\Ect
$$
Repeatedly applying the recurrence $\Gamma(x+1)=x\Gamma(x)$ for the Gamma function and using that $\Gamma\left(\frac12\right)=\sqrt{\pi}$, one arrives at the following expressions for the products of the individual Gamma functions making up the factor on the right hand side of (\Ecp):
$$
\spreadlines{4\jot}
\align
\prod_{i=1}^d\Gamma(i)&=H(d)\tag\Ecu
\\
\prod_{i=1}^d\Gamma\left(i+\frac12\right)&=
\frac{\pi^\frac d2\sqrt{E(d)}}{2^{d(d+1)}d!H(d)}\tag\Ecv
\\
\prod_{i=1}^d\Gamma\left(i-\frac12\right)&=
\frac{\pi^\frac d2\sqrt{E(d)}}{2^{d(d-1)}(2d)!H(d)}.\tag\Ecw
\endalign
$$
Using (\Ecs) and (\Ect) in (\Ecu)--(\Ecw), it follows that 
$$
\prod_{i=1}^d\frac{\Gamma^2(i)}{\Gamma\left(i-\frac12\right)\Gamma\left(i+\frac12\right)}
\sim
\frac{2^\frac{1}{12} e^\frac14}{A^3}d^{-\frac14},\ \ \ d\to\infty.\tag\Ecww
$$
The asymptotic equality in equation (\Ecf) in the statement of the theorem follows now by (\Ecp) and~(\Ecww).
Equation (\Ech) follows from the first equality in Corollary {\Tbf} and two applications of (\Ecee). This completes the proof of part~(a).

To prove (\Eci), we express $\omega([\circ\times]_a,[\times\circ]_b;2d+1)$ in terms of $\omega([\circ\times]_a,[\circ\times]_b;2d)$ as follows. We claim that
$$
\spreadlines{4\jot}
\align
\frac
{\omega([\circ\times]_a,[\times\circ]_b;2d+1)}
{\omega([\circ\times]_a,[\circ\times]_b;2d)}&=
\dfrac
{
\dfrac{\{(2d+2a+2b)\cdots(2d+2b+2)\}\{(2b-2)\cdots(2)\}}
{\{(2d+2a+2b-1)\cdots(2d+2b+1)\}\{(2b-1)\cdots(1)\}\phantom{|}}
}
{
\dfrac{\{(2d+2a)\cdots(2d+2)\}\{(2b-2)\cdots(2)\}\phantom{|}}
{\{(2d+2a-1)\cdots(2d+1)\}\{(2b-1)\cdots(1)\}}
},\tag\Ecx
\endalign
$$
where successive factors in the products in all curly braces decrease by two units from one factor to the next.

Indeed, use Corollary {\Tbf} to write both the numerator and the denominator on the left hand side above as an explicit product. Since the defect cluster at the denominator is obtained from the one at the numerator by moving the rightmost hole in slit $[\times\circ]_b$ to the position just to the left of that slit, almost all factors of the products in the previous sentence cancel out; the ones that do not are shown on the right hand side of (\Ecx).

%Simplify out the common factors in (\Ecx). 
Dividing each remaining factor by 2 to express the leftover expression in terms of Pochhammer symbols, and then using (\Ebpp) to convert the latter to Gamma functions, we obtain from (\Ecx) that
$$
\spreadlines{4\jot}
\align
\frac
{\omega([\circ\times]_a,[\times\circ]_b;2d+1)}
{\omega([\circ\times]_a,[\circ\times]_b;2d)}&=
\frac{(d+b+1)_a\left(a+\frac12\right)_a}{\left(d+b+\frac12\right)_a(d+1)_a}
\\
&=
\frac
{\Gamma(a+b+d+1)\Gamma\left(a+d+\frac12\right)\Gamma\left(b+d+\frac12\right)\Gamma(d+1)}
{\Gamma\left(a+b+d+\frac12\right)\Gamma\left(a+d+1\right)\Gamma\left(b+d+1\right)
\Gamma\left(d+\frac12\right)}.
\tag\Ecy
\endalign
$$
Dividing side by side equation (\Ecy) with its $d=0$ specialization we obtain
$$
\spreadlines{4\jot}
\align
\frac
{\omega([\circ\times]_a,[\times\circ]_b;2d+1)}
{\omega([\circ\times]_a,[\times\circ]_b;1)}&=
\frac
{\Gamma\left[\matrix a+b+d+1,a+d+\frac12,a+d+\frac12,d+1\\a+b+d+\frac12,a+d+1,b+d+1,d+\frac12\endmatrix\right]}
{\Gamma\left[\matrix a+b+1,a+\frac12,a+\frac12,1\\a+b+\frac12,a+1,b+1,\frac12\endmatrix\right]}%\\
%\times
\frac
{\omega([\circ\times]_a,[\circ\times]_b;2d)}
{\omega([\circ\times]_{a+b})},
\tag\Ecz
\endalign
$$
where in the interest of brevity we used the notation 
$$
\Gamma\left[\matrix a_1,\dotsc,a_k\\b_1,\dotsc,b_l\endmatrix\right]
=\frac
{\Gamma(a_1)\cdots\Gamma(a_k)}
{\Gamma(b_1)\cdots\Gamma(b_l)}.\tag\Ecza
$$
It readily follows from (\Ebp) that the first fraction on the right hand side of (\Ecz) has the limit
$$
\Gamma\left(\frac12\right)\frac{\Gamma(d+1)}{\Gamma\left(d+\frac12\right)}%\tag\Eczb
$$
as $a,b\to\infty$. Therefore, by taking the limit $a,b\to\infty$ in (\Ecz) and using that $\Gamma\left(\frac12\right)=\sqrt{\pi}$, we obtain
$$
\spreadlines{4\jot}
\align
\lim_{a,b\to\infty}
\frac
{\omega([\circ\times]_a,[\times\circ]_b;2d+1)}
{\omega([\circ\times]_a,[\times\circ]_b;1)}&=
\sqrt{\pi}
\frac{\Gamma(d+1)}{\Gamma\left(d+\frac12\right)}
%\\
%\times
\lim_{a,b\to\infty}
\frac
{\omega([\circ\times]_a,[\circ\times]_b;2d)}
{\omega([\circ\times]_{a+b})}.
\tag\Eczb
\endalign
$$
As by (\Ebp) one has
$$
\frac{\Gamma(d+1)}{\Gamma\left(d+\frac12\right)}\sim d^\frac12,\ \ \ d\to\infty,
$$
equations (\Eczb) and (\Ecf) imply (\Eci).
The proof is completed by noting that, just as it was the case for~(\Ech), equation (\Ecj) follows from Corollary {\Tbf} and Lemma {\Tcc}. \epf

The results stated in Theorem {\Tcd} can be extended to any finite number of slits. Define
$$
{\omega([\circ\times]_{a_1},\dotsc,[\circ\times]_{a_k};d_1,\dotsc,d_{k-1})}\tag\Eczc
$$
to be the correlation at the center (same as in \cite{\gd}) of the monomer cluster consisting of the fluctuating slit $[\circ\times]_{a_1}$ followed, after $d_1$ unaffected sites, by the  fluctuating slit $[\circ\times]_{a_2}$, and so on, ending, after $d_{k-1}$ unaffected sites, with the slit $[\circ\times]_{a_k}$. 

Then the same arguments that led to the above proof of Theorem {\Tcd} can be used to prove the following result.

\proclaim{Theorem \Tce} We have 
$$
\lim_{a_1,\dotsc,a_k\to\infty}\omega([\circ\times]_{a_1},\dotsc,[\circ\times]_{a_k};2d_1,\dotsc,2d_{k-1})
\sim
\frac{2^{\frac{k}{3}} e^{\frac{k}4}}{A^{3k}}(2d_1)^{-\frac14}\cdots(2d_k)^{-\frac14},\ \ \ d_1,\dotsc,d_k\to\infty.
\ \ \ \ \ \ \ \ \ \ \ \square
%\tag\Eczd
$$

\endproclaim

\mysec{4. The interaction of neutral slits of finite length}

In this section we study the asymptotics of the correlation of fluctuating slits of fixed length as the separation between them grows large. We obtain the following result.

\proclaim{Theorem \Tda} For any fixed integers $a,b\geq1$ we have, as $d\to\infty$, that
$$
\omega([\circ\times]_{a},[\circ\times]_{b};2d)
-\omega([\circ\times]_{a})\omega([\circ\times]_{b})
\sim
\frac{1}{\pi^{a+b}}E^2([\circ\times]_{a})E^2([\circ\times]_{b})\frac{ab}{4d^2},\tag\Eda
$$
and
$$
\omega([\circ\times]_{a},[\times\circ]_{b};2d+1)
-\omega([\circ\times]_{a})\omega([\times\circ]_{b})
\sim
-\frac{1}{\pi^{a+b}}E^2([\circ\times]_{a})E^2([\circ\times]_{b})\frac{ab}{4d^2}.\tag\Edaa
$$

On the other hand, for any non-negative integer $d$ we have
$$
\omega([\circ\times]_{a},[\circ\times]_{b};2d+1)
-\omega([\circ\times]_{a})\omega([\circ\times]_{b})
=
\omega([\circ\times]_{a},[\times\circ]_{b};2d)
-\omega([\circ\times]_{a})\omega([\times\circ]_{b})
=
0.\tag\Edaaa
$$

\endproclaim

\pf By Corollary {\Tbf} we have that
$$
\omega([\circ\times]_{a},[\circ\times]_{b};2d)
-\omega([\circ\times]_{a})\omega([\circ\times]_{b})
=
\frac{1}{\pi^{a+b}}E^2([\circ\times]_{a})E^2([\circ\times]_{b})
\left(E^2_*([\circ\times]_{a},[\circ\times]_{b})-1\right),
\tag\Edb
$$
where the star at the subscript indicates that only those factors in (\Ebb) 
that correspond to defects in different slits are taken into account. 

By the calculation that led to (\Ecc) we obtain that
$$
E^2_*([\circ\times]_{a},[\circ\times]_{b})
=
\prod_{i=1}^b\frac{(d+i)_a^2}{\left(d+i-\frac12\right)_a\left(d+i+\frac12\right)_a}.
\tag\Edc
$$
Therefore, we have
$$
\spreadlines{3\jot}
\align
E^2_*([\circ\times]_{a},[\circ\times]_{b})-1
&=
\prod_{i=1}^b\frac{(d+i)_a^2}{\left(d+i-\frac12\right)_a\left(d+i+\frac12\right)_a}-1
\\
&=
\frac
{\prod_{i=1}^b A_i-\prod_{i=1}^b B_i}
{\prod_{i=1}^b B_i},\tag\Edd
\endalign
$$
where
$$
A_i=(d+i)^2(d+i+1)^2\cdots(d+i+a-1)^2\tag\Ede
$$
and 
$$
B_i=\left\{(d+i)^2-\frac14\right\}\left\{(d+i+1)^2-\frac14\right\}\cdots\left\{(d+i+a-1)^2-\frac14\right\},\tag\Edf
$$
for $i=1,\dotsc,b$. Write the numerator in (\Edd) as
$$
A_1\cdots A_{b-1}(A_b-B_b)+A_1\cdots A_{b-2}(A_{b-1}-B_{b-1})B_b+\cdots
+(A_1-B_1)B_2\cdots B_b.\tag\Edg
$$
It is clear from (\Ede) and (\Edf) that both $A_i$ and $B_i$ are asymptotically equal to $d^{2a}$, while $A_i-B_i$ is asymptotically equal to $\frac{a}{4}d^{2a-2}$, as $d\to\infty$. It follows then from (\Edd) and (\Edg) that 
$$
E^2_*([\circ\times]_{a},[\circ\times]_{b})-1
\sim
\frac{ab}{4d^2},\ \ \ d\to\infty,
$$
which together with (\Edb) implies (\Eda). Equation (\Edaa) follows by a perfectly analogous argument, while (\Edaaa) is a direct consequence of Corollary {\Tbf}. \epf

\medskip
\flushpar
{\smc Remark \rf.} The three possible behaviors of attraction, repellance, or no interaction appear here just as they did in the previous section. However, now the lengths $2a$ and $2b$ of the slits do have an effect, as the main term in the asymptotics is proportional to $ab/\pi^{a+b}$.

\mysec{5. Growing slit length and separation at the same time: A Casimir force}

In this section we consider what is perhaps the most natural set-up, namely to let the length of the fluctuating slits and the separation between them grow at the same rate. The results we obtain have a surprising interpretation, discussed in Remark \rg.

\proclaim{Theorem \Tea} If $a$, $b$ and $d$ go to infinity so that $a\sim\alpha n$, $b\sim\beta n$ and $d\sim\delta n$ for some fixed $\alpha,\beta,\delta>0$, we have
$$
\frac
{\omega([\circ\times]_a,[\circ\times]_b;2d)}
{\omega([\circ\times]_{a})\,\omega([\circ\times]_{b})}
\to
%\frac{2^\frac{1}{12} e^\frac14}{A^3}
\left[\frac{(\alpha+\delta)(\beta+\delta)}
{\delta(\alpha+\beta+\delta)}\right]^\frac14
%\frac{1}{n^\frac14}
\tag\Eea
$$
and 
$$
\frac
{\omega([\circ\times]_a,[\times\circ]_b;2d+1)}
{\omega([\circ\times]_{a})\,\omega([\times\circ]_{b})}
\to
%\frac{2^\frac{1}{12} e^\frac14}{A^3}
\left[\frac{\delta(\alpha+\beta+\delta)}
{(\alpha+\delta)(\beta+\delta)}\right]^\frac14
%\frac{1}{n^\frac14}
.\tag\Eeb
$$
The ratios of correlations on the left hand sides of $(\Eea)$ and $(\Eeb)$ corresponding to the other parity of the distance between the fluctuating slits are equal to $1$ for all non-negative integers $a$, $b$ and $d$.

\endproclaim

\pf Rewrite (\Ecn) as
$$
\spreadlines{4\jot}
\align
\frac
{\omega([\circ\times]_a, [\circ\times]_b;2d)}
{\omega([\circ\times]_{a+b})}
&
=
\prod_{i=1}^d
%\left\{
\left[\frac{\Gamma(a+b+i)\Gamma(i)}
{\Gamma(a+i)\Gamma(b+i)}\right]^2
%\right.
%\\
%&\ \ \ \ \ \ \ \ \ \ \ \ \ \ \ \ \ \ \
%\left.
%\times
\frac
{\Gamma\left(a+i-\frac12\right)\Gamma\left(a+i+\frac12\right)
\Gamma\left(b+i-\frac12\right)\Gamma\left(b+i+\frac12\right)}
{\Gamma\left(a+b+i-\frac12\right)\Gamma\left(a+b+i+\frac12\right)
\Gamma\left(i-\frac12\right)\Gamma\left(i+\frac12\right)}
%\right\}
\\
&
=
\prod_{i=1}^d\frac{\Gamma^2(i)}{\Gamma\left(i-\frac12\right)\Gamma\left(i+\frac12\right)}
\prod_{i=a+1}^{a+d}\frac{\Gamma\left(i-\frac12\right)\Gamma\left(i+\frac12\right)}{\Gamma^2(i)}
\\
&
\times
\prod_{i=b+1}^{b+d}\frac{\Gamma\left(i-\frac12\right)\Gamma\left(i+\frac12\right)}{\Gamma^2(i)}
\prod_{i=a+b+1}^{a+b+d}\frac{\Gamma^2(i)}{\Gamma\left(i-\frac12\right)\Gamma\left(i+\frac12\right)}
\\
&
=
\frac
{P_d\,\dfrac{P_{a+b+d}}{P_{a+b}}}
{\dfrac{P_{a+d}}{P_a}\,   \dfrac{P_{b+d}}{P_b}}
=
\frac{P_a\, P_b\, P_d\, P_{a+b+d}}{P_{a+b}\, P_{a+d}\, P_{b+d}},
\tag\Eec
\endalign
$$
where we set
$$
P_n:=\prod_{i=1}^n\frac{\Gamma^2(i)}{\Gamma\left(i-\frac12\right)\Gamma\left(i+\frac12\right)}.
\tag\Eed
$$
The asymptotics of the product $P_n$ has been worked out in Section 3, where we saw that it is given by~(\Ecww). Using (\Ecww) for each factor on the right hand side of (\Eec) we obtain 
$$
\frac
{\omega([\circ\times]_a,[\circ\times]_b;2d)}
{\omega([\circ\times]_{a+b})}
\sim
\frac{2^\frac{1}{12} e^\frac14}{A^3}
\left[\frac{(\alpha+\beta)(\alpha+\delta)(\beta+\delta)}
{\alpha\beta\delta(\alpha+\beta+\delta)}\right]^\frac14
\frac{1}{n^\frac14}.\tag\Eee
$$
By Lemma {\Tcc} and (\Ecww) we have that
$$
\omega([\circ\times]_{n})
\sim
\frac{2^\frac{1}{12} e^\frac14}{A^3}
\frac{1}{2^n n^\frac14},\ \ \ n\to\infty,
\tag\Eef
$$
and therefore
$$
\frac
{\omega([\circ\times]_{a+b}}
{\omega([\circ\times]_{a}\omega([\circ\times]_{b}}
\sim
\frac{A^3}{2^\frac{1}{12} e^\frac14}
\left(\frac{\alpha\beta}{\alpha+\beta}\right)^\frac14 
n^\frac14,\ \ \ a\sim\alpha n,\, b\sim\beta n.
\tag\Eeg
$$
Multiplying together (\Eee) and (\Eeg) we obtain (\Eea).

To prove (\Eeb), note that by Lemma {\Tcc} and equation (\Ecy), we have
$$
\spreadlines{4\jot}
\align
&
\frac
{\omega([\circ\times]_a,[\times\circ]_b;2d+1)}
{\omega([\circ\times]_{a})\,\omega([\times\circ]_b)}
=
\frac
{\omega([\circ\times]_a,[\times\circ]_b;2d+1)}
{\omega([\circ\times]_a,[\circ\times]_b;2d)}
\frac
{\omega([\circ\times]_a,[\circ\times]_b;2d)}
{\omega([\circ\times]_{a})\,\omega([\times\circ]_{b})}
\\
&\ \ \ \ \ \ \ \ \ \ \ 
=
\frac
{\Gamma(a+b+d+1)\Gamma\left(a+d+\frac12\right)\Gamma\left(b+d+\frac12\right)\Gamma(d+1)}
{\Gamma\left(a+b+d+\frac12\right)\Gamma\left(a+d+1\right)\Gamma\left(b+d+1\right)
\Gamma\left(d+\frac12\right)}
\frac
{\omega([\circ\times]_a,[\circ\times]_b;2d)}
{\omega([\circ\times]_{a})\,\omega([\circ\times]_{b})}.\tag\Eeh
\endalign
$$
Applying four times formula (\Ebp) for the asymptotics of ratios of Gamma functions on the right
hand side of (\Eeh) and using (\Eea) one obtains (\Eeb). 

The last fact in the statement of the theorem follows from Corollary {\Tbf}, as for the remaining parities of the separating distance the two slits can be viewed as collections of dipoles of opposite parities. \epf

\medskip
\flushpar
{\smc Remark \rg.} In the statement of the above result we chose to normalize by the ``$n=\infty$'' position of the strings, because this leads to the simplest formulas. The variant with $\omega([\circ\times]_{a+b})$ as the normalizing denominator is expressed by equation (\Eee).

\topinsert
\centerline{\mypic{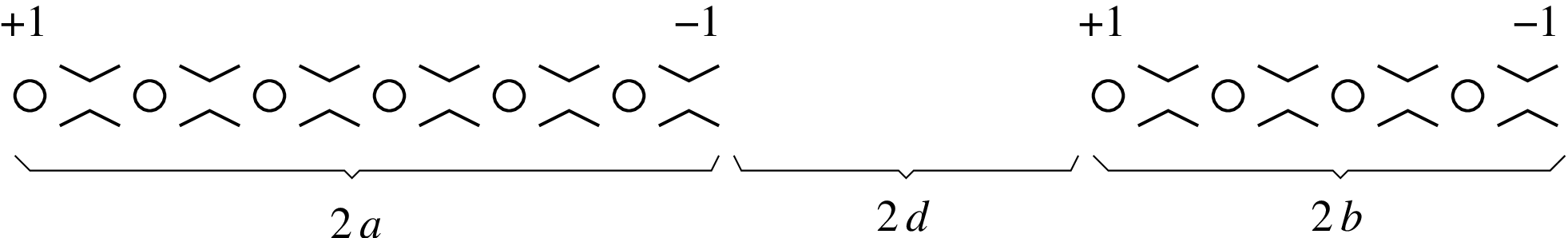}}
\medskip
\centerline{{\smc Figure~{\Fea}.} {\rm Interpretation of the expressions in Theorem {\Tea}.}}
\endinsert

The expressions that arise in Theorem {\Tea} have the following conceptual interpretation. Place a positive unit charge at each hole-end of a fluctuating slit, and a negative unit charge at each separation-end (this is illustrated in Figure {\Fea}). Form the Coulomb product corresponding to this system of 2D charges (see \cite{\gd,(2.10)}), omitting contributions coming from the ends of the same string (which cancel out due to the normalization), and {\it take its square root.} Then the resulting expression provides the answer to the asymptotic questions addressed in Theorem {\Tea}. It is as if the fluctuating strings would get polarized at the ends, with the ends interacting by a square root Coulomb law.

It is remarkable that this interpretation works for both (\Eea) and (\Eeb). Looking back at how (\Eeb) was deduced, one sees that this common interpretation of (\Eea) and (\Eeb) hinged upon the ratios of Gamma functions in (\Eeh) bringing in just the right factors when their asymptotics was considered. We believe that this is more than just a coincidence, and that this phenomenon may well govern the interaction of parallel fluctuating slits in any relative position. In particular we make the following conjecture.

Denote by $\omega_\parallel([\circ\times]_a,[\times\circ]_a;2d)$ the correlation of the fluctuating slits $[\circ\times]_a$ and $[\times\circ]_a$ when the latter is translated off $\ell$ so that it is positioned straight above the former, with $2d$ white horizontal diagonals of the square grid in between them (we consider the grid graph colored in chessboard fashion, with the vertices on $\ell$ white).

\proclaim{Conjecture {\Teb}} If $a$ and $d$ grow to infinity so that $a\sim\alpha n$ and $d\sim\delta n$ for some fixed $\alpha,\delta>0$, then we have
$$
\frac
{\omega_\parallel([\circ\times]_a,[\times\circ]_a;2d)}
{\omega([\circ\times]_a)\,\omega([\times\circ]_a)}
\to
\sqrt{1+\frac{\alpha^2}{\delta^2}}.\tag\Eei
$$

\endproclaim

Note that for $\delta/\alpha\to0$ this would imply that
$$
\frac
{\omega_\parallel([\circ\times]_a,[\times\circ]_a;2d)}
{\omega([\circ\times]_a)\,\omega([\times\circ]_a)}
\sim
\frac{\alpha}{\delta}.\tag\Eej
$$
Since $\alpha/\delta$ is also the limit of the ratio between the lengths of the parallel slits and the distance between them, (\Eej) could be viewed as a two dimensional analog of the Casimir force experienced by two parallel mirrors in vacuum placed very close to one another. In physics the Casimir force is explained by the quantum fluctuations of the vacuum, which is also the physical explanation for the electrostatic interaction of charged particles. Having a Casimir-like force arise purely from the randomness in a dimer system, just as we saw in \cite{\sc}\cite{\ec}\cite{\ef}\cite{\ov} the Coulomb force emerging, lends further support to the parallel between dimer systems with gaps and physical theories.

A comment is in order about the exception at the end of the statement of Theorem {\Tea}. Namely, one could wonder if the two parallel slits in Conjecture {\Teb} might be independent, just as they are when placed along $\ell$ and separated by an even number of sites. An analysis of parallel dipoles $[\circ\times]_1$ and $[\times\circ]_1$ shows that this independence happens only in one other direction besides both dipoles being on $\ell$ and separated by an even number of sites, namely when one dipole is situated straight above the other, with an even number of white diagonals in between. However, for the more general situation of $[\circ\times]_k$ and $[\times\circ]_k$, as soon as $k>1$ this independence ceases to hold. It is based on this and the interpretation presented in Remark~\rg that we were lead to formulate the above conjecture.

\mysec{6. A giant slit and a dipole}

In the previous sections we studied the asymptotics of the correlation of two long fluctuating slits. Another natural situation concerns the interaction of a long fluctuating slit and a short one. We discuss in this section the case when one slit has a fixed size (in the extreme case being just a dipole), while the other slit, as well as the distance between the slits, grows to infinity.

\proclaim{Theorem \Tfa} For any fixed positive integer $b$, as $a$ and $d$ approach infinity we have
$$
\frac
{\omega([\circ\times]_a,[\circ\times]_b;2d)}
{\omega([\circ\times]_a)\,\omega([\circ\times]_b)}
=
1+
\frac{b}{4}
\frac{a}{d(a+d)}
+O\left(\frac{1}{d^2}\right).
\tag\Efa
$$

\endproclaim

\pf Corollary {\Tbf} (with $l=0$) gives simple product expressions for both the numerator and denominator of the fraction on the left hand side of (\Efa). After simplifying out the common factors and converting the resulting Pochhammer symbols to Gamma functions, one arrives at the expression
$$
\spreadlines{4\jot}
\align
\frac
{\omega([\circ\times]_a,[\circ\times]_b;2d)}
{\omega([\circ\times]_a)\,\omega([\circ\times]_b)}
&=
\frac
{\Gamma\left(d+\frac12\right)\Gamma\left(d+\frac32\right)}
{\Gamma^2(d+1)}
\frac
{\Gamma^2(a+d+1)}
{\Gamma\left(a+d+\frac12\right)\Gamma\left(a+d+\frac32\right)}
\\
&
\cdot
\frac
{\Gamma\left(d+\frac32\right)\Gamma\left(d+\frac52\right)}
{\Gamma^2(d+2)}
\frac
{\Gamma^2(a+d+2)}
{\Gamma\left(a+d+\frac32\right)\Gamma\left(a+d+\frac52\right)}
\\
&
\ \ \ \ \ \ \ \ \ \ \ \ \ \ \ \ \ \ \ \ \ \ \ \ \ \ \ \ \ \ \ \ \ \ \vdots
\\
&
\cdot
\frac
{\Gamma\left(d+b-\frac12\right)\Gamma\left(d+b+\frac12\right)}
{\Gamma^2(d+b)}
\frac
{\Gamma^2(a+d+b)}
{\Gamma\left(a+d+b-\frac12\right)\Gamma\left(a+d+b+\frac12\right)}.
\tag\Efb
\endalign
$$
By (\Ebp), for any $1\leq i\leq b$ we have that 
$$
\spreadlines{4\jot}
\align
&
\frac
{\Gamma\left(d+i-\frac12\right)\Gamma\left(d+i+\frac12\right)}
{\Gamma^2(d+i)}
\frac
{\Gamma^2(a+d+i)}
{\Gamma\left(a+d+i-\frac12\right)\Gamma\left(a+d+i+\frac12\right)}
=
\\
&
\left(1-\frac{1}{8d}+O\left(\frac{1}{d^2}\right)\!\right)\!\!
\left(1+\frac{3}{8d}+O\left(\frac{1}{d^2}\right)\!\right)\!\!
\left(1+\frac{1}{8(a+d)}+O\left(\frac{1}{(a+d)^2}\!\!\right)\right)\!\!
\left(1-\frac{3}{8(a+d)}+O\left(\frac{1}{(a+d)^2}\!\!\right)\right)
\\
&\ \ \ \ \ \ \ \ \ \ \ \ \ \ 
=1+\frac{a}{4d(a+d)}+O\left(\frac{1}{d^2}\right).\tag\Efc
\endalign
$$
Multiplying together equations (\Efc) for $i=1,\dotsc,b$ we obtain (\Efa). \epf

For the special case when $b=1$, $a=n$ and $d$ grows to infinity so that $d\sim\delta n^\epsilon$ for some fixed $\delta,\epsilon>0$, we obtain that the $n\to\infty$ asymptotics of the correlation of a slit of length $2n$ with a dipole a distance $2\delta n^\epsilon$ from it is given by
$$
\spreadmatrixlines{3\jot}
\frac
{\omega([\circ\times]_a,[\circ\times]_1;2d)}
{\omega([\circ\times]_a)\,\omega([\circ\times]_1)}
-1
\sim
\left\{\matrix
\dfrac{1}{4\delta n^\epsilon}, & \ \ \ \ \ 0<\epsilon<1\\
\dfrac{1}{4\delta(1+\delta)n}, & \epsilon=1\\
\dfrac{1}{4\delta n^{2\epsilon-1}}, & \epsilon>1.
\endmatrix\right.
\tag\Efd
$$
Note the jump in the multiplicative constant when $\epsilon=1$; it is due to the fact that in that case (and that case only) the dominant part of $a+d$ is the sum of the dominant parts  of $a$ and $d$.

\mysec{7. The off center behavior of the correlation of defect clusters}

The correlation at the center is expected to be equal to the Fisher-Stephenson correlation defined via including the defects near the center of large squares, as the dimer statistics is not deformed by the effects of the boundary of the Aztec diamonds at their center (cf. \cite{\CEP}). However, at any point different from the center the dimer statistics {\it is} deformed. The occupation probabilities for individual dimers are given in \cite{\CEP}.
% formulas for the occupation probabilities for finite collections of dimers (also called dimer configurations) are conjectured in \cite{\CKP}. 
%In Section ?? we show that our results from this paper can be used to confirm the latter in some special cases. 
In this section we extend the scope to studying the correlation of arbitrary defect clusters (of not necessarily zero charge, as the dimers have) around an arbitrary point on the horizontal symmetry axis of the scaled Aztec diamond.

%As in \cite{\gd}, the Azted reactangle $AR_{m,n}$ is the graph whose vertices are the white squares, and whose edges connect precisely those pairs of white squares that are diagonally adjacent, in a $(2m+1)\times(2n+1)$ rectangular chessboard with black corners (which is placed so that its sides are parallel to the coordinate axes). If one labels the vertices on the horizontal symmetry axis of $AR_{2n,2n+k-l}$ by $1,2,\dotsc,2n+k-l$, then $AR_{2n,2n+k-l}(\{h_1,\dotsc,h_k\},\{s_1,\dotsc,s_l\})$ is the graph obtained from $AR_{2n,2n+k-l}$ by creating unit holes at the $h_i$'s, and separations at the $s_j$'s. 

Note that for any fixed $k$ and $l$ the scaling limit of the Aztec rectangles $AR_{2n,2n+k-l}$ is a square $S$, just as for Aztec diamonds (which correspond to $k=l$). Assume things are arranged so that $S$ has the interval $[-1,1]$ on the real line as its horizontal symmetry axis.

For any real number $-1<\alpha<1$ and any finite sets of integers $H=\{h_1,\dotsc,h_k\}$ and $S=\{s_1,\dotsc,s_l\}$ with $H\cap S=\emptyset$, define the {\it joint correlation $\tilde\omega_\alpha$ of having holes on $\ell$ at the elements of $H$ and separations on $\ell$ at the elements of $S$ in a window that scales to the point $\alpha$ on the horizontal symmetry axis of $S$} by
$$
\spreadlines{4\jot}
\align
&
\tilde\omega_\alpha(H,S)=\tilde\omega_\alpha(h_1,\dotsc,h_k;s_1,\dotsc,s_l)=
\\
&%\ \ \ \ \ \ \ \ \ \ \ \ \ \
\lim_{n\to\infty}\frac
{\M(AR_{2n,2n+k-l}(\{\lfloor n\alpha\rfloor+h_1,\dotsc,\lfloor n\alpha\rfloor+h_k\},
\{\lfloor n\alpha\rfloor+s_1,\dotsc,\lfloor n\alpha\rfloor+s_l\}))}
{\M(AR_{2n,2n+k-l}(\{\lfloor n\alpha\rfloor,\lfloor n\alpha\rfloor\!+\!1,\dotsc,\lfloor n\alpha\rfloor\!+\!k\!-\!1\},
\{\lfloor n\alpha\rfloor+k,\lfloor n\alpha\rfloor+k+1,\dotsc,\lfloor n\alpha\rfloor+k+l-1\}))},
\\
\tag\Ega
\endalign
$$
where the Aztec rectangles on the right hand side are translated so that they are symmetric about $\ell$, and their vertices on $\ell$ have coordinates $-n,-n+1,\dotsc,n+k-l-1$. 

\medskip
\flushpar
{\smc Remark \rh.} Note that $\tilde\omega_\alpha$ is an extension of the correlation $\tilde\omega$ of \cite{\gd, Section\, 5}, which corresponds to $\alpha=0$; we chose to extend this rather than the correlation $\bar\omega$ of \cite{\gd} because this makes the presentation simpler.

The behavior of the correlation $\tilde\omega_\alpha$ of an arbitrary collection of holes and separations under elementary moves (i.e., moving a hole one unit to the left, or moving a separation one unit to the left) is given by the following result, which is an $\alpha$-version of Lemma 5.1' of \cite{\gd}.

As in \cite{\gd}, for any two distinct integers $x$ and $y$ we denote by $\langle x, y\rangle$ the set of integers strictly in between $x$ and $y$, and define the {\it likes kernel} $L_D(x,y)$ and the {\it unlikes kernel} $U_D(x,y)$ by
$$
L_D(x,y):=
\spreadmatrixlines{3\jot}
\left\{\matrix
\dfrac{\Gamma\left(\frac{|x-y|-1}{2}\right)\Gamma\left(\frac{|x-y|+1}{2}\right)}
{\Gamma^2\left(\frac{|x-y|}{2}\right)},\text{\rm \ \ \ if $|\langle x,y\rangle\setminus D|$ is even}
\\
\dfrac{\Gamma\left(\frac{|x-y|}{2}\right)\Gamma\left(\frac{|x-y|}{2}+1\right)}
{\Gamma^2\left(\frac{|x-y|+1}{2}\right)},\text{\rm \ \ \ if $|\langle x,y\rangle\setminus D|$ is odd},
\endmatrix
\right.
\tag\Egb
$$
and
$$
U_D(x,y):=
\spreadmatrixlines{3\jot}
\left\{\matrix
\dfrac{\Gamma\left(\frac{|x-y|}{2}\right)\Gamma\left(\frac{|x-y|}{2}+1\right)}
{\Gamma^2\left(\frac{|x-y|+1}{2}\right)},\text{\rm \ \ \ if $|\langle x,y\rangle\setminus D|$ is even}
\\
\dfrac{\Gamma\left(\frac{|x-y|-1}{2}\right)\Gamma\left(\frac{|x-y|+1}{2}\right)}
{\Gamma^2\left(\frac{|x-y|}{2}\right)},\text{\rm \ \ \ if $|\langle x,y\rangle\setminus D|$ is odd}.
\endmatrix
\right.
\tag\Egc
$$

\proclaim{Proposition {\Tga} (Elementary Move in $\alpha$-window)} Let $a_1,\dotsc,a_k,b_1,\dotsc,b_l$
%$a_1<\dotsc<a_k$ and $b_1<\dotsc<b_l$ 
be distinct integers. 

Then if $a_i-1\notin\{a_1,\dotsc,a_k,b_1,\dotsc,b_l\}$, we have
$$
\!
\frac
{\tilde\omega_\alpha(a_1,\dotsc,a_{i-1},a_i,a_{i+1},\dotsc,a_k;b_1,\dotsc,b_l)}
{\tilde\omega_\alpha(a_1,\dotsc,a_{i-1},a_i-1,a_{i+1},\dotsc,a_k;b_1,\dotsc,b_l)}
=
\sqrt{\frac{1-\alpha}{1+\alpha}}
\frac
{\prod_{j:a_j<a_i}L_D(a_i,a_j)\prod_{j:b_j<a_i}U_D(a_i,b_j)}
{\prod_{j:a_j>a_i}L_{D'}(a_i-1,a_j)\prod_{j:b_j>a_i}U_{D'}(a_i-1,b_j)},
\tag\Egd
$$
where $D$ and $D'$ are the sets consisting of the locations of defects whose correlations are considered in the numerator and denominator on the left hand side, respectively.

Similarly, if $b_i-1\notin\{a_1,\dotsc,a_k,b_1,\dotsc,b_l\}$, we have
$$
\frac
{\tilde\omega_\alpha(a_1,\dotsc,a_k;b_1,\dotsc,b_{i-1},b_i,b_{i+1},\dotsc,b_l)}
{\tilde\omega_\alpha(a_1,\dotsc,a_k;b_1,\dotsc,b_{i-1},b_i-1,b_{i+1},\dotsc,b_l)}
=
\sqrt{\frac{1+\alpha}{1-\alpha}}
\frac
{\prod_{j:b_j<b_i}L_D(b_i,b_j)\prod_{j:a_j<b_i}U_D(b_i,a_j)}
{\prod_{j:b_j>b_i}L_{D'}(b_i-1,b_j)\prod_{j:a_j>b_i}U_{D'}(b_i-1,a_j)}
.\tag\Ege
$$

\endproclaim

\pf The stated equations can be proved by extending, step by step and in the same sequence, the arguments in the proof of Lemma 5.1' of \cite{\gd} --- which corresponds to $\alpha=0$ --- to the case of arbitrary $\alpha\in(-1,1)$. The main tool in the proof of \cite{\gd, Lemma\,5.1'} was to write the ratio of correlations, using their definition, as limits of ratios of numbers of perfect matchings of Aztec rectangles with defects along their symmetry axes, and use the $\Delta(\Cal O)\Delta(\Cal E)$ formula of \cite{\gd, Theorem\,1.1} to express the latter as ratios of type 
$\frac{\Delta(\Cal O)\Delta(\Cal E)}{\Delta(\Cal O')\Delta(\Cal E')}$. Since the systems of defects whose correlations are being compared are very close to one another (differing by a single elementary move), so are the sets $\Cal O$ and $\Cal O'$, and $\Cal E$ and $\Cal E'$, respectively. Therefore most factors in the ratio 
$\frac{\Delta(\Cal O)\Delta(\Cal E)}{\Delta(\Cal O')\Delta(\Cal E')}$ cancel out, and the remaining ones were seen in the proof of Lemma 5.1' of \cite{\gd} to lead precisely to the $\alpha=0$ specializations of the expressions on the right hand sides of (\Egd) and (\Ege).

The only difference is that some additional simplifications that occurred in the $n\to\infty$ limit for $\alpha=0$, no longer occur when $\alpha\neq0$. It is these no longer trivial contributions that end up producing the extra square root factors on the right hand sides of (\Egd) and (\Ege). We explain below how this comes about.

%The proof of Lemma 5.1' in \cite{\gd} was inductive on the number of defects. The induction step consisted of bringing in a new hole or separation to the right or to the left of the previous ones. 
%
%
%. The base case (that of no separations) was proved in its turn by an induction on the number of holes. We show below how the square root factors involving $\alpha$ arise in the base cases of these inductions (i.e., the case of the case of two holes), and in the two inductive steps. This will complete the proof.

The same arguments that led to \cite{\gd,(5.12)} prove that for any integers $a_1<a_2$ with an even number of integers in between we have\footnote{As in \cite{\gd}, all our arguments work just as well with $AR_{2n,2n+k-l}$ instead of $AR_{4n,4n+k-l}$, but then the indices of the Pochhammer symbols in (\Egf) and its analogs are $\lfloor\frac{n(1\pm\alpha)}{2}\rfloor$ rather than $\lfloor n(1\pm\alpha)\rfloor$; in order to keep the notation simpler, we stick to  $AR_{4n,4n+k-l}$ as the enclosing regions.}
$$
\spreadlines{4\jot}
\align
&
\frac
{\M(AR_{4n+2}(\{\lfloor 2n\alpha\rfloor+a_1,\lfloor 2n\alpha\rfloor+a_2\},\emptyset))}
{\M(AR_{4n+2}(\{\lfloor 2n\alpha\rfloor+a_1-1,\lfloor 2n\alpha\rfloor+a_2\},\emptyset))} 
\sim
\\
&\ \ \ \ \ \ \ \ \ \ \ \ \ \ 
\frac
{(1)_{\lfloor n(1+\alpha)\rfloor}}
{\left(\frac32\right)_{\lfloor n(1+\alpha)\rfloor}}
\frac
{\left(\frac32\right)_{\lfloor n(1-\alpha)\rfloor}}
{(1)_{\lfloor n(1-\alpha)\rfloor}}
\frac
{\left(\frac{d}{2}\right)_{\lfloor n(1-\alpha)\rfloor}\left(\frac{d}{2}+1\right)_{\lfloor n(1-\alpha)\rfloor}}
{\left(\frac{d+1}{2}\right)_{\lfloor n(1-\alpha)\rfloor}^2},\ \ \ n\to\infty,
\tag\Egf
\endalign
$$
where $d=a_2-a_1+1$. Indeed, the first two fractions on the right hand side of (\Egf) are the $\alpha$-versions of the two fractions in \cite{\gd, (5.8)}, and the third is the $\alpha$-version of \cite{\gd, (5.9)}.

Since $\alpha<1$, the large $n$ asymptotics of the third fraction on the right hand side of (\Egf) is clearly the same as that of its $\alpha=0$ specialization, and thus this part leads to the same contribution as in the $\alpha=0$ case.

Using formula (\Ebpp) to convert the Pochhammer symbols to ratios of Gamma functions, the product of the first two fractions on the right hand side of (\Egf) becomes
$$
\frac{\Gamma(\lfloor n(1+\alpha)\rfloor+1)}{\Gamma\left(\lfloor n(1+\alpha)\rfloor+\frac32\right)}
\frac{\Gamma\left(\lfloor n(1-\alpha)\rfloor+\frac32\right)}{\Gamma(\lfloor n(1-\alpha)\rfloor+1)}.\tag\Egg
$$
Application of (\Ebp) in the above expression readily shows that this approaches $\sqrt{\frac{1-\alpha}{1+\alpha}}$ as $n\to\infty$. Taking the limit as $n\to\infty$ in (\Egf) one obtains the proof of (\Egd) in this case. The case when there are an odd number of integers between $a_1$ and $a_2$ is handled the same way. 

The inductive step in the proof of Lemma 5.1' in \cite{\gd} is proved using equations \cite{\gd,(5.14)--(5.16)} for the case of bringing in a new hole, and equations \cite{\gd,(5.17)--(5.18)} for the case of bringing in a new separation. It is readily checked that the $\alpha$-versions of those calculations express the ratio of the correlations of the larger systems as the product between the ratio of the correlations of smaller systems and {\it the same} factor as in the $\alpha=0$ case. Indeed, this is seen by the same argument that showed that the $n\to\infty$ asymptotics of the third fraction on the right hand side of (\Egf) is the same for general $\alpha$ as for $\alpha=0$. Thus there are no further extra multiplicative factors in $\alpha$ for larger systems of defects, and equation (\Egd) follows. 

Equation (\Ege) is proved analogously. 
The origin of the multiplicative factor $\sqrt{\frac{1+\alpha}{1-\alpha}}$ on its right hand side resides in the fact that when a separation moves one unit to the left, in the corresponding $O$-$E$ patterns that amounts to a single $O$ (or $E$) moving one unit to the {\it right}, leading thus to the analog of (\Egf) for a moving separation having its first two fractions equal to the reciprocals of the ones displayed in (\Egf). \epf

\medskip
\flushpar
{\smc Remark \ri.} It follows from the proof of Proposition {\Tga} that the limit on the right hand side of (\Ega) exists. Indeed, the fraction whose limit is taken can readily be written as a finite product of similar fractions with the additional property that in each of them the systems of defects at the numerator and denominator are related by an elementary move. Limits of such fraction exist by the proof of Proposition~{\Tga}. Equations (\Egd) and (\Ege) can then be used to obtain an explicit, albeit complicated, product formula for the correlation~$\tilde\omega_\alpha$.

\medskip

If $O$ is a defect cluster on $\ell$ and $d\in\Z$, we denote by $d+O$ the translation of $O$ that moves all its constituent defects $d$ units to the right on $\ell$. Recall that the charge $\q(O)$ of the cluster $O$ is defined to equal the number of monomers in $O$ minus the number of separations in $O$.

\proclaim{Corollary \Tgb} Let $O_1$ and $O_2$ be arbitrary defect clusters. Then we have

$$
\lim_{d\to\infty}
\frac
{\tilde\omega_\alpha(O_1,d+1+O_2)}
{\tilde\omega_\alpha(O_1,d+O_2)}
=
\left(\frac{1-\alpha}{1+\alpha}\right)^\frac{\q(O_2)}{2}.
\tag\Egh
$$

\endproclaim

\pf Let $k$ be the number of defects in $O_2$, and for $i=0,1,\dotsc,k$, let $O_2^{(i)}$ be the defect cluster obtained from $d+1+O_2$ by shifting its $i$ leftmost defects one unit to the left. Then clearly $O_2^{(k)}=d+O_2$. Write
$$
\frac
{\tilde\omega_\alpha(O_1,d+1+O_2)}
{\tilde\omega_\alpha(O_1,d+O_2)}
=
\frac{\tilde\omega_\alpha(O_1,d+1+O_2)}{\tilde\omega_\alpha(O_1,O_2^{(1)})}
\frac{\tilde\omega_\alpha(O_1,O_2^{(1)})}{\tilde\omega_\alpha(O_1,O_2^{(2)})}
\cdots
\frac{\tilde\omega_\alpha(O_1,O_2^{(k-1)})}{\tilde\omega_\alpha(O_1,d+O_2)}.
\tag\Egi
$$
Apply Proposition {\Tga} to each fraction above. The square root factors involving $\alpha$ in (\Egd) and (\Ege) combine to give the expression on the right hand side of (\Egh). Thus it suffices to show that the product of the remaining factors approaches 1 as $d\to\infty$. To see this, note that the factors corresponding to pairs of defects within each $O_2^{(i)}$ cancel out even before taking the limit. Indeed, for each such pair of defects, there is a factor corresponding to the instance when the leftmost in the pair moves one unit to the left, and another when the rightmost in the pair moves one unit to the left; by Proposition {\Tga} these two factors are equal, and occur on opposite sides of the fraction bar. Therefore it is enough to show that the product of the rest of the factors approaches 1 as $d\to\infty$. This is so because each remaining factor corresponds to a pair of defects out of which one is in $O_1$ and the other in an $O_2^{(i)}$, and as $d\to\infty$ all ratios of products of Gamma functions on the right hand side in (\Egb) and (\Egc) are readily seen to approach 1. \epf

\medskip
\flushpar
{\smc Remark \rj.} Corollary {\Tgb} shows that for $\alpha>0$ the asymptotic behavior of $\tilde\omega_\alpha(O_1,d+O_2)$ as a function of the positive integer $d$ is exponential decay when $\q(O_2)>0$, and exponential blow-up when $\q(O_2)<0$. For $\alpha<0$, we get exponential decay when $\q(O_2)<0$, and exponential blow-up when $\q(O_2)>0$. 

It also follows from Corollary {\Tgb} that, quite surprisingly, when $\q(O_2)=0$, we have 
$$
\frac{\tilde\omega_\alpha(O_1,d+O_2)}{\tilde\omega_\alpha(O_1,O_2)}
=
\frac{\tilde\omega_0(O_1,d+O_2)}{\tilde\omega_0(O_1,O_2)}
=
\frac{\tilde\omega(O_1,d+O_2)}{\tilde\omega(O_1,O_2)},
\tag\Egj
$$
independently of $\alpha\in(-1,1)$, where $\tilde\omega$ is the correlation of \cite{\gd,\S5}. It is not hard to show that 
$$
\frac{\tilde\omega(O_1,d+O_2)}{\tilde\omega(O_1,O_2)}
=
\frac{\bar\omega(O_1,d+O_2)}{\bar\omega(O_1,O_2)},\tag\Egk
$$
where $\bar\omega$ is the correlation defined in \cite{\gd, \S2}.
It follows then from \cite{\gd,Theorem\,3.1} that for $\q(O_2)=0$ we have, for any $\alpha\in(-1,1)$, that
$$
\lim_{d\to\infty}\frac{\tilde\omega_\alpha(O_1,d+O_2)}{\tilde\omega_\alpha(O_1,O_2)}
=
\frac
{\bar\omega(O_1)\bar\omega(O_2)}
{\bar\omega(O_1,O_2)}.
\tag\Egl
$$

\medskip
\flushpar
{\smc Remark \rk.} %walls of charge
For $\alpha\neq0$, the above behavior of the correlation $\tilde\omega_\alpha(O_1,d+O_2)$ seems peculiar if one tries to understand it as an interaction between the defect clusters $O_1$ and $O_2$. The unusual thing is that, once $\alpha\neq0$ is fixed, this behavior is entirely determined by the charge of $O_2$, the moving cluster. This phenomenon can be understood in the light of the parallel to electrostatics developed in \cite{\sc}\cite{\ec}\cite{\ef}\cite{\ov} as follows.

Recall that the colors in the bipartition of the vertices of $AR_{2n,2n+k-l}(h_1,\dotsc,h_k;s_1,\dotsc,s_l)$ are such that the vertices on $\ell$ are white. Thus each monomer on $\ell$ is white. Recall that every separation can be viewed as the superposition of four trimers, each consisting of a white monomer on $\ell$ at the location of the separation, and two black neighboring monomers. Recall also that by the electrostatic picture that emerged from our earlier work, gaps in dimer systems on bipartite lattices (such as the square lattice considered currently) interact as electrical charges in 2D electrostatics, their charge being equal to the number of white monomers in the gap minus the number of black monomers in the gap. Therefore each monomer defect has charge $+1$, and each separation has charge $-1$. 

The new element in the current situation is that the boundary is making its presence felt when $\alpha\neq0$. More precisely, the situation is the following. The vertices of the square grid that are outside and to the left of the Aztec rectangle $AR_{2n,2n+k-l}(h_1,\dotsc,h_k;s_1,\dotsc,s_l)$, but have neighbors in it, are all white, and the same is true along the right boundary of the Aztec rectangle. Thus the portions of the grid that are to the left and to the right of the Aztec rectangle act on the defects on $\ell$ as enormous walls of positive charge, hugely repelling the monomers and hugely attracting the separations. Due to symmetry, when $\alpha=0$ their effects cancel, but as soon as we are off center (i.e., $\alpha$ becomes strictly positive or negative), they no longer do: Monomers are very strongly repelled towards the center, while separations are very strongly attracted to the left and right boundaries. 

The top and bottom portions of the boundary of the Aztec rectangle can be viewed similarly. 
The vertices of the square grid above the Aztec rectangle $AR_{2n,2n+k-l}(h_1,\dotsc,h_k;s_1,\dotsc,s_l)$ but having neighbors in it are all black, and the same is true along the bottom of the Aztec rectangle. Thus the portions of the grid that are above and below the Aztec rectangle act on the defects on $\ell$ as enormous negative charges, hugely attracting the monomers and hugely repelling the separations. Due to symmetry, the vertical components of these forces cancel out (in fact, the monomers and separations are required to stay on $\ell$ anyway), but the horizontal components do not. As a consequence, a monomer in an $\alpha$-window with $\alpha>0$, having longer portions of the top and bottom boundary to its left than to its right, will feel an attraction towards the center. Note that this effect of the top and bottom boundaries reinforces the previously discussed effect of the left and right boundaries. 

Proposition {\Tga} expresses quantitatively the cumulative effect of these competing electrical forces: Even moving one monomer of the cluster a single unit to the left in an $\alpha$-window with $\alpha>0$ will result in a configuration whose probability goes up by a factor of $\sqrt{\frac{1+\alpha}{1-\alpha}}$.

This way we see that in Corollary {\Tgb}, for $\alpha\neq0$, the behavior of $\tilde\omega_\alpha(O_1,d+O_2)$ as a function of $d$ is not due to the interaction of the moving cluster $O_2$ with the fixed cluster $O_1$ (for electrostatic interactions coming from a charge in $O_1$ and another in $d+O_2$ are essentially unchanged when $d$ is increased by 1), but it is instead caused by the interaction of $O_2$ with the boundary. A similar remark holds for Proposition~{\Tga}.

\medskip
The next result presents the relationship between the correlation $\tilde\omega_\alpha(O)$ of a defect cluster $O$ and the correlation $\tilde\omega_\alpha(1+O)$ of its translation one unit to the right. For $\alpha\neq0$ the effect of the boundary electrostatic forces discussed in the above remark is so powerful, that when the charge $\q(O)$ of the cluster is non-zero, even this one unit shift changes the value of the correlation $\tilde\omega_\alpha(O)$. However, when $\q(O)=0$, the electrostatic effects on the holes and separations turn out to exactly cancel out, so that $\tilde\omega_\alpha$ is translationally invariant for neutral defect clusters, for any $\alpha$.

\proclaim{Corollary \Tgc} For any distinct integers $h_1,\dotsc,h_k,s_1,\dotsc,s_l$ we have
$$
\tilde\omega_\alpha(h_1+1,\dotsc,h_k+1;s_1+1,\dotsc,s_l+1)
=
\left(\frac{1+\alpha}{1-\alpha}\right)^{\tfrac{\q(O)}{2}}
\tilde\omega_\alpha(h_1,\dotsc,h_k;s_1,\dotsc,s_l).
\tag\Egm
$$

\endproclaim

\pf Consider the sequence of $k+l$ defect clusters that transforms the defect cluster on the left hand side into the defect cluster on the right hand side by moving successive defects, one at a time, one unit to the left.  
Write the ratio of the two correlations in~(\Egm) as a product in the same manner as we did in~(\Egi). Apply Proposition {\Tga} to each of these fractions. The square root factors on the right hand sides of (\Egd) and (\Ege) combine to give the expression in $\alpha$ on the right hand side of (\Egm), while the other factors cancel out for the reason explained in the proof of Corollary {\Tgb}. 
\epf

Note that for neutral defect clusters the correlation $\tilde\omega_\alpha$ is especially natural, as the Aztec rectangles in its definition (\Ega) become simply Aztec diamonds.

Another consequence of Proposition {\Tga} is that if $O$ and $O'$ are two neutral defect clusters with the same ``center of charge'' (see (\Egn)), then the ratio $\tilde\omega_\alpha(O)/\tilde\omega_\alpha(O')$ is independent of $\alpha$.

\proclaim{Corollary \Tgd} Let $O=\{\circ_{h_1},\dotsc,\circ_{h_k},\times_{s_1},\dotsc,\times_{s_k}\}$ and $O'=\{\circ_{h'_1},\dotsc,\circ_{h'_k},\times_{s'_1},\dotsc,\times_{s'_k}\}$ be two neutral defect clusters. If the coordinates of the holes and separations in them satisfy
$$
h_1+\cdots+h_k-s_1-\cdots-s_k=h'_1+\cdots+h'_k-s'_1-\cdots-s'_k,\tag\Egn
$$
then we have that
$$
\frac
{\tilde\omega_\alpha(O)}
{\tilde\omega_\alpha(O')}
=
%\frac
%{\tilde\omega_0(O)}
%{\tilde\omega_0(O')}
%=
\frac
{\bar\omega(O)}
{\bar\omega(O')}\tag\Ego
$$
for any $\alpha\in(-1,1)$, where $\bar\omega$ is the correlation at the center of the Aztec diamonds defined in {\rm \cite{\gd,\S2}}.

\endproclaim

\pf Since $O$ and $O'$ have the same number of holes and the same number of separations, $O$ can be deformed into $O'$ by a finite sequence of elementary moves. As in (\Egi), $\tilde\omega_\alpha(O)/\tilde\omega_\alpha(O')$ is then the product of ratios of correlations of successive clusters in this deforming process. Apply Proposition {\Tga} to each of these fractions. In the resulting product expression, the only factors that depend on $\alpha$ are the square root factors on the right hand sides of (\Egd) and (\Ege). It is easy to see that the product of all the resulting square root factors is equal to 1 precisely when condition (\Egn) holds. The proof is completed by noting that for $\alpha=0$ the correlation $\tilde\omega_\alpha$ is the same as the correlation $\bar\omega$ of \cite{\gd,\S2}. \epf

\medskip
It would be interesting if there was a condition analogous to (\Egn) for dimers, so that if $D$ and $D'$ are two finite unions of dimers meeting this condition, then $\tilde\omega_\alpha(D)/\tilde\omega_\alpha(D')$ is independent of $\alpha$. 
%This would give some additional structure to Conjecture 13.5 of \cite{\CKP}. 

\medskip
%??Special case $\tilde\omega_\alpha(\circ_0\circ_1,\times_d\times_{d+1})$ how it's exp in $d$ and what it means in t.o. limits of inv Kast matrices in light of result from \cite{\KOS}?

%\bigskip
%\bigskip
%\bigskip

\mysec{8. The correlation of macroscopically separated defects in Aztec rectangles}

The purpose of this section is to determine the asymptotic behavior of the correlation of $k$ monomers and $l$ separations on the symmetry axis of a large Aztec rectangle, in the situation when the distances between defects, as well as the distances between them and the boundary, are comparable to the size of the Aztec rectangle. The asymptotic formulas we arrive at are stated in Theorems {\Thk} and {\Thn}.

We first present an explicit product formula for the number of perfect matchings of an Aztec rectangle with $k$ monomers in fixed positions on its symmetry axis $\ell$, in the case when the number of vertices on $\ell$ between any two consecutive monomers is even.

\proclaim{Proposition \Tha} For any integers $0\leq s_1<\cdots<s_k\leq n$ we have that
$$
\spreadlines{3\jot}
\align
&\!\!\!\!\!\!\!\!\!\!\!\!\!\!\!\!\!\!\!\!
\frac
{\M(AR_{2n,2n+k}(\{2s_1+1,2s_2+2,\dotsc,2s_k+k\},\emptyset))}
{\M(AR_{2n,2n+k}(\{1,2,\dotsc,k\},\emptyset))} 
=
\\
&\ \ \ \ \ \ 
\prod_{i=1}^{s_k}
\left[\frac{(1)_{i-1}\left(\frac32\right)_{n-i}}{\left(\frac32\right)_{i-1}(1)_{n-i}}\right]^2
\\
&\ \ 
\times
\prod_{i=1}^{s_{k-1}}
\left[
\frac
{(1)_{i-1}\left(\frac32\right)_{s_k-i}(s_k-i+2)_{n-s_k}}
{\left(\frac32\right)_{i-1}(1)_{s_k-i}\left(s_k-i+\frac32\right)_{n-s_k}}
\right]^2
\\
&\ \ 
\times
\prod_{i=1}^{s_{k-2}}
\left[
\frac
{(1)_{i-1}\left(\frac32\right)_{s_{k-1}-i}(s_{k-1}-i+2)_{s_k-s_{k-1}}\left(s_k-i+\frac52\right)_{n-s_k}}
{\left(\frac32\right)_{i-1}(1)_{s_{k-1}-i}\left(s_k-i+\frac32\right)_{n-s_k}(s_k-i+2)_{n-s_k}}
\right]^2
\\
&\ \ \ \ \ \ \ \ \ \ \ \ \ \ \ \ \ \ \ \ \ \ \ \ 
\vdots
\\
&\ \ 
\times
\prod_{i=1}^{s_{1}}
\left[
\frac
{(1)_{i-1}\left(\frac32\right)_{s_2-i}(s_2-i+2)_{s_3-s_2}\left(s_3-i+\frac52\right)_{s_4-s_3}\cdots
\left(s_k-i+\frac{k+2}{2}\right)_{n-s_k}}
{\left(\frac32\right)_{i-1}(1)_{s_2-i}\left(s_2-i+\frac32\right)_{s_3-s_2}(s_3-i+2_{s_4-s_3}\cdots
\left(s_k-i+\frac{k+1}{2}\right)_{n-s_k}}
\right]^2.\tag\Eha
\endalign
$$

\endproclaim

\medskip
\flushpar
Note that, due to forced edges in its perfect matchings, the graph at the denominator has the same number of perfect matchings as $AD_{2n}$, i.e., $2^{n(2n+1)}$. Thus (\Eha) provides an explicit formula for the number of perfect matchings of $AR_{2n,2n+k}(\{2s_1+1,2s_2+2,\dotsc,2s_k+k\},\emptyset)$.

\medskip
\pf We prove formula (\Eha) by induction on $k$. We claim that for any integer $i$ between 1 and $n$ we have
$$
\frac{\M(AR_{2n,2n+1}(\{2i+1\},\emptyset))}{\M(AR_{2n,2n+1}(\{2i\},\emptyset))}
=
\frac{(1)_{i-1}\left(\frac32\right)_{n-i}}{\left(\frac32\right)_{i-1}(1)_{n-i}}.
\tag\Ehb
$$
To see this, apply Theorem 1.1 of \cite{\gd} to the numerator and denominator on the left hand side above, but with each element $j$ of $\Cal O$ and $\Cal E$ replaced by an indeterminate $a_j$. One readily sees that, after simplifications, this yields the expression
$$
\frac
{(a_{2i}-a_{2})(a_{2i}-a_{4})\cdots(a_{2i}-a_{2i-2})(a_{2i}-a_{2i+3})(a_{2i}-a_{2i+5})\cdots(a_{2i}-a_{2n+1})}
{(a_{2i+1}-a_{2})(a_{2i+1}-a_{4})\cdots(a_{2i+1}-a_{2i-2})(a_{2i+1}-a_{2i+3})(a_{2i+1}-a_{2i+5})\cdots(a_{2i+1}-a_{2n+1})}.
\tag\Ehc
$$
Specializing back $a_j=j$ for all indices, the above expression is easily seen to be equal to the one on the right hand side of (\Ehb).

A similar argument shows that 
$$
\frac{\M(AR_{2n,2n+1}(\{2i\},\emptyset))}{\M(AR_{2n,2n+1}(\{2i-1\},\emptyset))}
=
\frac{(1)_{i-1}\left(\frac32\right)_{n-i}}{\left(\frac32\right)_{i-1}(1)_{n-i}},
\tag\Ehd
$$
for $i=1,\dotsc,n$. Indeed, the resulting analog of (\Ehc) is
$$
\frac
{(a_{2i-1}-a_{1})(a_{2i-1}-a_{3})\cdots(a_{2i-1}-a_{2i-3})(a_{2i-1}-a_{2i+2})(a_{2i-1}-a_{2i+4})\cdots(a_{2i-1}-a_{2n})}
{(a_{2i}-a_{1})(a_{2i}-a_{3})\cdots(a_{2i}-a_{2i-3})(a_{2i}-a_{2i+2})(a_{2i}-a_{2i+4})\cdots(a_{2i}-a_{2n})}.
\tag\Ehe
$$
Since each index in the above expression is exactly one unit less than the corresponding index in (\Ehc), when setting $a_j=j$ for all $j$ one obtains the same expression as before.

Multiplying together equations (\Ehb) and (\Ehd) for $i=1,\dotsc,s$, one obtains
$$
\frac{\M(AR_{2n,2n+1}(\{2s+1\},\emptyset))}{\M(AR_{2n,2n+1}(\{1\},\emptyset))}
=
\prod_{i=1}^s\left[\frac{(1)_{i-1}\left(\frac32\right)_{n-i}}{\left(\frac32\right)_{i-1}(1)_{n-i}}\right]^2,
\tag\Ehf
$$
which proves the base case $k=1$ of the induction.

For the induction step, assume that (\Eha) holds for $k-1$ monomers. Then we have
$$
\spreadlines{4\jot}
\align
&\!\!\!\!\!\!\!\!\!\!\!\!\!\!\!\!\!\!\!\!
\frac
{\M(AR_{2n,2n+k}(\{1,2s_2+2,2s_3+3,\dotsc,2s_k+k\},\emptyset))}
{\M(AR_{2n,2n+k}(\{1,2,\dotsc,k\},\emptyset))} 
=
\\
&\ \ 
\frac
{\M(AR_{2n,2n+k-1}(\{2s_2+1,2s_3+2,\dotsc,2s_k+k-1\},\emptyset))}
{\M(AR_{2n,2n+k-1}(\{1,2,\dotsc,k-1\},\emptyset))} 
=
\\
&\ \ \ \ \ \ 
\prod_{i=1}^{s_k}
\left[\frac{(1)_{i-1}\left(\frac32\right)_{n-i}}{\left(\frac32\right)_{i-1}(1)_{n-i}}\right]^2
\\
&\ \ 
\times
\prod_{i=1}^{s_{k-1}}
\left[
\frac
{(1)_{i-1}\left(\frac32\right)_{s_k-i}(s_k-i+2)_{n-s_k}}
{\left(\frac32\right)_{i-1}(1)_{s_k-i}\left(s_k-i+\frac32\right)_{n-s_k}}
\right]^2
\\
&\ \ 
\times
\prod_{i=1}^{s_{k-2}}
\left[
\frac
{(1)_{i-1}\left(\frac32\right)_{s_{k-1}-i}(s_{k-1}-i+2)_{s_k-s_{k-1}}\left(s_k-i+\frac52\right)_{n-s_k}}
{\left(\frac32\right)_{i-1}(1)_{s_{k-1}-i}\left(s_k-i+\frac32\right)_{n-s_k}(s_k-i+2)_{n-s_k}}
\right]^2
\\
&\ \ \ \ \ \ \ \ \ \ \ \ \ \ \ \ \ \ \ \ \ \ \ \ 
\vdots
\\
&\ \ 
\times
\prod_{i=1}^{s_{2}}
\left[
\frac
{(1)_{i-1}\left(\frac32\right)_{s_3-i}(s_3-i+2)_{s_4-s_3}\left(s_4-i+\frac52\right)_{s_5-s_4}\cdots
\left(s_k-i+\frac{k+1}{2}\right)_{n-s_k}}
{\left(\frac32\right)_{i-1}(1)_{s_3-i}\left(s_3-i+\frac32\right)_{s_4-s_3}(s_4-i+2_{s_5-s_4}\cdots
\left(s_k-i+\frac{k}{2}\right)_{n-s_k}}
\right]^2.\tag\Ehg
\endalign
$$
Indeed, the first equality follows because the hole at position 1 forces $2n$ edges along the left boundary
of the Aztec rectangles on the left hand side of (\Ehg) to be present in all their perfect matchings; after removing the vertices matched by these forced edges, the resulting graphs are precisely the ones involved in the middle expression in (\Ehg). The second equality follows by the induction hypothesis.

Next, we claim that for any $1\leq i<s_2$ we have
$$
\spreadlines{3\jot}
\align
&
\frac
{\M(AR_{2n,2n+k}(\{2i+1,2s_2+2,\dotsc,2s_k+k\},\emptyset))}
{\M(AR_{2n,2n+k}(\{2i,2s_2+2,\dotsc,2s_k+k\},\emptyset))} 
=
\\
&\ \ \ \ \ \ \ \ \ \ \ \ \ \ \ \ 
\frac
{(1)_{i-1}\left(\frac32\right)_{s_2-i}(s_2-i+2)_{s_3-s_2}\left(s_3-i+\frac52\right)_{s_4-s_3}\cdots
\left(s_k-i+\frac{k+2}{2}\right)_{n-s_k}}
{\left(\frac32\right)_{i-1}(1)_{s_2-i}\left(s_2-i+\frac32\right)_{s_3-s_2}(s_3-i+2_{s_4-s_3}\cdots
\left(s_k-i+\frac{k+1}{2}\right)_{n-s_k}}.
\tag\Ehh
\endalign
$$
This is a generalization of (\Ehb), and can be obtained by the same method. Namely, apply Theorem 1.1 of \cite{\gd} to the numerator and denominator on the left hand side above, with each element $j$ in $\Cal O$ and $\Cal E$ replaced by an indeterminate $a_j$. Then, after simplifications, the resulting expression is
$$
\align
\frac
{\displaystyle 
\prod_{1\leq j<2i+1\atop j \text{\ even}}(a_{2i}-a_j)
\prod_{2i+1<j<2s_2+2\atop j \text{\ odd}}(a_{2i}-a_j)     
\prod_{2s_2+2<j<2s_3+3\atop j \text{\ even}}(a_{2i}-a_j)
\cdots
\prod_{2s_{k}+k<j\leq 2n+k\atop j\equiv k \,(\text{mod}\,2)}(a_{2i}-a_j)}
{\displaystyle 
\prod_{1\leq j<2i+1\atop j \text{\ even}}(a_{2i+1}-a_j)
\prod_{2i+1<j<2s_2+2\atop j \text{\ odd}}(a_{2i+1}-a_j)     
\prod_{2s_2+2<j<2s_3+3\atop j \text{\ even}}(a_{2i+1}-a_j)
\cdots
\prod_{2s_{k}+k<j\leq 2n+k\atop j\equiv k \,(\text{mod}\,2)}(a_{2i+1}-a_j)}.
\\
\tag\Ehi
\endalign
$$
It is readily seen that, upon substituting $a_j=j$ for all $j$, the above expressions specializes to the right hand side of (\Ehh), and thus proves it.
 
A similar approach proves that 
$$
\spreadlines{3\jot}
\align
&
\frac
{\M(AR_{2n,2n+k}(\{2i,2s_2+2,\dotsc,2s_k+k\},\emptyset))}
{\M(AR_{2n,2n+k}(\{2i-1,2s_2+2,\dotsc,2s_k+k\},\emptyset))} 
=
\\
&\ \ \ \ \ \ \ \ \ \ \ \ \ \ \ \ 
\frac
{(1)_{i-1}\left(\frac32\right)_{s_2-i}(s_2-i+2)_{s_3-s_2}\left(s_3-i+\frac52\right)_{s_4-s_3}\cdots
\left(s_k-i+\frac{k+2}{2}\right)_{n-s_k}}
{\left(\frac32\right)_{i-1}(1)_{s_2-i}\left(s_2-i+\frac32\right)_{s_3-s_2}(s_3-i+2_{s_4-s_3}\cdots
\left(s_k-i+\frac{k+1}{2}\right)_{n-s_k}},
\tag\Ehj
\endalign
$$
for all $1\leq i<s_2$. Multiplying together equalities (\Ehh) and (\Ehj) for $i=1,\dotsc,s_1$, one arrives at 
$$
\spreadlines{3\jot}
\align
&
\frac
{\M(AR_{2n,2n+k}(\{2s_1+1,2s_2+2,\dotsc,2s_k+k\},\emptyset))}
{\M(AR_{2n,2n+k}(\{1,2s_2+2,\dotsc,2s_k+k\},\emptyset))} 
=
\\
&\ \ \ \ \ \ \ \ \ \ \ \ \ \ \ \ 
\prod_{i=1}^{s_1}
\left[
\frac
{(1)_{i-1}\left(\frac32\right)_{s_2-i}(s_2-i+2)_{s_3-s_2}\left(s_3-i+\frac52\right)_{s_4-s_3}\cdots
\left(s_k-i+\frac{k+2}{2}\right)_{n-s_k}}
{\left(\frac32\right)_{i-1}(1)_{s_2-i}\left(s_2-i+\frac32\right)_{s_3-s_2}(s_3-i+2_{s_4-s_3}\cdots
\left(s_k-i+\frac{k+1}{2}\right)_{n-s_k}}
\right]^2.
\tag\Ehk
\endalign
$$
Multiplication of (\Ehg) and (\Ehk) yields (\Eha), which completes the induction step and hence the proof. 
\epf

In order to carry out the asymptotic analysis of the formula in Proposition {\Eha}, it will be useful to rewrite it first in an equivalent form. 

For any positive integer $s$, and any positive integer or half-integer $a$, define the product $P_a(s)$ by
$$
P_a(s):=\prod_{i=1}^s\frac{\Gamma^2(i+a)}{\Gamma\left(i+a-\frac12\right)\Gamma\left(i+a+\frac12)\right)}.\tag\Ehl
$$
Define the product $Q(s,n)$, for $s$ a positive integer and $n$ a positive integer or half-integer, by
$$
Q(s,n):=\prod_{i=1}^s \frac{\Gamma(i)\Gamma(n-i)}{\Gamma\left(i+\frac12\right)\Gamma\left(n-i-\frac12\right)}.\tag\Ehm
$$
The equivalent form of (\Eha) that we will use for our asymptotic analysis is the following.

\proclaim{Corollary \Thb} We have that
$$
\spreadlines{3\jot}
\align
&\!\!\!\!\!\!\!\!\!\!\!\!\!\!\!\!\!\!\!\!\!\!\!\!\!\!\!\!\!\!\!\!\!\!\!\!\!\!\!\!\!\!\!\!\!\!\!\!\!\!\!\!\!\!
%\sqrt{
\frac
{\M(AR_{2n,2n+k}(\{2s_1+1,2s_2+2,\dotsc,2s_k+k\},\emptyset))}
{\M(AR_{2n,2n+k}(\{1,2,\dotsc,k\},\emptyset))}
%} 
=
\\
&\!\!\!\!\!
\left\{
\,Q\left(s_k,n+\frac32\right)
\right.
\\
\cdot
&\,Q(s_{k-1},n+2)\,\frac{P_{\frac12}(s_k)}{P_{\frac12}(s_k-s_{k-1})}
\\
\cdot
&\,Q\left(s_{k-2},n+\frac52\right)
\frac{P_{\frac12}(s_{k-1})}{P_{\frac12}(s_{k-1}-s_{k-2})}
\frac{P_1(s_k)}{P_1(s_k-s_{k-2})}
\\
&\ \ \ \ \ \ \ \ \ \ \ \ \ \ \ \ 
\vdots
\\
\cdot
&
\left.
\,Q\left(s_{1},n+\frac{k+2}{2}\right)
\frac{P_{\frac12}(s_{2})}{P_{\frac12}(s_{2}-s_{1})}
\frac{P_1(s_3)}{P_1(s_3-s_{1})}
\cdots
\frac{P_{\frac{k-1}{2}}(s_{k})}{P_{\frac{k-1}{2}}(s_{k}-s_{1})}
\right\}^2.
\tag\Ehn
\endalign
$$

\endproclaim

\pf Using formula (\Ebpp) to express the Pochhammer symbols in terms of Gamma functions, the expression on the right hand side of (8.1) becomes
$$
\spreadlines{3\jot}
\align
&
\prod_{j=1}^k\prod_{i=1}^{s_j}
\left[
\frac{\Gamma(i)\Gamma\left(n-i+\frac{k-j+3}{2}\right)}
{\Gamma\left(i+\frac12\right)\Gamma\left(n-i+\frac{k-j+2}{2}\right)}
\frac{\Gamma^2\left(s_{j+1}-i+\frac32\right)}
{\Gamma(s_{j+1}-i+1)\Gamma(s_{j+1}-i+2)}
\frac{\Gamma^2(s_{j+2}-i+2)}
{\Gamma\left(s_{j+2}-i+\frac32\right)\Gamma\left(s_{j+2}-i+\frac52\right)}
\cdots
\right.
\\
&\ \ \ \ \ \ \ \ \ \ \ \ \ \ \ \ \ \ \ \ \ \ \ \ \ \ \ \ \ \ \ \ \ \ \ \ \ \ \ \ \ \ \ \ \ \ \ \ \ \ \ \ \ \ \
\left.
\cdot
\frac{\Gamma^2\left(s_{k}-i+\frac{k-j+2}{2}\right)}
{\Gamma\left(s_{k}-i+\frac{k-j+1}{2}\right)\Gamma\left(s_{k}-i+\frac{k-j+3}{2}\right)}
\right]^2.
\tag\Eho
\endalign
$$
By the definition (\Ehm) of $Q$ we have
$$
\prod_{i=1}^{s_j}
\frac{\Gamma(i)\Gamma\left(n-i+\frac{k-j+3}{2}\right)}
{\Gamma\left(i+\frac12\right)\Gamma\left(n-i+\frac{k-j+2}{2}\right)}
=
Q\left(s_j,n+\frac{k-j+3}{2}\right),
$$
which accounts for the $Q$-factors in (\Ehn).

Note also that we have
$$
\spreadlines{4\jot}
\align
\prod_{i=1}^{s_{k-1}}\frac{\Gamma^2\left(s_k-i+\frac32\right)}{\Gamma(s_k-i+1)\Gamma(s_k-i+2)}
=
\frac
{\displaystyle \prod_{i=1}^{s_k}\frac{\Gamma^2\left(s_k-i+\frac32\right)}{\Gamma(s_k-i+1)\Gamma(s_k-i+2)}}
{\displaystyle \prod_{i=s_{k-1}+1}^{s_k}\frac{\Gamma^2\left(s_k-i+\frac32\right)}{\Gamma(s_k-i+1)\Gamma(s_k-i+2)}}
&=
\frac
{\displaystyle \prod_{i=1}^{s_k}\frac{\Gamma^2\left(i+\frac12\right)}{\Gamma(i+1)\Gamma(i+2)}}
{\displaystyle \prod_{i=1}^{s_k-s_{k-1}}\frac{\Gamma^2\left(i+\frac12\right)}{\Gamma(i+1)\Gamma(i+2)}}
\\
&=
\frac{P_{\frac12}(s_k)}{P_{\frac12}(s_k-s_{k-1})},
\endalign
$$
where the last equality holds by the definition (\Ehl) of $P$.
Rewriting all the remaining products of (\Eho) in this way, one is lead to (\Ehn). \epf

Next we work out the asymptotics of the products $P$ and $Q$.

\proclaim{Lemma \Thc} For any fixed positive integer $a$ we have that
$$
P_a(s)\sim
\frac{2^\frac{1}{12}e^\frac{1}{4}}{A^3}
\left(\prod_{i=1}^a\frac{\Gamma\left(i-\frac12\right)\Gamma\left(i+\frac12\right)}{\Gamma^2(i)}\right)
\frac{1}{s^\frac14},\ \ \ s\to\infty,\tag\Ehp
$$
and
$$
P_{a+\frac12}(s)\sim
\frac{A^3}{2^\frac{1}{12}e^\frac{1}{4}\pi^\frac12}
\left(\prod_{i=1}^a\frac{\Gamma(i)\Gamma(i+1)}{\Gamma^2\left(i+\frac12\right)}\right)
\frac{1}{s^\frac14},\ \ \ s\to\infty,\tag\Ehq
$$
where $A$ is the Glaisher-Kinkelin constant $(\Ecg)$.

\endproclaim

\pf Write
$$
P_{a+\frac12}(s)=\frac
{\displaystyle \prod_{i=1}^{a+s}\frac{\Gamma^2\left(i+\frac12\right)}{\Gamma(i)\Gamma(i+1)}}
{\displaystyle \prod_{i=1}^{a}\frac{\Gamma^2\left(i+\frac12\right)}{\Gamma(i)\Gamma(i+1)}},\tag\Ehr
$$
and note that the asymptotics of the product at the numerator was worked out in \cite{\gd, Proposition\,7.1}. Then (\Ehq) follows by (\Ehr) and \cite{\gd, (7.1)}.

To prove (\Ehp), write 
$$
P_{a}(s)=\frac
{\displaystyle \prod_{i=1}^{a+s}\frac{\Gamma^2(i)}{\Gamma\left(i-\frac12\right)\Gamma\left(i+\frac12\right)}}
{\displaystyle \prod_{i=1}^{a}\frac{\Gamma^2(i)}{\Gamma\left(i-\frac12\right)\Gamma\left(i+\frac12\right)}},
\tag\Ehs
$$
and note that
$$
\spreadlines{3\jot}
\align
\prod_{i=1}^n\frac{\Gamma^2\left(i+\frac12\right)}{\Gamma(i)\Gamma(i+1)}
&=
\frac{\Gamma\left(\frac32\right)\Gamma\left(\frac32\right)}{\Gamma(1)\Gamma(2)}
\frac{\Gamma\left(\frac52\right)\Gamma\left(\frac52\right)}{\Gamma(2)\Gamma(3)}
\cdots
\frac{\Gamma\left(\frac{n-1}{2}\right)\Gamma\left(\frac{n-1}{2}\right)}{\Gamma(n-1)\Gamma(n)}
\frac{\Gamma\left(\frac{n+1}{2}\right)\Gamma\left(\frac{n+1}{2}\right)}{\Gamma(n)\Gamma(n+1)}
\\
&=
\frac
{\displaystyle \frac{\Gamma\left(n+\frac12\right)}{\Gamma(n+1)}}
{\displaystyle \frac{\Gamma\left(\frac12\right)}{\Gamma(1)}}
\prod_{i=1}^n\frac{\Gamma\left(i-\frac12\right)\Gamma\left(i+\frac12\right)}{\Gamma^2(i)}.
\tag\Ehrr
\endalign
$$
Therefore the asymptotics of the product at the numerator on the right hand side of (\Ehs) follows from (\Ehrr) and the asymptotics of the product on the left hand side of (\Ehrr), which as we noted above is given by \cite{\gd, (7.1)}. This, together with (\Ehs), leads to (\Ehp). \epf

\proclaim{Lemma \Thd} For any fixed positive integer $k$ and rational number $\alpha\in(0,1)$, we have
$$
\left[Q\left(s,n+\frac{k}{2}\right)\right]^2\sim\frac{e^\frac14}{2^\frac{5}{12}A^3n^\frac14}
\frac{1}{\alpha^{\alpha n+\frac14}(1-\alpha)^{(1-\alpha)n+\frac14+\frac{k-3}{2}}},\ \ \ s=\alpha n,\ n\to\infty,
\tag\Ehss
$$
where it is understood that $n$ takes on only values for which $\alpha n\in\Z$, so that the left hand side is defined.

\endproclaim

\pf Due to the particular combinations of Gamma functions that arise, the calculations needed to deduce (\Ehss) differ slightly for the cases of even and odd $k$. We present below the details for the case of odd $k$.

%For any positive integer $n$ define
Recall our notation from Section 3:
$$
H(n)=0!\,1!\cdots(n-1)!\tag\Eht
$$ 
and
$$
E(n)=2!\,4!\cdots(2n)!,\tag\Ehu
$$
for any positive integer $n$.
It is routine to check that 
$$
Q\left(s,n+\frac32\right)=\frac{1}{2^{2s(n-s)}}\frac{H(s+1)H(n-s+1)}{H(n+1)}\frac{H(s)H(n-s)}{H(n)}\frac{E(n)}{E(s)E(n-s)}.\tag\Ehv
$$
%By the definition (\Ecg) of the Glaisher-Kinkelin constant $A$, we have that
Recall from Section 3 the asymptotic formulas
$$
H(n)\sim
\frac{(2\pi)^\frac{n}{2}n^{\frac{n^2}{2}-\frac{1}{12}}}{Ae^{\frac{3n^2}{4}-\frac{1}{12}}},\ \ \ n\to\infty
\tag\Ehw
$$
%Writing
%$$
%E^2(n)=2^n(0!\,1!\cdots(2n-1)!)n!\,(2n)!,
%$$
%one obtains from (\Ehg) and Stirling's approximation for the factorial that
and
$$
E^2(n)\sim 
\frac
{\displaystyle \pi^{n+1}2^{2n^2+4n+\frac{17}{12}}n^{2n^2+3n+\frac{11}{12}}}
{\displaystyle A\,e^{3n^2+3n-\frac{1}{12}}},\ \ \ n\to\infty.
\tag\Ehx
$$
Substituting $s=\alpha n$ in (\Ehv) and using the asymptotic formulas (\Ehw) and (\Ehx), we obtain, after simplifications, that
$$
\left[Q\left(\alpha n,n+\frac32\right)\right]^2\sim
\frac{e^\frac14}{2^\frac{5}{12}A^3\,n^\frac14}\frac{1}{\alpha^{\alpha n+\frac14}(1-\alpha)^{(1-\alpha)n+\frac14}},\ \ \ n\to\infty,
\tag\Ehy
$$ 
which checks (\Ehss) for $k=3$.

We can deduce the general odd $k$ case from the above by determining the relative change in $Q$ caused by incrementing $n$ by one unit. We have
$$
\frac{Q(s,n+1)}{Q(s,n)}
=
\frac{\Gamma(n)}{\Gamma\left(n-\frac12\right)}
\frac{\Gamma\left(n-s-\frac12\right)}{\Gamma(n-s)}
\sim
n^\frac12(n-s)^{-\frac12}
\to\sqrt{\frac{1}{1-\alpha}}, \ \ \ s=\alpha n,\ n\to\infty,
\tag\Ehz
$$
the second relation being true by (\Ebp). By repeated application of (\Ehz), (\Ehy) implies (\Ehs) for any odd $k$. 

The case of even $k$ follows by an analogous argument. \epf

We are now ready to present the asymptotics of the correlation of $k$ monomers when all separations between consecutive monomers are even.

\proclaim{Proposition \The} Let $0<\alpha_1<\cdots\alpha_k<1$ be rational numbers. Then if $s_i=\alpha_i n$, $i=1,\dotsc,k$, are integers, we have that
$$
\spreadlines{3\jot}
\align
&
\frac
{\M(AR_{2n,2n+k}(\{2s_1+1,2s_2+2,\cdots,2s_k+k\},\emptyset))}
{\M(AR_{2n,2n+k}(\{1,2,\cdots,k\},\emptyset))}
\sim
\\
&\ \ \ \ \ \ \ \ \ \ \
\left(\frac{e^\frac14}{2^\frac{5}{12}A^3\,n^\frac14}\right)^k
\prod_{i=1}^k\frac{1}{\alpha_i^{\alpha_i n+\frac14+\frac{i-1}{2}}(1-\alpha_i)^{(1-\alpha_i)n+\frac14+\frac{k-i}{2}}}
{\prod_{1\leq i<j\leq k}(\alpha_j-\alpha_i)^\frac12}, \ \ \ n\to\infty.
\tag\Ehza
\endalign
$$

\endproclaim

\pf Write the ratio on the left hand side using the formula provided by Corollary {\Thb}. Apply Lemmas~{\Thc} and {\Thd} to the $P$ and $Q$ factors in (\Ehn). The $k$ applications of formula (\Ehs) for the asymptotics of the $Q$ factors account for the multiplicative constant and almost all of the first product in (\Ehza), the omitted factors being $\alpha_i^{(i-1)/2}$. The latter, as well as the second product in (\Ehza), are precisely the overall contribution of the expressions (\Ehp)--(\Ehq) that result from the asymptotics of the $P$ factors in (\Ehn). Indeed, all the $P$ factors in (\Ehn) occur in pairs, with factors in the same pair sharing the same index, and being on opposite sides of the fraction bar; therefore all but the $s$-parts of the expressions (\Ehp)--(\Ehq) cancel out. It is readily checked that the leftover factors combine in the way we stated, which completes the proof. \epf

The next step towards the general result of this section --- which will address the case of an arbitrary finite collection of monomers and separations on $\ell$ --- is to extend Theorem {\The} to the case when the $k$ monomers have arbitrary positions on $\ell$, and not necessarily with an even number of sites between consecutive~monomers.

In order to accomplish this we will need the following ``finite version'' of the elementary move lemma of~\cite{\gd} (i.e., a version that takes into account the presence of the boundary of the enclosing Aztec rectangles).

\proclaim{Lemma {\Thf} (Elementary move in Aztec rectangles with $k$ holes)} Let $1\leq a_1<\cdots<a_k\leq 2n+k$ be integers, and assume $a_{i-1}<a_i-1$. Define the subset $S_{n,i}(a_1,\dotsc,a_k)$ of $\{1,2\dotsc,2n+k\}$ as follows. Starting from coordinate $a_i$, move towards the right on the integer coordinates on $\ell$, skipping over one existing node $($i.e., a node at which there is no monomer$)$ at each time. Next, starting from coordinate $a_i-1$, move towards the left on the integer coordinates on $\ell$, skipping over one existing node at each time. Let $S_{n,i}(a_1,\dotsc,a_k)$ be the set of coordinates that have been visited this way, not including coordinates $a_i$ and $a_i-1$ $($see Figure ${\Fha}$ for an illustration$)$. 

Then we have
$$
\frac
{\M(AR_{2n,2n+k}(\{a_1,\dotsc,a_{i-1},a_i,a_{i+1},\dotsc,a_k\},\emptyset)}
{\M(AR_{2n,2n+k}(\{a_1,\dotsc,a_{i-1},a_i-1,a_{i+1},\dotsc,a_k\},\emptyset)}
=
\prod_{i\in S_{n,i}(a_1,\dotsc,a_k)}\frac{|(a_i-1)-j|}{|a_i-j|}.
\tag\Ehzb
$$

\endproclaim

\topinsert
\centerline{\mypic{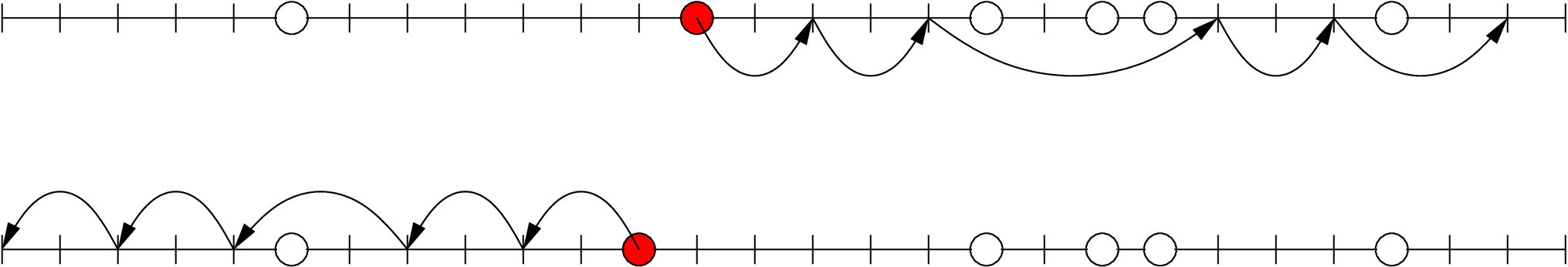}}
\medskip
\centerline{{\smc Figure~8.1.} {\rm The set $S_{11,2}(6,13,18,20,21,25)$ is the union of the sites with an arrow pointing to them.}}
\endinsert

\pf Apply Theorem 1.1 of \cite{\gd} to both the numerator and denominator on the left hand side above. Due to the fact that the defect distributions are nearly identical, most of the resulting factors occur both at the numerator and at the denominator, and thus cancel out. It is readily checked that the leftover factors combine to give precisely the expression on the right hand side of (\Ehzb). \epf

\proclaim{Corollary \Thg} If $1\leq a_1<\cdots<a_n\leq 2n+k$ are so that between $a_j$ and $a_{j+1}$  there are an even number of integers, for $j=1,\dotsc,n-1$, and $a_{i-1}<a_i-1$, then we have
$$
\spreadlines{5\jot}
\align
&
\frac
{\M(AR_{2n,2n+k}(\{a_1,\dotsc,a_{i-1},a_i,a_{i+1},\dotsc,a_k\},\emptyset)}
{\M(AR_{2n,2n+k}(\{a_1,\dotsc,a_{i-1},a_i-1,a_{i+1},\dotsc,a_k\},\emptyset)}
=
\\
&\ \ \ \ \ \ \ \ 
\frac
{\displaystyle \Gamma\left(\frac{a_i-2-\delta_{a_1,\,\text{odd}}}{2}\right)}
{\displaystyle \Gamma\left(\frac{a_i-1-\delta_{a_1,\,\text{odd}}}{2}\right)}
\frac
{\displaystyle \Gamma\left(\frac{2n+k-a_i+1+\delta_{a_k+k,\,\text{odd}}}{2}\right)}
{\displaystyle \Gamma\left(\frac{2n+k-a_i+\delta_{a_k+k,\,\text{odd}}}{2}\right)}
\\
&\ \ \ \ 
\times
\frac
{\displaystyle
 \frac{\displaystyle \Gamma^2\left(\frac{a_{i+1}-a_i+2}{2}\right)}
      {\displaystyle \Gamma\left(\frac{a_{i+1}-a_i+1}{2}\right)\Gamma\left(\frac{a_{i+1}-a_i+3}{2}\right)}
 \cdots
 \frac{\displaystyle \Gamma^2\left(\frac{a_{k}-a_i+2}{2}\right)}
      {\displaystyle \Gamma\left(\frac{a_{k}-a_i+1}{2}\right)\Gamma\left(\frac{a_{k}-a_i+3}{2}\right)}}
{\displaystyle
 \frac{\displaystyle \Gamma^2\left(\frac{a_{i}-a_1}{2}\right)}
      {\displaystyle \Gamma\left(\frac{a_{i}-a_1-1}{2}\right)\Gamma\left(\frac{a_{i}-a_1+1}{2}\right)}
 \cdots
 \frac{\displaystyle \Gamma^2\left(\frac{a_{i}-a_{i-1}}{2}\right)}
      {\displaystyle \Gamma\left(\frac{a_{i}-a_{i-1}-1}{2}\right)\Gamma\left(\frac{a_{i}-a_{i-1}+1}{2}\right)}},
\tag\Ehzc
\endalign
$$
where $\delta_{s,\,\text{odd}}$ is defined to be $1$ for odd $s$, and $0$ for even $s$.

\endproclaim

\pf Express the ratio on the left hand side using the product formula of Lemma {\Thf}. Both at the numerator and at the denominator of (\Ehzb), the factors can be grouped into runs in which consecutive members are integers at distance 2 from one another. Dividing at the top and at the bottom by $2^{|S_{n,i}(a_1,\dotsc,s_k)|}$, these are turned into runs of consecutive integers or consecutive half-integers. The assumption that there are an even number of integers between $a_i$ and $a_{i+1}$ for all $i$ makes it possible to write out explicitly the resulting products in terms of Pochhammer symbols. The expression obtained this way at the denominator is
$$
\spreadlines{4\jot}
\align
&
(1)_\frac{a_{i+1}-a_i-1}{2}
\left(\frac{a_{i+1}-a_i+2}{2}\right)_\frac{a_{i+2}-a_{i+1}-1}{2}
\left(\frac{a_{i+2}-a_i+2}{2}\right)_\frac{a_{i+3}-a_{i+2}-1}{2}
\cdots
\\
&\ \ \ \ \ \ \ \ \ \ \ \ \ \ \ \ 
\times
\left(\frac{a_{k-1}-a_i+2}{2}\right)_\frac{a_{k}-a_{k-1}-1}{2}
\left(\frac{a_{k}-a_i+2}{2}\right)_\frac{2n+k-a_k-\delta_{a_k+k,\,\text{odd}}}{2}
\\
&
\times
2\left(\frac12\right)_{\frac{a_i-a_{i-1}-1}{2}}
\left(\frac{a_{i}-a_{i-1}+1}{2}\right)_\frac{a_{i-1}-a_{i-2}-1}{2}
\left(\frac{a_{i}-a_{i-2}+1}{2}\right)_\frac{a_{i-2}-a_{i-3}-1}{2}
\cdots
\\
&\ \ \ \ \ \ \ \ \ \ \ \ \ \ \ \ 
\times
\left(\frac{a_{i}-a_{2}+1}{2}\right)_\frac{a_{2}-a_{1}-1}{2}
\left(\frac{a_{i}-a_{1}+1}{2}\right)_\frac{a_1-2-\delta_{a_1,\,\text{odd}}}{2}.
\tag\Ehzd
\endalign
$$
Using formula (\Ebpp) to express the Pochhammer symbols in terms of Gamma functions, the above expression becomes
$$
\spreadlines{4\jot}
\align
&
\frac{\Gamma\left(\frac{a_{i+1}-a_i+1}{2}\right)}{\Gamma(1)}
\frac{\Gamma\left(\frac{a_{i+2}-a_i+1}{2}\right)}{\Gamma\left(\frac{a_{i+1}-a_i+2}{2}\right)}
\frac{\Gamma\left(\frac{a_{i+3}-a_i+1}{2}\right)}{\Gamma\left(\frac{a_{i+2}-a_i+2}{2}\right)}
\cdots
%\\
%&\ \ \ \ \ \ \ \ \ \ \ \ \ \ \ \ 
%\times
\frac{\Gamma\left(\frac{a_{k}-a_i+1}{2}\right)}{\Gamma\left(\frac{a_{k-1}-a_i+2}{2}\right)}
\frac{\Gamma\left(\frac{2n+k-a_i+1+\delta_{a_k+k,\,\text{odd}}}{2}\right)}{\Gamma\left(\frac{a_{k}-a_i+2}{2}\right)}
\\
&
\times
\frac{2\,\Gamma\left(\frac{a_i-a_{i-1}}{2}\right)}{\Gamma\left(\frac12\right)}
\frac{\Gamma\left(\frac{a_{i}-a_{i-2}}{2}\right)}{\Gamma\left(\frac{a_{i}-a_{i-1}+1}{2}\right)}
\frac{\Gamma\left(\frac{a_{i}-a_{i-3}}{2}\right)}{\Gamma\left(\frac{a_{i}-a_{i-2}+1}{2}\right)}
\cdots
\frac{\Gamma\left(\frac{a_{i}-a_{1}}{2}\right)}{\Gamma\left(\frac{a_{i}-a_{2}+1}{2}\right)}
\frac{\Gamma\left(\frac{a_{i}-1-\delta_{a_k+k,\,\text{odd}}}{2}\right)}{\Gamma\left(\frac{a_{i}-a_{1}+1}{2}\right)}.
\tag\Ehze
\endalign
$$
A similar calculation shows that the analogous expression that results at the numerator is
$$
\spreadlines{4\jot}
\align
&
\frac{\Gamma\left(\frac{a_{i+1}-a_i+2}{2}\right)}{\Gamma\left(\frac32\right)}
\frac{\Gamma\left(\frac{a_{i+2}-a_i+2}{2}\right)}{\Gamma\left(\frac{a_{i+1}-a_i+3}{2}\right)}
\frac{\Gamma\left(\frac{a_{i+3}-a_i+2}{2}\right)}{\Gamma\left(\frac{a_{i+2}-a_i+3}{2}\right)}
\cdots
%\\
%&\ \ \ \ \ \ \ \ \ \ \ \ \ \ \ \ 
%\times
\frac{\Gamma\left(\frac{a_{k}-a_i+2}{2}\right)}{\Gamma\left(\frac{a_{k-1}-a_i+3}{2}\right)}
\frac{\Gamma\left(\frac{2n+k-a_i+2+\delta_{a_k+k,\,\text{odd}}}{2}\right)}{\Gamma\left(\frac{a_{k}-a_i+3}{2}\right)}
\\
&
\times
\frac{2\,\Gamma\left(\frac{a_i-a_{i-1}-1}{2}\right)}{\Gamma(1)}
\frac{\Gamma\left(\frac{a_{i}-a_{i-2}-1}{2}\right)}{\Gamma\left(\frac{a_{i}-a_{i-1}}{2}\right)}
\frac{\Gamma\left(\frac{a_{i}-a_{i-3}-1}{2}\right)}{\Gamma\left(\frac{a_{i}-a_{i-2}}{2}\right)}
\cdots
\frac{\Gamma\left(\frac{a_{i}-a_{1}-1}{2}\right)}{\Gamma\left(\frac{a_{i}-a_{2}}{2}\right)}
\frac{\Gamma\left(\frac{a_{i}-2-\delta_{a_k+k,\,\text{odd}}}{2}\right)}{\Gamma\left(\frac{a_{i}-a_{1}}{2}\right)}.
\tag\Ehzf
\endalign
$$
The ratio of (\Ehze) and (\Ehzd) yields the expression on the right hand side of (\Ehzc). \epf

%Recall that the vertices of $AR_{2n,2n+k}$ that are on $\ell$ are labeled from left to right by $1,2,\dotsc,2n+k$. 
%For any such vertex $j$, define $L_j$ (resp., $R_j$) to be the number of vertices on $\ell$ strictly to the left (resp., to the right) of $j$; i.e., we set
%$$
%\align
%L_j&:=j-1\tag\Ehzg
%\\
%R_j&:=2n+k-j\tag\Ehzh
%\endalign
%$$

We are now ready to present the case of $k$ monomers of arbitrary coordinates.
%in the situation when the distances between the monomers are fixed rational multiples 
%of the width of the Aztec rectangle.

\proclaim{Proposition \Thh} Let $0<\alpha_1<\cdots\alpha_k<2$ be rational numbers, and let $c_1,\dotsc,c_k$ be fixed integers. Then if $s_i=\alpha_i n$, $i=1,\dotsc,k$, are integers, we have, as $n\to\infty$, that
$$
\spreadlines{3\jot}
\align
%&
\frac
{\M(AR_{2n,2n+k}(\{s_1+c_1,s_2+c_2,\cdots,s_k+c_k\},\emptyset))}
{\M(AR_{2n,2n+k}(\{1,2,\cdots,k\},\emptyset))}
\sim
%\\
%&\ \ \ \ \ \ \ \ \ \ \
\left(\frac{e^\frac14}{2^\frac{5}{12}A^3\,n^\frac14}\right)^k
\frac
{\displaystyle \prod_{1\leq i<j\leq k}\sqrt{\frac{\alpha_j}{2}-\frac{\alpha_i}{2}}}
{\displaystyle \prod_{i=1}^k \left(\sqrt{\frac{\alpha_i}{2}}\,\right)^{L_i+\frac12}\left(\sqrt{1-\frac{\alpha_i}{2}}\,\right)^{R_i+\frac12}},
\\
\tag\Ehzi
\endalign
$$
where $L_i$ and $R_i$ are the number of vertices on $\ell\cap AR_{2n,2n+k}$ that are strictly to the left and right of $s_i+c_i$, respectively.

\endproclaim

\pf The case when $s_i+c_i=2s'_i+i$ for some $s'_i\in\Z$, $i=1,\dotsc,k$ --- which we will refer to as the even separation case for short ---
%the number of vertices on $\ell$ between any two consecutive monomers (as well as the number of vertices on $\ell$ to the left of the leftmost monomer) is even 
constitutes the subject of Proposition~{\The}. 

If the coordinates $a_1,\dotsc,a_k$ of the monomers are in the even separation case, Corollary {\Thg} allows us to deduce the case of coordinates $a_1,\dotsc,a_{i-1},a_i-1,a_{i+1},\dotsc,a_k$ as follows. By (\Ebp) we have that 
$$
\frac{\Gamma^2(x+1)}{\Gamma\left(x-\frac12\right)\Gamma\left(x+\frac12\right)}
\to 1,\ \ \ x\to\infty.
$$
Therefore the third fraction on the right hand side of (\Ehzc) approaches 1 as $n\to\infty$. On the other hand,~(\Ebp) also implies that the limit of the product of the first two fractions on the right hand side of~(\Ehzc) is
$$
\frac
{\displaystyle \Gamma\left(\frac{a_i-2-\delta_{a_1,\,\text{odd}}}{2}\right)}
{\displaystyle \Gamma\left(\frac{a_i-1-\delta_{a_1,\,\text{odd}}}{2}\right)}
\frac
{\displaystyle \Gamma\left(\frac{2n+k-a_i+1+\delta_{a_k+k,\,\text{odd}}}{2}\right)}
{\displaystyle \Gamma\left(\frac{2n+k-a_i+\delta_{a_k+k,\,\text{odd}}}{2}\right)}
\sim
a_i^{-\frac12}(2n+k-a_i)^\frac12
\to
\sqrt{\frac{2-\alpha_i}{\alpha_i}},\ \ \ n\to\infty.
\tag\Ehzj
$$
The above observations yield
$$
\frac
{\M(AR_{2n,2n+k}(\{a_1,\dotsc,a_{i-1},a_i,a_{i+1},\dotsc,a_k\},\emptyset)}
{\M(AR_{2n,2n+k}(\{a_1,\dotsc,a_{i-1},a_i-1,a_{i+1},\dotsc,a_k\},\emptyset)}
\to
\sqrt{\frac{2-\alpha_i}{\alpha_i}},\ \ \ n\to\infty.
\tag\Ehzk
$$
The case of coordinates $a_1,\dotsc,a_{i-1},a_i-1,a_{i+1},\dotsc,a_k$ follows from the even separation case $a_1,\dotsc,a_k$ using (\Ehzk). 

To cover the general case, note that in Corollary {\Thg} the only reason the coordinates $a_1,\dotsc,a_k$ were assumed to be in the even separation case was in order to be able to write down explicitly, and in a simpler way, the arguments of the Pochhammer symbols. It follows from the proof of Corollary {\Thg} that a formula analogous to (\Ehzc) exists for any coordinates $a_1,\dotsc,a_k$; depending on the parities of the $a_i$'s, the arguments of the resulting Gamma functions will be shifted by some finite multiples of $\frac12$, but the resulting formula has the same structure as the expression on the right hand side of (\Ehzc). But then the arguments employed above to deduce (\Ehzk) work without change, and imply that (\Ehzk) holds in fact for coordinates $a_1,\dotsc,a_k$ of arbitrary parity. 

Note that any configuration $a_1,\dotsc,a_k$ can be reached from a configuration in the even separation case, by doing a finite number of elementary moves (i.e., moving a single monomer one unit to the left). By (\Ehzk), the effect of each such move on the asymptotics of the left hand side of (\Ehzi) is to multiply it by $\sqrt{\frac{\alpha_i}{2-\alpha_i}}$, if the moved monomer was the $i$th one. Since the right hand sides of (\Ehzi) corresponding to $a_1,\dotsc,a_k$ and $a_1,\dotsc,a_{i-1},a_i-1,a_{i+1},\dotsc,a_k$ are related by the same multiplicative factor (as the above elementary move reduces $L_i$ by one unit and increases $R_i$ by one unit), (\Ehzi) follows by repeated applications of (\Ehzk). \epf

To address the case of a general distribution of defects on $\ell$ we will need the following result.

\proclaim{Lemma \Thi \ (Elementary move in Aztec rectangles with $k$ holes and $l$ separations)} 
Let $1\leq a_1<\cdots<a_k\leq 2n+k-l$ and $1\leq b_1<\cdots<b_l\leq 2n+k-l$ be distinct integers. 

$(${\text{\rm a}}$)$. Define the subset $S_{n,i}(a_1,\dotsc,a_k;b_1,\dotsc,b_l)$ of $\{1,2\dotsc,2n+k-l\}$ as follows. Starting from coordinate $a_i$, move towards the right on the integer coordinates on $\ell$, skipping over one existing node $($i.e., a node at which there is no monomer$)$ at each time, with the convention that separations are considered ``doubled'' nodes: If we land on the first node of a separation $s$, the second node in that separation counts when we do the next jump, thus taking us to the first node to the right of $s$ at which we have no monomer; if we land on the second node of a separation $s$ $($this happens precisely when the previous skipped over node was the first node of $s$$)$, we have to skip over the first existing node $($either a no-defect node, or the first node of a separation$)$ to the right of $s$. 

Next, starting from coordinate $a_i-1$, move towards the left on the integer coordinates on $\ell$, skipping over one existing node at each time, using the convention from the previous paragraph when a separation is encountered. Let $S_{n,i}(a_1,\dotsc,a_k;b_1,\dotsc,b_l)$ be the set of coordinates that have been visited this way, not including coordinates $a_i$ and $a_i-1$ $($an example is illustrated in Figure ${\Fhb}$$)$.

Then we have
$$
\frac
{\M(AR_{2n,2n+k-l}(\{a_1,\dotsc,a_{i-1},a_i,a_{i+1},\dotsc,a_k\},\{b_1,\dotsc,b_l\})}
{\M(AR_{2n,2n+k-l}(\{a_1,\dotsc,a_{i-1},a_i-1,a_{i+1},\dotsc,a_k\},\{b_1,\dotsc,b_l\})}
=
\prod_{i\in S_{n,i}(a_1,\dotsc,a_k;b_1,\dotsc,b_l)}\frac{|(a_i-1)-j|}{|a_i-j|}.
\tag\Ehzl
$$

$(${\text{\rm b}}$)$. Define the subset $S'_{n,i}(a_1,\dotsc,a_k;b_1,\dotsc,b_l)$ of $\{1,2\dotsc,2n+k-l\}$ as follows.
Starting from the leftmost of the two nodes corresponding to the separation at 
$b_i$, move towards the right on the integer coordinates on $\ell$ according to the rule in part 
$(${\text{\rm a}}$)$.
%applying the convention about separations also at the very beginning of the process $($as we are now starting from a separation$)$. 
Then, starting from the rightmost of the two nodes corresponding to the separation at $b_i-1$ $($where the separation formerly at $b_i$ has moved$)$, move towards the left on the integer coordinates on $\ell$, skipping over one existing node at each time, using the same convention for separations as before.
%and applying it at the starting node $b_i-1$ $($where the separation formerly at $b_i$ has moved$)$ as well. 
Let $S'_{n,i}(a_1,\dotsc,a_k;b_1,\dotsc,b_l)$ be the set of coordinates that have been visited this way, not including coordinates~$b_i$ and $b_i-1$ $($see Figure ${\Fhc}$ for an example$)$.

Then we have
$$
\frac
{\M(AR_{2n,2n+k-l}(\{a_1,\dotsc,a_k\},\{b_1,\dotsc,b_{i-1},b_i,b_{i+1},\dotsc,b_l\})}
{\M(AR_{2n,2n+k-l}(\{a_1,\dotsc,a_k\},\{b_1,\dotsc,b_{i-1},b_i-1,b_{i+1},\dotsc,b_l\})}
=
\prod_{i\in S'_{n,i}(a_1,\dotsc,a_k;b_1,\dotsc,b_l)}\frac{|b_i-j|}{|(b_i-1)-j|}.
\tag\Ehzm
$$

\endproclaim

\topinsert
\centerline{\mypic{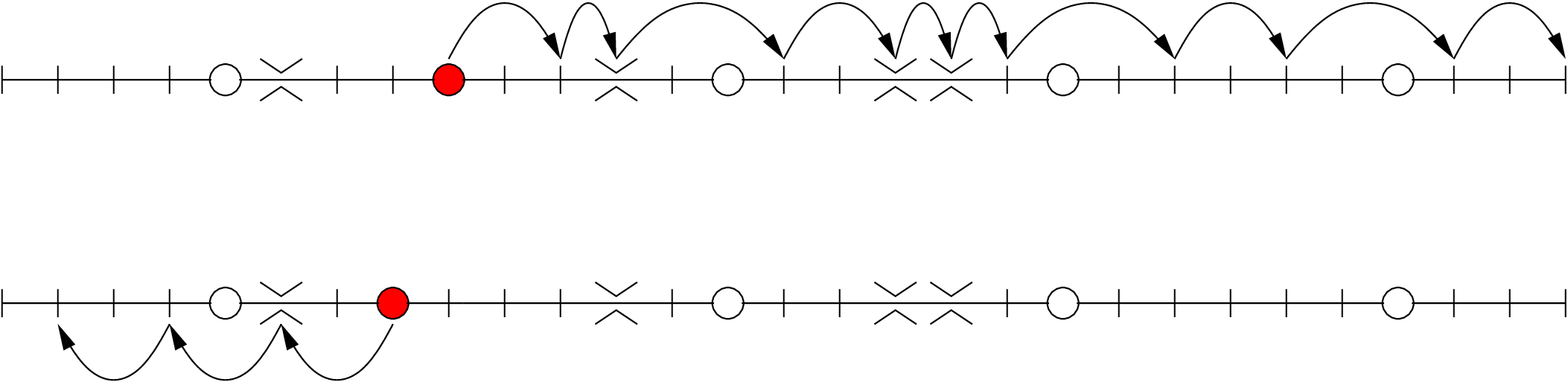}}
\medskip
\centerline{{\smc Figure~8.2.} {\rm The set $S_{14,2}(5,9,14,20,26;6,12,17,18)$.}}
\endinsert

\topinsert
\centerline{\mypic{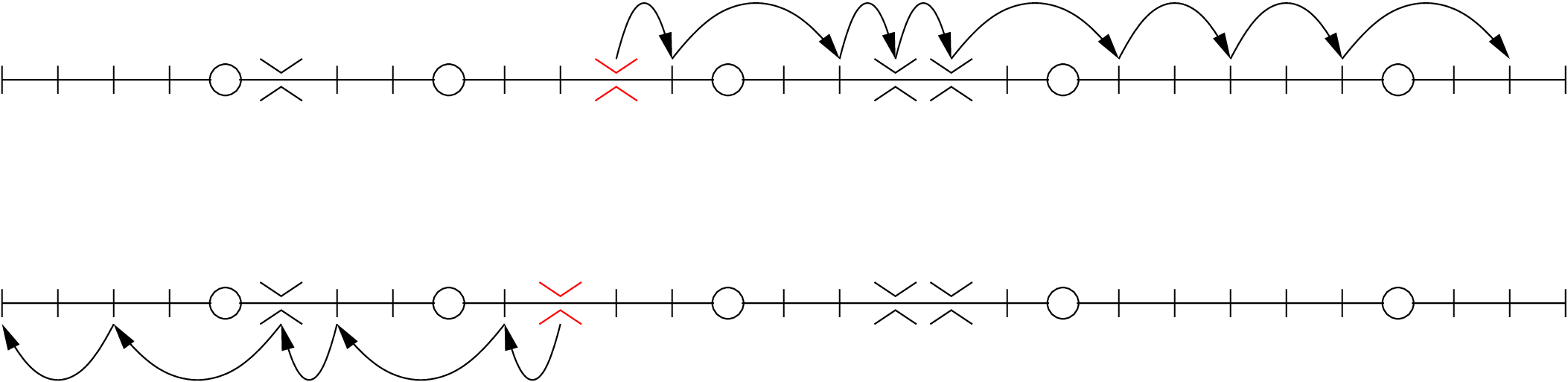}}
\medskip
\centerline{{\smc Figure~8.3.} {\rm The set $S'_{14,2}(5,9,14,20,26;6,12,17,18)$.}}
\endinsert

\pf As in the proof of Lemma {\Thf}, apply Theorem 1.1 of \cite{\gd} to both the numerator and denominator on the left hand side above. Since the two defect distributions are nearly identical, most of the resulting factors occur both at the numerator and at the denominator, and thus cancel out. It is easy to check that the remaining factors combine to give precisely the expression on the right hand sides of (\Ehzl) and~(\Ehzm).~\epf

\proclaim{Corollary \Thj \ (Changing a monomer into a separation)} We have
$$
\spreadlines{4\jot}
\align
&
\frac
{\displaystyle
\frac
{\M(AR_{2n,2n+k-l}(\{a_1,a_2,\dotsc,a_k\},\{b_1,\dotsc,b_l\})}
{\M(AR_{2n,2n+k-l}(\{a_1-1,a_2,\dotsc,a_k\},\{b_1,\dotsc,b_l\})}
}
{\displaystyle
\frac
{\M(AR_{2n+2,2n+k-l}(\{a_2,\dotsc,a_k\},\{a_1,b_1,\dotsc,b_l\})}
{\M(AR_{2n+2,2n+k-l}(\{a_2,\dotsc,a_k\},\{a_1-1,b_1,\dotsc,b_l\})}
}
=
\\
&\ \ \ \ \ \ \ \ \ \ \ \ \ \ \ \ 
\frac{(a_1-1)-(2n+k-l)}{a_1-(1)}
\prod_{j=2}^k\frac{a_1-a_j}{(a_1-1)-a_j}
\prod_{j=1}^l\frac{(a_1-1)-b_j}{a_1-b_j}.
\tag\Ehzn
\endalign
$$

\endproclaim

\pf Apply Lemma {\Thi} to express the numerator and denominator of the left hand side of (\Ehzn) as simple products. It is readily seen that the two resulting products are almost identical; the four factors that do not cancel out when taking their ratio lead to the expression on the right hand side of (\Ehzn). \epf

By repeated application of Corollary {\Thj}, one can bring the leftmost defects (which is a monomer in the numerator, and a separation in the denominator) all the way to the boundary of the Aztec rectangle. This allows one to express the case of $k$ monomers and $l$ separations (for short, the $(k,l)$-case) in terms of the $(k+1,l-1)$- and $(k,l-1)$-case. Doing the asymptotic analysis for the resulting products of Gamma functions one obtains the following general result.

%Since for $k$ monomers and $l$ separations on $\ell$ the enclosing Aztec rectangles are of the form $AR_{2n,2n+k-l}$, we modify the definitions (\Ehzg)-(\Ehzh) of the number of vertices on $\ell$ strictly to the left and right of vertex $j$ to 
%$$
%\align
%L_j&:=j-1\tag\Ehzo
%\\
%R_j&:=2n+k-l-j\tag\Ehzp
%\endalign
%$$

\proclaim{Theorem \Thk} Let $0<\alpha_1<\cdots<\alpha_k<2$ and $0<\beta_1<\cdots<\beta_l<2$ be distinct rational numbers, and let $c_1,\dotsc,c_k$ and $d_1,\dotsc,d_l$ be fixed integers. Then if $s_i=\alpha_i n$, $i=1,\dotsc,k$, and $t_i=\beta_i n$, $i=1,\dotsc,l$, are integers, we have, as $n\to\infty$, that
$$
\spreadlines{4\jot}
\align
&\!\!\!\!\!\!\!\!\!\!\!\!\!\!\!\!
\frac
{\M(AR_{2n,2n+k-l}(\{s_1+c_1,s_2+c_2,\dotsc,s_k+c_k\},\{t_1+d_1,t_2+d_2,\dotsc,t_l+d_l\}))}
{\M(AD_{2n})}
\sim
\\
&\!\!\!\!
\left(\frac{e^\frac14}{2^\frac{5}{12}A^3\,n^\frac14}\right)^{k+l}
\frac
{\displaystyle \prod_{j=1}^l \left(\sqrt{\frac{\beta_j}{2}}\,\right)^{L'_j+\frac12}\left(\sqrt{1-\frac{\beta_j}{2}}\,\right)^{R'_j+\frac12}}
{\displaystyle \prod_{i=1}^k \left(\sqrt{\frac{\alpha_i}{2}}\,\right)^{L_i+\frac12}\left(\sqrt{1-\frac{\alpha_i}{2}}\,\right)^{R_i+\frac12}}
%\\
%&\ \ \ \ \ \ \ \ \
%\times
\frac
{\displaystyle \prod_{1\leq i<j\leq k}\sqrt{\frac{\alpha_j}{2}-\frac{\alpha_i}{2}}
               \prod_{1\leq i<j\leq l}\sqrt{\frac{\beta_j}{2}-\frac{\beta_i}{2}}}
{\displaystyle \prod_{i=1}^k\prod_{j=1}^l\sqrt{\left|\frac{\alpha_i}{2}-\frac{\beta_j}{2}\right|}},
\tag\Ehzo
\endalign
$$
where $L_i$ and $R_i$ are the number of vertices on $\ell\cap AR_{2n,2n+k-l}$ that are strictly to the left and right of $s_i+c_i$, respectively, and $L'_j$ and $R'_j$ are the number of vertices on $\ell\cap AR_{2n,2n+k-l}$ that are strictly to the left and right of $t_j+d_j$, respectively. \epf

\endproclaim 

Note that for $l=0$ this specializes to Proposition {\Thh}, as, due to forced edges, the denominator on the left hand side of (\Ehzi) is equal to $\M(AD_{2n})$.

\medskip
\flushpar
{\smc Remark \rl.} It follows from (\Ehzo) (specifically, from the first of the three factors on its right hand side) that both creating an additional hole on $\ell$, and creating an additional separation on $\ell$ (while maintaining the same height $2n$ for the Aztec rectangles), cause the order of the number of perfect matchings of the Aztec rectangle with defects to be multiplied by $n^{-\frac14}$.
%(and furthermore, the multiplicative constant is also the same). 
When a new hole is created, the width of the Aztec rectangle increases by one, but the new hole introduces a microscopic boundary, whose restricting effect prevails over this increase in width, resulting in an overall decrease of order $n^{-\frac14}$. When a new separation is created, the width of the Aztec rectangle decreases by one, and this prevails over the added freedom around the new separation, resulting in the same overall decrease of order $n^{-\frac14}$ in the number of perfect matchings.

The second and third factors on the right hand side of (\Ehzo) also have natural interpretations. The third factor in (\Ehzo) is easily recognized, in view of \cite{\gd, Theorem\,3.1}, as being almost the same as the Coulomb interaction between defects in the bulk, but with the change that here instead of the distances between the defects we take the {\it relative distances} between them, compared to the width of the enclosing Aztec rectangles.

The remaining factor on the right hand side of (\Ehzo) --- the middle one --- gives the interaction of the defects with the boundary. It is interesting that the exponents of the square root factors are exactly equal to the ``number'' of sites on $\ell\cap AR_{2n,2n+k-l}$ to the left and right of the corresponding defect, if the defect in question is considered to contribute $\frac12$ to both these counts --- as if it had split in two.

It is also interesting that in the interaction with the boundary, the defects turn out to act independently: The middle factor in (\Ehzo) is equal to the product of the corresponding factors in the individual interaction of each defect with the boundary.

Formula (\Ehzo) shows that monomers and separations that are macroscopically separated in an Aztec rectangle interact in a satisfyingly symmetric way, that places them on equal footing.

The independence mentioned above can be stated in an especially simple form in the case when $k$ and~$l$ have opposite parities.

\proclaim{Corollary \Thl} If $k$ and $l$ have opposite parities, with the same assumptions as in Theorem ${\Thk}$, we have that
$$
\spreadlines{4\jot}
\align
&
\lim_{n\to\infty}
\frac
{\M(AR_{2n,2n+k-l}(\{s_1+c_1,s_2+c_2,\cdots,s_k+c_k\},\{t_1+d_1,t_2+d_2,\cdots,t_l+d_l\}))}
{\prod_{i=1}^k\M(AR_{2n+k-l-1,2n+k-l}(\{s_i+c_i\},\emptyset))\prod_{j=1}^l\M(AR_{2n+k-l+1,2n+k-l}(\emptyset,\{t_j+d_j\}))}
=
\\
&\ \ \ \ \ \ \ \ \ \ \ \ \ \ \ \ 
\frac
{\displaystyle \prod_{1\leq i<j\leq k}\sqrt{\frac{\alpha_j}{2}-\frac{\alpha_i}{2}}
               \prod_{1\leq i<j\leq l}\sqrt{\frac{\beta_j}{2}-\frac{\beta_i}{2}}}
{\displaystyle \prod_{i=1}^k\prod_{j=1}^l\sqrt{\left|\frac{\alpha_i}{2}-\frac{\beta_j}{2}\right|}}.
\tag\Ehzp
\endalign
$$

\endproclaim

\pf Use the special cases $k=1$, $l=0$ and $k=0$, $l=1$ of Theorem {\Thk} to obtain the asymptotics of each factor in the products at the denominator on the left hand side above. When dividing the right hand side of (\Ehzo) by the product of the resulting expressions, all factors but the ones on the right hand side of (\Ehzp) cancel out. \epf

%- Figure top 52/XIII

%- Figure bot 55/XIII

%- Figure top 56/XIII

%Add remark saying that interaction with boundary is so strong that one can't say what the result is when scaled coordinates approach given points in (0,2) --- *how* the points are approached has an effect!

What about the situation when the $\alpha_i$'s and $\beta_j$'s are not necessarily rational? Due to the very fast growth of the function $d^d$ which gives the interaction with the boundary (see the middle fraction on the right hand side of (\Ehzo)), simply replacing $\alpha n$ by $\lfloor \alpha n\rfloor$ in the statement of Theorem {\Thk} would not leave the multiplicative constant in (\Ehzo) unchanged. A good way to address this case is to use the following standard fact from the theory of Diophantine approximations\footnote{ This was pointed out to the author by Michael Larsen.}.

\proclaim{Lemma \Thm} Let $\alpha_1,\dotsc,\alpha_k$ be real numbers.
% at least one of which is irrational. 
Then for infinitely many integers $q$ there exists a $k$-tuple of integers $(p_1,\dotsc,p_k)$ so that
$$
\left|\alpha_i-\frac{p_i}{q}\right|<\frac{1}{q^{1+\frac{1}{k}}}
\tag\Ehzq
$$
for $i=1,\dotsc,k$.

\endproclaim

\pf Let $n$ be a positive integer. Consider the $n^k+1$ points 
$$
(\{\alpha_1 i\},\dotsc,\{\alpha_k i\}),
$$
$i=1,\dotsc,n^k+1$, in the hypercube $[0,1)^k$ ($\{a\}$ denotes the fractional part of the real number $a$). Partition this hypercube into $n^k$ hypercubes of side-length $\frac{1}{n}$. By the pigeonhole principle, two of the above points must be in the same small hypercube. It follows that there exist integers $p_1,\dotsc,p_k$ and some integer $q$ between 1 and $n^k$ so that
$$
\alpha_i q=p_i+\epsilon_i,
$$
$i=1,\dotsc,k$, with $|\epsilon_i|<\frac{1}{n}$ for all $i$. It follows that
$$
\left|\alpha_i-\frac{p_i}{q}\right|<\frac{1}{qn},
\tag\Ehzr
$$
for $i=1,\dotsc,k$. Using that $\frac{1}{n}\leq\frac{1}{q^\frac{1}{k}}$ we arrive at (\Ehzq). According to (\Ehzr), the precision of approximating any irrational $\alpha_i$ by $\frac{p_i}{q}$ can be improved indefinitely by increasing $n$. It follows that as $n\to\infty$ in this process, we must also have $q\to\infty$. 
On the other hand, if all $\alpha_i$'s are rational, the left hand sides of (\Ehzq) can be simultaneously made equal to zero for any $q$ that is a common multiple of their denominators. This completes the proof. \epf

We can now phrase the statement of Theorem {\Thk} for arbitrary $\alpha_i$'s and $\beta_j$'s.

\proclaim{Theorem \Thn} Let $0<\alpha_1<\cdots<\alpha_k<2$ and $0<\beta_1<\cdots<\beta_l<2$ be distinct real numbers.
% at least one of which is irrational. 
Let $n,s_1,\dotsc,s_k$ and $t_1,\dotsc,t_l$ be positive integers so that 
$$
\left|\alpha_i-\frac{s_i}{n}\right|<\frac{1}{n^{1+\frac{1}{k+l}}},
\tag\Ehzs
$$
$i=1,\dotsc,k$, and 
$$
\left|\beta_j-\frac{t_j}{n}\right|<\frac{1}{n^{1+\frac{1}{k+l}}},
\tag\Ehzt
$$
$j=1,\dotsc,l$ $($the existence of such $n$, $s_i$'s and $t_j$'s is guaranteed by Lemma ${\Thm}$$)$.

Then as $n\to\infty$ we have that
$$
\spreadlines{4\jot}
\align
&\!\!\!\!\!\!\!\!\!\!\!\!\!\!\!\!
\frac
{\M(AR_{2n,2n+k-l}(\{s_1,\dotsc,s_k\},\{t_1,\dotsc,t_l\}))}
{\M(AD_{2n})}
\sim
\\
&\!\!\!\!
\left(\frac{e^\frac14}{2^\frac{5}{12}A^3\,n^\frac14}\right)^{k+l}
\frac
{\displaystyle \prod_{j=1}^l \left(\sqrt{\frac{\beta_j}{2}}\,\right)^{L'_j+\frac12}\left(\sqrt{1-\frac{\beta_j}{2}}\,\right)^{R'_j+\frac12}}
{\displaystyle \prod_{i=1}^k \left(\sqrt{\frac{\alpha_i}{2}}\,\right)^{L_i+\frac12}\left(\sqrt{1-\frac{\alpha_i}{2}}\,\right)^{R_i+\frac12}}
%\\
%&\ \ \ \ \ \ \ \ \
%\times
\frac
{\displaystyle \prod_{1\leq i<j\leq k}\sqrt{\frac{\alpha_j}{2}-\frac{\alpha_i}{2}}
               \prod_{1\leq i<j\leq l}\sqrt{\frac{\beta_j}{2}-\frac{\beta_i}{2}}}
{\displaystyle \prod_{i=1}^k\prod_{j=1}^l\sqrt{\left|\frac{\alpha_i}{2}-\frac{\beta_j}{2}\right|}},
\tag\Ehzu
\endalign
$$
where $L_i$ and $R_i$ are the number of vertices on $\ell\cap AR_{2n,2n+k-l}$ that are strictly to the left and right of $s_i$, respectively, and $L'_j$ and $R'_j$ are the number of vertices on $\ell\cap AR_{2n,2n+k-l}$ that are strictly to the left and right of $t_j$, respectively. 

\endproclaim

\pf There is one new argument we need in addition to those that proved Theorem {\Thk}. We present it below in the case $k=1$, $l=0$; it extends straightforwardly to the general case.

By Corollary {\Thb} and (\Ehv) we have that
$$
\spreadlines{4\jot}
\align
\frac
{\M(AR_{2n,2n+1}(\{2s+1\},\emptyset))}
{\M(AD_{2n})}
&=
\left[Q\left(s,n+\frac32\right)\right]^2
\\
&=
\left[
\frac{1}{2^{2s(n-s)}}
\frac{H(s+1)H(n-s+1)}{H(n+1)}
\frac{H(s)H(n-s)}{H(n)}
\frac{E(n)}{E(s)E(n-s)}
\right]^2.
\tag\Ehzv
\endalign
$$
Using (\Ehw) and (\Ehx), one readily sees that (\Ehy) holds for large $s$, $n-s$ and $n$ also before setting $s=\alpha n$; i.e., that as $s,n-s,n\to\infty$, we have
$$
\spreadlines{4\jot}
\align
\left[Q\left(s,n+\frac32\right)\right]^2
&=
\left[
\frac{1}{2^{2s(n-s)}}
\frac{H(s+1)H(n-s+1)}{H(n+1)}
\frac{H(s)H(n-s)}{H(n)}
\frac{E(n)}{E(s)E(n-s)}
\right]^2
\\
&%\ \ \ \ \ \ \ \ 
\sim
\frac{e^\frac14}{2^\frac{5}{12}A^3 n^\frac14}
\frac{n^{n+\frac12}}{s^{s+\frac14}(n-s)^{n-s+\frac14}} 
=
\frac{e^\frac14}{2^\frac{5}{12}A^3 n^\frac14}
\frac{(2n)^{n+\frac12}}{(2s)^{s+\frac14}(2n-2s)^{n-s+\frac14}}
\\
&%\ \ \ \ \ \ \ \
\sim
\frac{(2n)^{n+\frac12}}{(2s+1)^{s+\frac14}(2n-2s-1)^{n-s+\frac14}}.
\tag\Ehzw
\endalign
$$
We claim that for $s$ and $n$ going to infinity so that (\Ehzs) holds, i.e., in our case,
$$
\left|\alpha-\frac{2s+1}{n}\right|<\frac{1}{n^2},
\tag\Ehzx
$$
one has
$$
\frac{(2n)^{n+\frac12}}{(2s+1)^{s+\frac14}(2n-2s-1)^{n-s+\frac14}}
\sim
\frac{1}
{\left(\sqrt{\frac{\alpha}{2}}\right)^{2s+\frac12}
\left(\sqrt{1-\frac{\alpha}{2}}\right)^{2n-2s+\frac12}}.
\tag\Ehzy
$$
Indeed, write the left hand side above as
$$
\frac
{1}
{\left(\frac{2s+1}{2n}\right)^{s+\frac14}\left(1-\frac{2s+1}{2n}\right)^{n-s+\frac14}}.
$$
Note that since
$$
\frac{\left(\frac{2s+1}{2n}\right)^{s+\frac14}}{\left(\frac{\alpha}{2}\right)^{s+\frac14}}
=
\left[1+\frac{2}{\alpha}\left(\frac{2s+1}{2n}-\frac{\alpha}{2}\right)\right]^{s+\frac14},
$$
it follows that, for $s,n\to\infty$ as in (\Ehzx), we have
$$
\left(\frac{2s+1}{2n}\right)^{s+\frac14}
\sim
\left(\frac{\alpha}{2}\right)^{s+\frac14}.\tag\Ehzz
$$
A similar argument shows that 
$$
\left(1-\frac{2s+1}{2n}\right)^{s+\frac14}
\sim
\left(1-\frac{\alpha}{2}\right)^{s+\frac14},\tag\Ehzza
$$
for $s,n\to\infty$ as in (\Ehzx). The above two asymptotic equalities imply (\Ehzy), thus checking the case $k=1$, $l=0$. 

The only difference in the general case is that the exponent 2 at the denominator on the right hand side of (\Ehzx) is replaced by $1+\frac{1}{k+l}$. But this is still enough to obtain the analogs of (\Ehzz) and (\Ehzza), which imply the statement of the theorem in the same fashion as above. \epf

\mysec{9. Large clusters consisting of consecutive monomers}

Section 8 showed that if one has a finite number of monomers that can move about on $\ell$, they experience a tremendous force from the boundaries pushing them towards the center, which overwhelms their tendency to repel one another. But what if the number of monomers is comparable to the width of the Aztec rectangle? This section considers the case of large monomer clusters consisting of runs of consecutive monomers.

Denote by $C_{k,p}$ the cluster consisting of monomers $\circ_{2k+1},\circ_{2k+2},\dotsc,\circ_{2k+2p}$; we will sometimes call such a cluster a {\it bar of charge} (for the sake of notational simplicity, we restrict our analysis to the case when the monomer clusters have even lengths, and start on an odd index node; our arguments can straightforwardly be extended to the general situation). 

We start by presenting an exact formula for the number of perfect matchings of an Aztec rectangle containing two such monomer clusters.

\topinsert
\centerline{\mypic{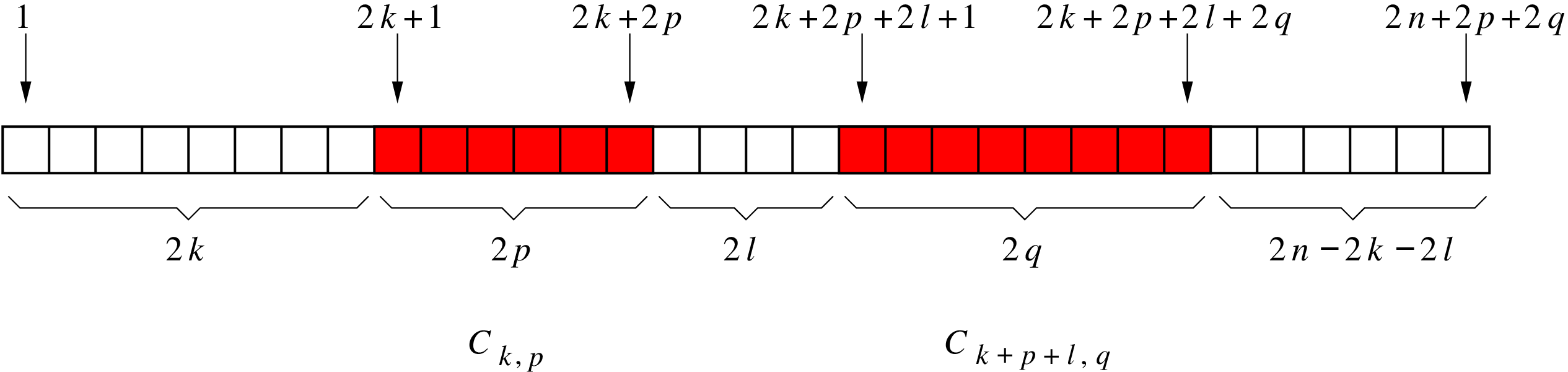}}
\medskip
\centerline{{\smc Figure~{\Fia}.} {\rm Illustration of the two bars of charge for $n=9$, $k=4$, $l=2$, $p=3$, $q=4$.}}
\endinsert

\proclaim{Proposition \Tia} For any non-negative integers $k$, $l$, $p$, $q$ with $k+l\leq n$, we have 
$$
\spreadlines{4\jot}
\align
&
\M(AR_{2n,2n+2p+2q}(C_{k,p},C_{k+p+l,q}))
=
2^{n(2n+1)}
\prod_{i=1}^k\left[\frac{\Gamma(i)\Gamma(n+p+q-i+1)}{\Gamma(p+i)\Gamma(n+q-i+1)}
%\right]^2
%\prod_{i=1}^k
%\left[
\frac{\Gamma(p+l+i)\Gamma(q+l+i)}{\Gamma(l+i)\Gamma(p+q+l+i)}\right]^2
\\
&\ \ \ \ \ \ \ \ \ \ \ \ \ \ \ \ \ \ \ \ \ \ \ \ \ \ \ \ \ \ \ \ \ \ \ \ \ \ \ \ \ \ \ \ \ \ \ \ \ \ 
\times
\prod_{i=1}^{k+l}\left[\frac{\Gamma(i)\Gamma(n+q-i+1)}{\Gamma(q+i)\Gamma(n-i+1)}\right]^2.
\tag\Eia
\endalign
$$

\endproclaim

(The case $n=9$, $k=4$, $l=2$, $p=3$, $q=4$ is illustrated in Figure {\Fia}.)

\pf Applying \cite{\gd,Theorem\,1.1} to the numerators and denominators on the left hand sides below, it follows, after simplifications, that one has
$$
\frac
{\M(AR_{2n,2n+2p+2q}(C_{i,p},C_{k+p+l,q}))}
{\M(AR_{2n,2n+2p+2q}(-1+C_{i,p},C_{k+p+l,q}))}
=
\frac
{(1)_{k-i}(p+1)_{l+i-1}(p+l+i+q)_{n-k-l}}
{(p+1)_{k-i}(1)_{l+i-1}(l+i+q)_{n-k-l}}
\tag\Eib
$$
and
$$
\frac
{\M(AR_{2n,2n+2p+2q}(-1+C_{i,p},C_{k+p+l,q}))}
{\M(AR_{2n,2n+2p+2q}(C_{i-1,p},C_{k+p+l,q}))}
=
\frac
{(1)_{k-i}(p+1)_{l+i-1}(p+l+i+q)_{n-k-l}}
{(p+1)_{k-i}(1)_{l+i-1}(l+i+q)_{n-k-l}},
\tag\Eic
$$
for all $1\leq i\leq n$
(for a defect cluster $O$, $-1+O$ denotes the defect cluster obtained from $O$ by translating all its constituent defects one unit to the left). Multiplying together equations (\Eib)--(\Eic) for $i=1,\dotsc,k$ yields
$$
\frac
{\M(AR_{2n,2n+2p+2q}(C_{k,p},C_{k+p+l,q}))}
{\M(AR_{2n,2n+2p+2q}(C_{0,p},C_{k+p+l,q}))}
=
\prod_{i=1}^k
\left[\frac
{(1)_{k-i}(p+1)_{l+i-1}(p+l+i+q)_{n-k-l}}
{(p+1)_{k-i}(1)_{l+i-1}(l+i+q)_{n-k-l}}
\right]^2.
\tag\Eid
$$
Due to forced edges, $AR_{2n,2n+2p+2q}(C_{0,p},C_{k+p+l,q})$ and $AR_{2n,2n+2q}(C_{k+l,q})$ have the same number of perfect matchings:
$$
\M(AR_{2n,2n+2p+2q}(C_{0,p},C_{k+p+l,q}))=\M(AR_{2n,2n+2q}(C_{k+l,q})).
\tag\Eie
$$
The same approach that gave (\Eib)--(\Eic) also shows that
$$
\frac
{\M(AR_{2n,2n+2q}(C_{i,q}))}
{\M(AR_{2n,2n+2q}(-1+C_{i,q}))}
=
\frac
{(1)_{k-i}(q+1)_{n-k+i-1}}
{(q+1)_{k-i}(1)_{n-k+i-1}}
\tag\Eif
$$
and
$$
\frac
{\M(AR_{2n,2n+2q}(-1+C_{i,q}))}
{\M(AR_{2n,2n+2q}(C_{i-1,q}))}
=
\frac
{(1)_{k-i}(q+1)_{n-k+i-1}}
{(q+1)_{k-i}(1)_{n-k+i-1}},
\tag\Eig
$$
for all $1\leq i\leq n$. Multiplying together equations (\Eif) and (\Eig) for $i=1,\dotsc,k$, we obtain that
$$
\frac
{\M(AR_{2n,2n+2q}(C_{k,q}))}
{\M(AR_{2n,2n+2q}(C_{0,q}))}
=
\prod_{i=1}^{k}\left[
\frac
{(1)_{k-i}(q+1)_{n-k+i-1}}
{(q+1)_{k-i}(1)_{n-k+i-1}}
\right]^2.
\tag\Eih
$$
Using the fact that, due to forced edges, $AR_{2n,2n+2q}(C_{0,q})$ has the same number of perfect matchings as $AR_{2n}$ (namely, $2^{n(2n+1)}$), it follows from equations (\Eid), (\Eie) and (\Eih) that
$$
\spreadlines{4\jot}
\align
&
\M(AR_{2n,2n+2p+2q}(C_{k,p},C_{k+p+l,q}))
=
\\
&\ \ \ \ \ \ \ \ 
2^{n(2n+1)}
\prod_{i=1}^k
\left[\frac
{(1)_{k-i}(p+1)_{l+i-1}(p+l+i+q)_{n-k-l}}
{(p+1)_{k-i}(1)_{l+i-1}(l+i+q)_{n-k-l}}
\right]^2
\prod_{i=1}^{k+l}\left[
\frac
{(1)_{k+l-i}(q+1)_{n-k-l+i-1}}
{(q+1)_{k+l-i}(1)_{n-k-l+i-1}}
\right]^2.
\tag\Eii
\endalign
$$
Using formula (\Ebpp) to express the Pochhammer symbols above in terms of Gamma functions one obtains~(\Eia). \epf

\proclaim{Proposition \Tib} For any non-negative integers $k$, $l$, $p$, $q$ with $k+l\leq n$, we have 
$$
\spreadlines{4\jot}
\align
&
\frac
{\M(AR_{2n,2n+2p+2q}(C_{k,p},C_{k+p+l,q}))}
{\M(AD_{2n})}
=
\\
&\ \ \ \ 
\left[
\frac
{H(k)H(l)H(p)H(q) H(n-k-l) H(k+l+p)H(l+p+q) H(n+q-k) H(n+p+q) }
{H(n)H(k+p)H(l+p)H(l+q) H(n+q-k-l) H(k+l+p+q) H(n+p+q-k) }
\right]^2
\tag\Eij
\\
&
\sim
\frac{e^\frac13}{A^4}
\frac
{k^{k^2-\frac16} l^{l^2-\frac16} p^{p^2-\frac16} q^{q^2-\frac16} (n-k-l)^{(n-k-l)^2-\frac16}(k+l+p)^{(k+l+p)^2-\frac16}(l+p+q)^{(l+p+q)^2-\frac16}}
{n^{n^2-\frac16}(k+l)^{(k+l)^2-\frac16}(l+p)^{(l+p)^2-\frac16}(l+q)^{(l+q)^2-\frac16}(n+q-k-l)^{(n+q-k-l)^2-\frac16}}
\\
&\ \ \ 
\times
\frac
{(n+q-k)^{(n+q-k)^2-\frac16}(n+p+q)^{(n+p+q)^2-\frac16}}
{(k+l+p+q)^{(k+l+p+q)^2-\frac16}(n+p+q-k)^{(n+p+q-k)^2-\frac16}},
\tag\Eik
\endalign
$$
where $H(n)$ is given by $(\Ecq)$,  $A$ is the Glaisher-Kinkelin constant $(\Ecg)$, and in $(\Eik)$ the integers $n$, $k$, $l$, $p$ and $q$ are growing to infinity in a way that makes $n-k-l$ grow to infinity as well.

\endproclaim

\pf Define the products
$$
\spreadlines{3\jot}
\align
P_1(n,k,p,q)&:=
\prod_{i=1}^k
\frac{\Gamma(i)\Gamma(n+p+q-i+1)}{\Gamma(p+i)\Gamma(n+q-i+1)},\tag\Eil
\\
P_2(n,k,l,p,q)&:=
\prod_{i=1}^k
\frac{\Gamma(p+l+i)\Gamma(q+l+i)}{\Gamma(l+i)\Gamma(p+q+l+i)}\tag\Eim
\endalign
$$
and
$$
P_3(n,k,l,p,q):=
\prod_{i=1}^{k+l}\frac{\Gamma(i)\Gamma(n+q-i+1)}{\Gamma(q+i)\Gamma(n-i+1)}.
\tag\Ein
$$
Then the statement of Proposition {\Tia} can be rewritten as
$$
\frac
{\M(AR_{2n,2n+2p+2q}(C_{k,p},C_{k+p+l,q}))}
{\M(AD_{2n})}
=
(P_1(n,k,p,q)P_2(n,k,l,p,q)P_3(n,k,l,p,q))^2.
\tag\Eio
$$
The products $P_1$, $P_2$ and $P_3$ can be expressed in terms of the hyperfactorial function $H(n)$ of (\Ecq) as follows:
$$
\spreadlines{3\jot}
\align
P_1(n,k,p,q)&=\frac{H(k)H(p)}{H(k+p)}\frac{H(n+p+q)H(n+q-k)}{H(n+q)H(n+p+q-k)}\tag\Eip
\\
P_2(n,k,l,p,q)&=\frac{H(l)H(k+l+p)}{H(k+l)H(l+p)}\frac{H(k+l+q)H(l+p+q)}{H(l+q)H(k+l+p+q)}\tag\Eiq
\\
P_3(n,k,l,p,q)&=\frac{H(k+l)H(n-k-l)}{H(n)}\frac{H(q)H(n+q)}{H(k+l+q)H(n-k-l+q)}.\tag\Eir
\endalign
$$
Plugging in the above expressions into (\Eio) one obtains equation (\Eij).

Recall that by (\Ecg) we have
$$
H(n)\sim
\frac{e^\frac{1}{12}}{A}
\frac{(2\pi)^\frac{n}{2}n^{\frac{n^2}{2}-\frac{1}{12}}}{e^{\frac{3n^2}{4}}},\ \ \ n\to\infty.
\tag\Eis
$$
It is apparent that in the expression (\Eij), the sum of the arguments of the $H$'s at the numerator is equal to the sum of the arguments of the $H$'s at the denominator. Remarkably, the sum of the {\it squares} of the arguments on top is also equal to the sum of the squares of the arguments on the bottom. This implies that the exponents of $2\pi$, as well as the exponents of $e$ --- and also almost all of the exponents of $n$ --- obtained when applying formula (\Eis) for the asymptotics of $H$ in (\Eij) cancel out. It is readily checked that the leftover factors combine to yield precisely the expression on the right hand side of (\Eik). This completes the proof. \epf

As in Section 8, we present first the corresponding scaling limit in the case when the lengths of the clusters are fixed rational multiples of the width of the Aztec rectangle.

\proclaim{Theorem \Tic} Let $0<\alpha,\beta,\gamma,\delta<1$ be rational numbers with $\alpha+\beta<1$. Then if $k=\alpha n$, $l=\beta n$, $p=\gamma n$ and $q=\delta n$ are integers, we have as $n\to\infty$ that
$$
\frac
{\M(AR_{2n,2n+2p+2q}(C_{k,p},C_{k+p+l,q}))}
{\M(AD_{2n})}
\sim
\frac{e^\frac13}{A^4 n^\frac13}
\prod_{i,j\in I,\,i<j}|i-j|^{\epsilon_{i,j}\left(|i-j|^2n^2-\frac{1}{6}\right)},
\tag\Eit
$$
where 
$$
I:=\{0,\alpha,\alpha+\gamma,\alpha+\beta+\gamma,\alpha+\beta+\gamma+\delta,1+\gamma+\delta\},
\tag\Eiu
$$
the constant $A$ is the Glaisher-Kinkelin constant $(\Ecg)$, and
$$
\epsilon_{i,j}:=\left\{\matrix 1,\ \ \ \text{\rm if there are on even number of elements of $I$ between $i$ and $j$}\\
                             -1,\ \ \text{\rm if there are on odd number of elements of $I$ between $i$ and $j$.}
\endmatrix\right.
\tag\Eiv
$$

\endproclaim

\pf Written out explicitly, (\Eit) states that
$$
\spreadlines{3\jot}
\align
&
\frac
{\M(AR_{2n,2n+2p+2q}(C_{k,p},C_{k+p+l,q}))}
{\M(AD_{2n})}
\sim
\frac{e^\frac13}{A^4 n^\frac13}
\alpha^{\alpha^2 n^2 -\frac{1}{6}}
\beta^{\beta^2 n^2 -\frac{1}{6}}
\gamma^{\gamma^2 n^2 -\frac{1}{6}}
\delta^{\delta^2 n^2 -\frac{1}{6}}
\\
&%\ \ \ \ \ \ \ \ \ \ \ \ \ \ \ \ 
\times
\frac
{
(\alpha+\beta+\gamma)^{(\alpha+\beta+\gamma)^2 n^2-\frac{1}{6}}
(\beta+\gamma+\delta)^{(\beta+\gamma+\delta)^2 n^2-\frac{1}{6}}
}
{
(\alpha+\gamma)^{(\alpha+\gamma)^2 n^2-\frac{1}{6}}
(\beta+\gamma)^{(\beta+\gamma)^2 n^2-\frac{1}{6}}
(\beta+\delta)^{(\beta+\delta)^2 n^2-\frac{1}{6}}
}
\\
&%\ \ \ \ \ \ \ \ \ \ \ \ \ \ \ \ 
\times
\frac
{
(1+\gamma+\delta)^{(1+\gamma+\delta)^2 n^2-\frac{1}{6}}
(1+\delta-\alpha)^{(1+\delta-\alpha)^2 n^2-\frac{1}{6}}
(1-\alpha-\beta)^{(1-\alpha-\beta)^2 n^2-\frac{1}{6}}
}
{
(\alpha+\beta+\gamma+\delta)^{(\alpha+\beta+\gamma+\delta)^2 n^2-\frac{1}{6}}
(1+\gamma+\delta-\alpha)^{(1+\gamma+\delta-\alpha)^2 n^2-\frac{1}{6}}
(1+\delta-\alpha-\beta)^{(1+\delta-\alpha-\beta)^2 n^2-\frac{1}{6}}
}.
\tag\Eiw
\endalign
$$
Replacing $k$, $l$, $p$ and $q$ by $\alpha n$, $\beta n$, $\gamma n$ and $\delta n$, respectively, the right hand side of (\Eik) becomes precisely the expression on the right hand side of (\Eiw). The statement of the theorem follows thus by Proposition~{\Tib}.~\epf

\medskip
\flushpar
{\smc Remark \rmm.} The form of the expression on the right hand side of (\Eit) suggests a possible physical interpretation along the following lines. It is as if in the dimer system on the Aztec rectangle with the two monomer clusters there is an effective polarization at the endpoints of its intersection with $\ell$ and at the endpoints of the monomer clusters, resulting in unit charges of alternating signs placed at those locations, interacting among themselves pairwise according to the fantastically strong interaction specified by the factor in the product on the right hand side of (\Eit), which note operates {\it oppositely} from the usual way: Here likes attract, and unlikes repel! 

\medskip
In Theorem {\Thn} of Section 8 we presented the exact asymptotics of the number of perfect matchings of an Aztec rectangle with a fixed number of defects on $\ell$, when they occupy arbitrary positions in the scaling limit, of rational or irrational coordinates. By contrast, in the present context, if some of $\alpha$, $\beta$, $\gamma$ and $\delta$ are irrational, it turns out that the asymptotics of the left hand side of (\Eit) is given by a modification of the expression on the right hand side of (\Eit) in which, in the factor in front of the product, both the constant and the power of $n$ depend on how well $\alpha$, $\beta$, $\gamma$ and $\delta$ can be approximated by rationals. Due to this, in the next result the multiplicative constant is not given explicitly, but instead asymptotic bounds are provided for it. We phrase the statement in terms of the free energy per site of the dimer system, i.e., the logarithm of the number of perfect matchings, divided by the number of vertices of the graph.

\proclaim{Theorem \Tid} Let $\alpha,\beta,\gamma,\delta>0$ be fixed real numbers with $\alpha+\beta<1$. Let $(k_n)_{n\geq1}$, $(l_n)_{n\geq1}$, $(p_n)_{n\geq1}$ and $(q_n)_{n\geq1}$ be sequences of positive integers so that $\left|\alpha-\frac{k_n}{n}\right|<\frac{1}{n^\frac54}$, $\left|\beta-\frac{l_n}{n}\right|<\frac{1}{n^\frac54}$, $\left|\gamma-\frac{p_n}{n}\right|<\frac{1}{n^\frac54}$ and $\left|\delta-\frac{q_n}{n}\right|<\frac{1}{n^\frac54}$ $($the existence of such sequences is a consequence of Lemma ${\Thm}$$)$.

%$k_n/n\to\alpha$, $l_n/n\to\beta$, $p_n/n\to\gamma$ and  $q_n/n\to\delta$, as~$n\to\infty$. 

Then if 
$G_n=AR_{2n,2n+2p_n+2q_n}(C_{k_n,p_n},C_{k_n+p_n+l_n,q_n})$, we have that
$$
\spreadlines{4\jot}
\align
&
\frac{1}{|V(G_n)|}\ln\M(G_n)
=
\frac14\ln 2 
+\frac18\sum_{i,j\in I,\, i<j}\epsilon_{i,j}|i-j|^2\ln|i-j|+O\left(\frac{1}{n^\frac54}\right),
\tag\Eix
\endalign
$$
where $I=\{0,\alpha,\alpha+\gamma,\alpha+\beta+\gamma,\alpha+\beta+\gamma+\delta,1+\gamma+\delta\}$, and the sign $\epsilon_{i,j}$ is given by $(\Eiv)$.

\endproclaim

\pf We claim that under the hypotheses of the theorem we have, for all $n$ large enough, that
$$
 e^{-4k_nn^{-\frac14}}
\leq
\frac
{\left(\frac{k_n}{n}\right)^{k_n^2}}
{\alpha^{k_n^2}}
\leq e^{4k_nn^{-\frac14}}.
\tag\Eiy
$$
Indeed, we have
$$
\spreadlines{4\jot}
\align
\frac
{\left(\frac{k_n}{n}\right)^{k_n^2}}
{\alpha^{k_n^2}}
=
\frac
{\left(\alpha+\frac{k_n}{n}-\alpha\right)^{k_n^2}}
{\alpha^{k_n^2}}
=
\left[
1+\frac{1}{\alpha}\left(\frac{k_n}{n}-\alpha\right)\right]^{k_n^2}
=
\left(
1+\frac{\epsilon}{\alpha n^\frac54}\right)^{k_n^2}
&\leq
\left[
\left(1+\frac{1}{\alpha n^\frac54}\right)^{n^\frac54}
\right]^{\frac{k_n}{n}\frac{k_n}{n^\frac14}}
\\
&\leq
\left(e^\frac{2}{\alpha}\right)^{2\alpha k_n n^{-\frac14}}
\endalign
$$
for $n$ large enough, where we used that, by hypothesis, $\alpha-\frac{k_n}{n}=\frac{\epsilon}{n^\frac54}$ for some $\epsilon$ with $|\epsilon|<1$. This proves the right inequality in (\Eiy); the left inequality follows by the same calculation, using $\epsilon>-1$.

Inequalities (\Eiy) and their analogs allow us to give asymptotic bounds on the left hand side of (\Eij), starting from the exact asymptotics (\Eik), as follows. Divide both the numerator and the denominator in (\Eik) by $n$ raised to a power equal to the sum of the exponents at the numerator (which, recall, is the same as the sum of the exponents at the denominator). For each resulting fraction of the form $\left(\frac{k}{n}\right)^{k^2}$ use the corresponding analog of inequalities (\Eiy). It follows that
$$
\spreadlines{3\jot}
\align
&
\frac
{\M(AR_{2n,2n+2p_n+2q_n}(C_{k_n,p_n},C_{k_n+p_n+l_n,q_n}))}
{\M(AD_{2n})}
=
B
\alpha^{\alpha^2 n^2 }
\beta^{\beta^2 n^2 }
\gamma^{\gamma^2 n^2 }
\delta^{\delta^2 n^2 }
\\
&%\ \ \ \ \ \ \ \ \ \ \ \ \ \ \ \ 
\times
\frac
{
(\alpha+\beta+\gamma)^{(\alpha+\beta+\gamma)^2 n^2}
(\beta+\gamma+\delta)^{(\beta+\gamma+\delta)^2 n^2}
}
{
(\alpha+\gamma)^{(\alpha+\gamma)^2 n^2}
(\beta+\gamma)^{(\beta+\gamma)^2 n^2}
(\beta+\delta)^{(\beta+\delta)^2 n^2}
}
\\
&%\ \ \ \ \ \ \ \ \ \ \ \ \ \ \ \ 
\times
\frac
{
(1+\gamma+\delta)^{(1+\gamma+\delta)^2 n^2}
(1+\delta-\alpha)^{(1+\delta-\alpha)^2 n^2}
(1-\alpha-\beta)^{(1-\alpha-\beta)^2 n^2}
}
{
(\alpha+\beta+\gamma+\delta)^{(\alpha+\beta+\gamma+\delta)^2 n^2}
(1+\gamma+\delta-\alpha)^{(1+\gamma+\delta-\alpha)^2 n^2}
(1+\delta-\alpha-\beta)^{(1+\delta-\alpha-\beta)^2 n^2}
},
\tag\Eiz
\endalign
$$
with the factor $B$ satisfying
$$
e^{\lambda n^\frac34}\leq B\leq e^{\mu n^\frac34}
\tag\Eiza
$$
for all large enough $n$, where $\lambda$ and $\mu$ are some real numbers depending only on $\alpha$, $\beta$, $\gamma$ and $\delta$. The statement of the theorem follows now by taking the logarithm in (\Eiz), dividing by the number of vertices of $AR_{2n,2n+2p_n+2q_n}(C_{k_n,p_n},C_{k_n+p_n+l_n,q_n})$ (which is asymptotically the same as the number of vertices of $AD_{2n}$, i.e., $4n(2n+1)$), and taking the limit as $n\to\infty$. \epf

A natural question is the following. In view of our physical interpretation, the two monomer clusters are long bars of positive charge, which repel one another and are also repelled by the boundary of the Aztec rectangle. One therefore expects that for any fixed $\gamma$ and $\delta$ (the bar charge magnitudes), there will be a unique equilibrium position $(\alpha_0,\beta_0)$ of the bars, where their mutual repelling is precisely compensated by the repelling from the boundary. The result below shows that this is indeed the case.

Define the function $F$ by
$$
\spreadlines{4\jot}
\align
&
F=F(\alpha,\beta,\gamma,\delta):=
%\frac14\ln 2 +\frac18\{
\alpha^2\ln\alpha+\beta^2\ln\beta+\gamma^2\ln\gamma+\delta^2\ln\delta
+(1-\alpha-\beta)^2\ln(1-\alpha-\beta)
\\
&\ \ \ \ \ \ \ \ \ \ \ \ \ \ \ \ \ \ \ \ \ \ \ \ \ \ \ \ \ \ \ \ \ \ \ \ \ \ \ \ \ \ \ \ \ \ 
+(\alpha+\beta+\gamma)^2\ln(\alpha+\beta+\gamma)
+(\beta+\gamma+\delta)^2\ln(\beta+\gamma+\delta)
\\
&\ \ \ \ \ \ \ \ \ \ \ \ \ \ \ \ \ \ \ \ \ \ \ \ \ \ \ \ \ \ \ \ \ \ \ \ \ \ \ \ \ \ \ \ \ \ 
+(1-\alpha-\beta+\gamma+\delta)^2\ln(1-\alpha-\beta+\gamma+\delta)
\\
&\ \ \ \ \ \ \ \ \ \ \ \ \ \ \ \ \ \ \ \ \ \ \ \ \ \ \ \ \ \ \ \ \ \ \ \ \ \ \ \ \ \ \ \ \ \ 
+(1-\alpha-\beta)^2\ln(1-\alpha-\beta)
\\
&\ \ \ \ \ \ \ \ \ \ \ \ \ \ \ \ \ \ \ \ \ \ \ \ \ \ \ \ \ \ \ \ \ \ \ \ \ \ \ \ \ \ \ \ \ \ 
-(\alpha+\gamma)^2\ln(\alpha+\gamma)-(\beta+\gamma)^2\ln(\beta+\gamma)-(\gamma+\delta)^2\ln(\gamma+\delta)
\\
&\ \ \ \ \ \ \ \ \ \ \ \ \ \ \ \ \ \ \ \ \ \ \ \ \ \ \ \ \ \ \ \ \ \ \ \ \ \ \ \ \ \ \ \ \ \ 
-(1-\alpha-\beta+\delta)^2\ln(1-\alpha-\beta+\delta)
\\
&\ \ \ \ \ \ \ \ \ \ \ \ \ \ \ \ \ \ \ \ \ \ \ \ \ \ \ \ \ \ \ \ \ \ \ \ \ \ \ \ \ \ \ \ \ \ 
-(\alpha+\beta+\gamma+\delta)^2\ln(\alpha+\beta+\gamma+\delta)
\\
&\ \ \ \ \ \ \ \ \ \ \ \ \ \ \ \ \ \ \ \ \ \ \ \ \ \ \ \ \ \ \ \ \ \ \ \ \ \ \ \ \ \ \ \ \ \ 
-(1-\alpha+\gamma+\delta)^2\ln(1-\alpha+\gamma+\delta).
%\}.
\tag\Eizb
\endalign
$$
It turns out that for any fixed $\gamma,\delta>0$, the function $F$ has a unique maximum in the range $\alpha,\beta>0$, $\alpha+\beta<1$. 

\proclaim{Corollary \Tie\ (Equilibrium of two bars of charge)} Let $\gamma$ and $\delta$ be fixed positive real numbers, and denote by $(\alpha_0,\beta_0)$ the unique maximum of the function $(\Eizb)$ in the domain $\alpha,\beta>0$, $\alpha+\beta<1$. Let $(\alpha,\beta)\neq(\alpha_0,\beta_0)$ be some other point in that domain.

Let $(k_n^0)_{n\geq1}$, $(l_n^0)_{n\geq1}$, $(k_n)_{n\geq1}$, $(l_n)_{n\geq1}$, $(p_n)_{n\geq1}$ and $(q_n)_{n\geq1}$ be sequences of positive integers so that $\left|\alpha_0-\frac{k_n^0}{n}\right|<\frac{1}{n^\frac76}$, $\left|\beta_0-\frac{l_n^0}{n}\right|<\frac{1}{n^\frac76}$,
$\left|\alpha-\frac{k_n}{n}\right|<\frac{1}{n^\frac76}$, $\left|\beta-\frac{l_n}{n}\right|<\frac{1}{n^\frac76}$, $\left|\gamma-\frac{p_n}{n}\right|<\frac{1}{n^\frac76}$ and $\left|\delta-\frac{q_n}{n}\right|<\frac{1}{n^\frac76}$ 
$($again, the existence of such sequences is a consequence of Lemma ${\Thm}$$)$.

Then we have that
$$
\frac
{\M(AR_{2n,2n+2p_n+2q_n}(C_{k_n,p_n},C_{k_n+p_n+l_n,q_n}))}
{\M(AR_{2n,2n+2p_n+2q_n}(C_{k_n^0,p_n},C_{k_n^0+p_n+l_n^0,q_n}))}
=
O\left(e^{-\lambda n^2}\right),
\tag\Eizc
$$
for some positive real number $\lambda$.

\endproclaim

The likelihood of a position different from the equilibrium position is thus exponentially small in $n^2$.

\smallpagebreak

\pf By two applications of (\Eiz), we obtain that
$$
\frac
{\M(AR_{2n,2n+2p_n+2q_n}(C_{k_n,p_n},C_{k_n+p_n+l_n,q_n}))}
{\M(AR_{2n,2n+2p_n+2q_n}(C_{k_n^0,p_n},C_{k_n^0+p_n+l_n^0,q_n}))}
=
\frac
{B e^{n^2 F(\alpha,\beta,\gamma,\delta)}}
{B_0 e^{n^2 F(\alpha_0,\beta_0,\gamma,\delta)}},
\tag\Eizd
$$
where both $B$ and $B_0$ satisfy inequalities of the form (\Eiza). Since 
$F(\alpha,\beta,\gamma,\delta)<F(\alpha_0,\beta_0,\gamma,\delta)$, the statement follows. \epf

%* Remark about interpretation of formula (p.18, c.XIV)

%!! Rephrase Thm 9.2 (maybe even Thm 8.11?? --- yes) for arb. alfa, etc (incl irr) --- see *-obs on p.17/c.XIV (this is required if we want to present equilibrium result --- unique solution (alfa0,beta0) most likely irrational!

%* Equilibrium position

%* How all other positions doubly exponentially approach 0

The above arguments are straightforwardly extended to the case of any finite number of monomer clusters consisting of consecutive monomers. We obtain the following result.

\proclaim{Theorem \Tif} $(${\rm a}$)$. Let $\alpha_1,\dotsc,\alpha_s>0$ and $\gamma_1,\dotsc,\gamma_s>0$ be fixed real numbers with $\alpha_1+\cdots+\alpha_s<1$. Let $(k_n^{(1)})_{n\geq1},\dotsc,(k_n^{(s)})_{n\geq1}$ and $(p_n^{(1)})_{n\geq1},\dotsc,(p_n^{(s)})_{n\geq1}$ be sequences of positive integers so that 
$\left|\alpha_i-\frac{k_n^{(i)}}{n}\right|<\frac{1}{n^{1+\frac{1}{2s}}}$ and 
$\left|\gamma_i-\frac{p_n^{(i)}}{n}\right|<\frac{1}{n^{1+\frac{1}{2s}}}$, for $i=1,\dotsc,s$ 
$($the existence of such sequences is a consequence of Lemma ${\Thm}$$)$.

Then if 
$$
G_n=AR_{2n,2n+2p_n^{(1)}+\cdots+2p_n^{(s)}}(C_{k_n^{(1)},p_n^{(1)}},C_{k_n^{(1)}+p_n^{(1)}+k_n^{(2)},p_n^{(2)}},\dotsc,C_{k_n^{(1)}+p_n^{(1)}+\cdots+k_n^{(s-1)}+p_n^{(s-1)}+k_n^{(s)},p_n^{(s)}}),
$$
we have that
$$
\spreadlines{4\jot}
\align
&
\frac{1}{|V(G_n)|}\ln\M(G_n)
=
\frac14\ln 2 
+\frac18\sum_{i,j\in I,\, i<j}\epsilon_{i,j}|i-j|^2\ln|i-j|+O\left(\frac{1}{n^{1+\frac{1}{2s}}}\right),
\tag\Eize
\endalign
$$
where $I=\{0,\alpha_1,\alpha_1+\gamma_1,\dotsc,\alpha_1+\cdots+\alpha_s+\gamma_1+\cdots+\gamma_{s-1},\alpha_1+\cdots+\alpha_s+\gamma_1+\cdots+\gamma_s,1+\gamma_1+\cdots+\gamma_s\}$, and the sign $\epsilon_{i,j}$ is given by $(\Eiv)$. 

$(${\rm b}$)$. For any fixed $\gamma_1,\dotsc,\gamma_s>0$, the $2s$-variable analog of the function $F$ of $(\Eizb)$ has a unique maximum $(\alpha_1^0,\dotsc,\alpha_s^0)$ in the domain $\alpha_1>0,\dotsc,\alpha_s>0$, $\alpha_1+\cdots+\alpha_s<1$, and this gives the equilibrium position of the $s$ bars of charge in the sense of Corollary ${\Tie}$. \epf

\endproclaim

%* k consecutive-monomer-clusters

\mysec{10. Concluding remarks} 

In this paper we studied in detail the interaction of monomer and separation defects in Aztec rectangles in the case when they are situated on the horizontal symmetry axis. The results we found can be phrased in terms of simple quantities specifying the location of the defects, such as the number of sites to their left or right, and the distance between the defects relative to the width of the Aztec rectangle. A natural question concerns the case when the monomers and separations have arbitrary positions. Of special interest would be the form of the asymptotic formulas corresponding to the general situation, in particular the quantities corresponding to the number of sites to the left or right of a defect, and to the inter-defect distance relative to width.

In Section 9 we saw that there is a unique equilibrium position for a finite number of ``bars of charge'' (i.e., clusters consisting of contiguous monomers). It would be interesting to study this problem in the case when the monomers are not bound together in clusters, but free to move anywhere on $\ell$, and to understand the shape of the ``globular cluster'' arising in the scaling limit.

Another interesting future direction is to phrase in the present context the analog of the Baik et al (see \cite{\Baiktwo}) problem of probability distributions of fixed parallel lozenges along a lattice line, in lozenge tilings of hexagons on the triangular lattice. The new ingredient is that now there are two kinds of charges (corresponding to monomers and separations), and this should lead to a more complex behavior.

%* Interesting problem on bott. p.21, c.XIV. 

%\newpage

\mysec{References}
{\openup 1\jot \frenchspacing\raggedbottom
\roster

\myref{\Baiktwo}
  J. Baik, T. Kriecherbauer, K. McLaughlin and P. Miller, ``Discrete orthogonal polynomials. Asymptotics and applications,'' {\it Annals of Math Studies,} Princeton University Press, Princeton, NJ, 2007.

%\myref{\ri}
%  M. Ciucu, Rotational invariance of quadromer correlations on the hexagonal lattice, {\it Adv. in Math.} {\bf 191} 
%(2005), 46-77.

\myref{\sc}
  M. Ciucu, A random tiling model for two dimensional electrostatics, {\it Mem. Amer. Math. Soc.} {\bf 178} (2005),
no. 839, 1--106.

%\myref{\ppone}
%  M. Ciucu, A random tiling model for two dimensional electrostatics, {\it Mem. Amer. Math. Soc.} {\bf 178} (2005),
%no. 839, 107--144.

\myref{\ec}
  M. Ciucu, The scaling limit of the correlation of holes on the triangular lattice with periodic boundary 
conditions, {\it Mem. Amer. Math. Soc.} {\bf 199} (2009), 1--100.

\myref{\ef}
  M. Ciucu, The emergence of the electrostatic field as a Feynman sum in random tilings with holes, {\it  Trans. Amer. Math. Soc.} {\bf 362}  (2010), 4921--4954.

\myref{\ov}
  M. Ciucu, Dimer packings with gaps and electrostatics, {\it Proc. Natl. Acad. Sci. USA}  {\bf 105}  (2008),  no. 8, 2766--2772.

\myref{\gd}
  M. Ciucu, The interaction of collinear gaps of arbitrary charge in a two dimensional dimer system, {\it Comm. Math. Phys.}, accepted, in press.

\myref{\CEP}
  H. Cohn, N. Elkies, and J. Propp, Local statistics for random domino tilings of the 
Aztec diamond, {\it Duke Math. J.} {\bf 85} (1996), 117-166.

%\myref{\CKP}
%  H. Cohn, R. Kenyon, and J. Propp, A variational principle for domino tilings, {\it J. Amer. Math. Soc.}  {\bf 14}
%(2001), 297--346

%\myref{\CLP}
%  H. Cohn, M. Larsen, and J. Propp, The shape of a typical boxed plane partition, 
%{\it New York J. of Math.} {\bf 4} (1998), 137--165.
%

\myref{\FS} 
  M. E. Fisher and J. Stephenson, Statistical mechanics of dimers on a plane 
lattice. II. Dimer correlations and monomers, {\it Phys. Rev. (2)} {\bf 132} (1963),
1411--1431.

\myref{\MF} 
  M. E. Fisher, {\it Private Communication.}

%\myref{\FisherIsing}
%  M. E. Fisher, On the dimer solution of planar Ising models, {\it J. Math. Phys.} {\bf 7} (1966), 1776--1781.

\myref{\Glaish}
  J. W. L. Glaisher, On Certain Numerical Products in which the Exponents Depend Upon the Numbers, {\it  Messenger Math.} {\bf 23} (1893), 145--175.

%\myref{\QED} 
%  R. P. Feynman, ``QED: The strange theory of light and matter,'' Princeton University Press, Princeton, New Jersey, 
%1985.

%\myref{\Hart} 
%  R. E. Hartwig, Monomer pair correlations, {\it J. Mathematical Phys.} {\bf 7}
%(1966), 286--299.

%\myref{\SM}
%  W. Krauth and  R. Moessner, Pocket Monte Carlo algorithm for classical doped dimer models,
%{\it Physical Review B} {\bf 67} (2003), 064503.

%\myref{\MRR}
%  W. H. Mills, D. P. Robbins, and H. Rumsey, Alternating sign matrices and
%descending plane partitions, {\it J. Comb. Theory Ser. A} {\bf 34} (1983), 
%340--359.

%\myref{\ZI}
%  J. B. Zuber and C. Itzykson, Quantum field theory and the two-dimensional Ising model, {\it Phys. Rev. D} {\bf 15} (1977), 2875--2884.

\myref{\KOS}
  R. Kenyon, A. Okounkov and S. Sheffield, Dimers and amoebae, {\it Ann. of Math.(2)}  163, no. 3 (2006), 
1019--1056.

\myref{\Olver}
  F. W. J. Olver, Asymptotics and special functions, Academic Press, New York, 1974.

\endroster\par}

\enddocument